\documentclass[final,3p,times]{elsarticle}

\usepackage{amssymb}
\usepackage{amsthm}

\journal{Physic Reports}
\biboptions{sort&compress}

\begin{document}

\begin{frontmatter}



\title{Mesoscopic coherence in light scattering from cold, optically dense\\ and disordered atomic systems}


\author{D.V.~Kupriyanov}\ead{kupr@dk11578.spb.edu} \author{I.M.~Sokolov}

\address{Department of Theoretical Physics, St.-Petersburg State Polytechnic University\\ 195251, St.-Petersburg, Russia}

\author{M.D.~Havey}

\address{Department of Physics, Old Dominion University, Norfolk, VA 23529}

\begin{abstract}
Coherent effects manifested in light scattering from cold, optically dense and disordered atomic systems are reviewed from a primarily theoretical point of view. Development of the basic theoretical tools is then elaborated through several physical atomic physics based processes which have been at least partly explored experimentally. These include illustrations drawn from the coherent backscattering effect, random lasing in atomic gases, quantum memories and light-atoms interface assisted by the light trapping mechanism. Current understanding and challenges associated with the transition to high atomic densities and cooperativity in the scattering process is also discussed in some detail.
\end{abstract}


\end{frontmatter}

$\copyright$ 2017.  This manuscript version is made available under the CC-BY-NC-ND 4.0 License.

\tableofcontents


\section{Introduction}\label{Section I}
Cold atomic systems are quite challenging and promising objects for research in quantum optics and atomic physics.  This circumstance grows broadly out of a wide range of applications and fundamental explorations that can be found in such research areas as atomic and molecular spectroscopy, metrology, quantum simulations, and quantum information science.  Each of these subjects in itself is worthy of one or more comprehensive reviews.  The optical responses of non-linear coherent controlled media were the subject of intensive discussions a decade ago \cite{Fleischhauer}. At present the light-matter interface resulting in quantum memories, quantum entanglement, and quantum non-demolition measurements forms the background for various challenging implications of the physical protocols based on coherently controlled and optically dense atomic systems \cite{PolSorHam,Simon}. Surprisingly, but practically independently of this, the various coherent mechanisms affecting light diffusion, with emphasis on the role of disorder in light transport and in context of localization phenomena and wave dynamics in random media have been intensively studied during the last two decades \cite{KSSH06,Labeyrie08,MullerDelande}.

In this review we are motivated to follow how these frontier research directions studying the same physical objects, but sometime with an aim towards different goals, have a natural tendency for integrating the various physical ideas and approaches. Indeed, in nearly all aspects research in the above areas depends on interaction of the physical systems with electromagnetic radiation, either to prepare the system, to interrogate it, or in the case of several mutually coherent applied fields to manipulate the bare physical properties for possible practical applications.  Even in the least complex case of a single very weak field, or a single photon resonantly probing a gas of two state systems, there are subtleties to the processes that can occur, these being reflected in the collective and cooperative responses of the system. For this reason, we have been motivated to review the essential theoretical and experimental tools that have been used, or developed, over the past two decades to study collective and cooperative effects in cold and essentially atomic gases.  In this, although principally focus on theoretical approaches here, we have at the same time, and in most cases, stayed close to physical phenomena which have been observed in the laboratory. It is intriguing that the discussed phenomena may partly have classical precursors but the intrinsic quantum nature of atomic systems and the interaction process dramatically modify their observable behavior as well as the dependencies on external parameters. These include the coherent backscattering effect, random lasing, manipulation of light diffusion through electromagnetically induced transparency, and quantum memories based on diffusive light transport.

Another area of motivation comes from recognizing that an atomic gas (even if, as in our discussion, not quantum degenerate) at low temperatures, combined with interacting coherent fields, is essential a many body physics problem; the general optical properties of aggregated and mesoscopically scaled matter  differ significantly from those of individual atomic scatterers.  In the case of a very dilute gas of atoms the response of the system to applied and weak resonant, or nonresonant, monochromatic electromagnetic radiation can be relevantly understood via the entire Green's function formalism. This eventually states a self-consistent macroscopic approach based on Maxwell theory and on the concept of subsequent multiple scattering from spatial inhomogeneities.

However, upon increasing the atomic density, the quantum optics of such systems modifies such a classical vision and develops correlated characteristics in the interaction process. This is especially true in cold and nearly motionless atomic gases, where coherence shared between applied and scattered fields, and the atoms comprising the samples, can survive environmental or thermal decoherence for substantial periods of time. We are then also interested in the evolution of the quantum optics of a cold atomic gas as the physical density of the gas increases, up to the point where strong atom-atom interactions can dominate the physics. The crucial peculiarity of the light-atom interaction in this case is that both the far field radiative coupling and near field static interaction mediate the cooperative dynamics of an atomic ensemble. Remarkable progress of the past decades in developing numerical methods and approaches in quantum electrodynamics gives us a tool for exploiting the microscopic simulations for light scattering from atomic samples of macroscopic size. The last section of this review is dedicated in part to aspects of this subject.

\section{Basic approaches, assumptions and definitions}\label{Section II}
\setcounter{equation}{0}
\setcounter{figure}{0}

\subsection{The long-wavelength dipole approximation}\label{Section II.A}
Most of the calculations of interaction processes of the electromagnetic field with atomic systems in the optical domain are performed in the dipole approximation. The dipole approximation can be rigorously introduced via the Power, Zienau and Woolley canonical/unitary gauge transformation \cite{PowerZienauWoolley,Woolley}. For practical applications the general and rather cumbersome formalism can be simplified with the assumption that the atomic size is much smaller than a typical wavelength of the field modes actually contributing to the interaction process. Such a long-wavelength dipole approximation, see \cite{ChTnDRGr} for derivation details, leads to the following interaction Hamiltonian for an atomic system consisting of $N$ dipole-type scatterers interacting with the quantized electromagnetic field
\begin{eqnarray}
\hat{H}_{\mathrm{int}}&=&-\sum_{a=1}^{N}%
\hat{\mathbf{d}}^{(a)}\hat{\mathbf{E}}(\mathbf{r}_a)\;+\;\hat{H}_{\mathrm{self}},%
\nonumber\\%
\hat{H}_{\mathrm{self}}&=&\sum_{a=1}^{N}\frac{2\pi}{{\cal V}}\sum_{s}\left(\mathbf{e}_s\cdot\hat{\mathbf{d}}^{(a)}\right)^2.%
\label{2.1}%
\end{eqnarray}
The first and most important term is characteristically interpreted as the interaction of an $a$-th atomic dipole $\mathbf{d}^{(a)}$ with the electric field $\hat{\mathbf{E}}(\mathbf{r})$ at the point of the dipole location. However, as strictly defined in the dipole gauge, the latter quantity performs the microscopic displacement field and it can be expressed by a standard expansion in the basis of plane waves $s\equiv{\mathbf{k},\mathbf{e}_{k\alpha}}$, where $\alpha=1,2$ enumerates two orthogonal transverse polarization vectors for each $\mathbf{k}$, which we assume here as linear and real
\begin{eqnarray}
\hat{\mathbf{E}}(\mathbf{r})&\equiv&\hat{\mathbf{E}}^{(+)}(\mathbf{r})\;%
+\;\hat{\mathbf{E}}^{(-)}(\mathbf{r})%
\nonumber\\%
&=&\sum_{s}\left(\frac{2\pi\hbar\omega_s}{{\cal V}}\right)^{1/2}%
\left[i\mathbf{e}_s a_s\,\mathrm{e}^{i\mathbf{k}_s\mathbf{r}}%
-i\mathbf{e}_s a_s^{\dagger}\,\mathrm{e}^{-i\mathbf{k}_s\mathbf{r}}\right]%
\nonumber\\%
&=&\hat{\mathbf{E}}_{\bot}(\mathbf{r})+\sum_{b=1}^{N}\frac{4\pi}{{\cal V}}%
\sum_{s}\mathbf{e}_s(\mathbf{e}_s\cdot\hat{\mathbf{d}}^{(b)})\,%
\mathrm{e}^{i\mathbf{k}_s(\mathbf{r}\,-\,\mathbf{r}_b)}.%
\label{2.2}%
\end{eqnarray}%
Here $a_s$ and $a_s^{\dagger}$ are the annihilation and creation operators for the $s$-th field mode and the quantization scheme includes periodic boundary conditions on the quantization volume ${\cal V}$. The third line shows the difference between the actual transverse electric field denoted as $\hat{\mathbf{E}}_{\bot}(\mathbf{r})$ and the displacement field. The second term in Eq.(\ref{2.1}) appears as a non-converging self energy of the dipoles. This term is often omitted in the interaction Hamiltonian since it does not principally affect the dipole dynamics, particularly when the difference between the transverse electric and displacement fields is small. The vector potential operator is given by
\begin{eqnarray}
\hat{\mathbf{A}}(\mathbf{r})&\equiv&\hat{\mathbf{A}}^{(+)}(\mathbf{r})\;%
+\;\hat{\mathbf{A}}^{(-)}(\mathbf{r})%
\nonumber\\%
&=&\sum_{s}\left(\frac{2\pi\hbar c^2}{\omega_s{\cal V}}\right)^{1/2}%
\left[\mathbf{e}_s a_s\,\mathrm{e}^{i\mathbf{k}_s\mathbf{r}}\,%
+\,\mathbf{e}_s a_s^{\dagger}\,\mathrm{e}^{-i\mathbf{k}_s\mathbf{r}}\right]%
\label{2.3}%
\end{eqnarray}
and this operator is not modified by the transformation from the Coulomb to the dipole gauge, see
\cite{ChTnDRGr}.

Let us make a few remarks concerning the applicability of the dipole approach and how it is related to the level of disorder in the atomic system. The difference between the transverse electric and displacement fields becomes important at distances comparable with the atomic size. For a low density atomic ensemble, when the atoms are separated by a radiation zone such that $n_0\lambdabar^3\ll 1$, $n_0$ is the density of atoms, $\lambdabar=\lambda/2\pi$ and $\lambda$ is the light wavelength, the difference between the above quantities in the interaction operator can be safely ignored. In terms of disorder, classification of such a dilute atomic gas performs a homogeneous and slightly disordered configuration of the dipole scatterers, where the transport length for light diffusion $l_{\mathrm{tr}}$ is much longer than the light bar-wavelength $\lambdabar$. The description of the light transport in such a medium can be explicitly described in terms of the Green's function formalism and in the self-consistent macroscopic Maxwell theory approach.

An alternate situation occurs when $n_0\lambdabar^3> 1$ brings the system to the strongly disordered regime defined through the Ioffe-Regel condition $l_{\mathrm{tr}}\sim\lambdabar$, which was originally pointed out in \cite{IoffeRegel}, and which we simplify here for the particular case of cold atomic systems. For such a dense configuration the exact definitions (\ref{2.1}), (\ref{2.2}) become important. One should be quite careful with substituting (\ref{2.2}) to (\ref{2.1}) and then to any operator equations because any atomic variable commutes with the operator of the displacement field but not always with the operator of the transverse electric field. In a dense system the self-energy term cannot be ignored and in some calculational schemes can manifest itself in the dynamics of atomic variables in cooperation with another singular term, namely, with the self-contact dipole interaction. The latter appears when $\mathbf{r}=\mathbf{r}_a=\mathbf{r}_b$ for interaction of a specific $a$-th dipole in (\ref{2.1}) with the longitudinal field created by the same dipole in the second term of the last line in Eq.(\ref{2.2}).

Throughout our discussion, oriented towards existing experimental capabilities, we will mostly deal with the dilute atomic configurations for which the above difficulties can be smoothed and are not actually important. But we come back to the problem in the last part of this review in Section \ref{Section V}, where we consider the situation with a large degree of disorder, cooperativity and local field effects. In \ref{Appendix.A} we have also discussed some facets of the dipole gauge with respect to atomic subsystems.

\subsection{The light correlation function}

The intrinsic characteristics of light associated with its energy transfer and coherence are expressed by its correlation function
\begin{equation}
D_{\mu\nu}^{(E)}(\mathbf{r},t;\mathbf{r}',t')\ =\ \left\langle \hat{E}_{\nu}^{(-)}(\mathbf{r}',t')\,%
\hat{E}_{\mu}^{(+)}(\mathbf{r},t)\right\rangle.%
\label{2.4}%
\end{equation}
Here the angle brackets denote the statistical averaging of the Heisenberg field operators $E_{\mu}^{(+)}(\mathbf{r},t)$ and $E_{\nu}^{(-)}(\mathbf{r}',t')$, with frequency and polarization components given by (\ref{2.2}). Having in mind, in this section, mostly a dilute atomic system with low density, we shall consider this correlation function similarly defined both outside and inside the medium and associate the field operators with the transverse electric field components. The rigorous and detailed analysis presented by Glauber in his seminal papers \cite{Glauber} introduces the general set of correlation functions of higher orders, which are measurable by coincidence spectroscopy and by homodyne detection techniques. Here we restrict our consideration only to the first order correlation function defined by Eq.(\ref{2.4}).

If the optical field locally performs a stationary plane wave freely propagating in the direction $\mathbf{n}$ then the spectral energy distributed with it can be expressed by a Fourier transform over the time difference $\tau=t-t'$
\begin{eqnarray}
I(\mathbf{n},\mathbf{r},\omega)&=&\frac{c}{2\pi}\sum_{\mu=x,y}%
\int_{-\infty}^{\infty}d\tau\,\mathrm{e}^{i\omega\tau}
\; D_{\mu\mu}^{(E)}(\mathbf{r},\bar{t}+\tau/2;\mathbf{r},\bar{t}-\tau/2).%
\label{2.5}%
\end{eqnarray}
Here $\bar{t}$ is understood as a time argument dependence which vanishes under steady state conditions. For time dependent processes it is more convenient and natural to introduce the Poynting vector associated with the light energy transported in the direction $\mathbf{n}$. This can be done via tracing the correlation function over its polarization components $\mu=x,y$ in the plane orthogonal to $\mathbf{n}$ and considering it at coincident spatial points and times.
\begin{equation}
I(\mathbf{n},\mathbf{r},t)\ =\ \frac{c}{2\pi}\sum_{\mu=x,y}%
D_{\mu\mu}^{(E)}(\mathbf{r},t;\mathbf{r},t).%
\label{2.6}%
\end{equation}
In some situations, see Section \ref{2.F.1} below, these two representations can be mixed with the Wigner-type density matrix, which considers the instantaneous part of the field energy flux for a certain part of its spectrum as distributed near the point $\mathbf{r}$ and propagating in the direction $\mathbf{n}$. The relevant expression for the combined quantity $I(\mathbf{n},\mathbf{r},\omega,t)$ with $\bar{t}\to t$ would be given by Eq.(\ref{2.5}), preserving time dependence in the right hand side.

Even in a dilute but optically dense medium randomly scattering light over all directions, the original incident plane light wave will be fractured into many packets. These fragments will propagate and be re-scattered by atoms more or less independently. Following such a natural randomization, the light propagation would be visualized as a diffusion process and at any spatial point at a certain time there would be contributions (\ref{2.6}) arriving from any directions. This process would develop incoherently and is expected to be described by a diffusion type master equation for the local light energy. In applications and in the literature there are numerous examples of light transport master equations, see \cite{Chandrasekhar,Rossum}, which operate with the time dependent local intensity. A general feature of the diffusive transport is that the initial coherent properties of light seem completely lost and the emerging waves would be unpolarized and incoherent such that only the outgoing energy flux would be a physically important characteristic for the diffusion process. However, as we will see, without thermal losses this naive point of view cannot be exactly true and there exist manifestations of the wave nature and coherence in the multiple scattering process.

\subsection{The Green's function formalism}

Convenient expansions for the correlation function (\ref{2.4}) can be subsequently built up via its perturbation theory analysis. Let us transform the original definition (\ref{2.4}), performed in the Heisenberg representation, to the interaction representation
\begin{equation}
D_{\mu\nu}^{(E)}(\mathbf{r},t;\mathbf{r}',t')\ =\ \left\langle%
\tilde{T}\!\left[\!S^{\dagger}\hat{E}_{\nu}^{(0,-)}(\mathbf{r}',t')\!\right]%
T\!\left[\!S\hat{E}_{\mu}^{(0,+)}(\mathbf{r},t)\!\right]\right\rangle,%
\label{2.7}%
\end{equation}
where the chronological operators $T$ and $\tilde{T}$ respectively order and anti-order in time the field and atomic operators in the brackets. The evolution operator $S$ is given by
\begin{equation}
S\ =\ S(\infty,-\infty)\ =\ T\exp\left[-\frac{i}{\hbar}\int_{-\infty}^{\infty}\hat{H}_{\mathrm{int}}^{(0)}(t)\,dt\right]%
\label{2.8}.%
\end{equation}
Here and below we superscript by "$0$" the time-dependent operators in the interaction representation. The important step in introducing the representation (\ref{2.7}) is that initial conditions are moved to the infinite past as is typically assumed in scattering theory. That allows us to approach and specify Eq.(\ref{2.7}) by the limits $t,t'\to -\infty$ where it reproduces the correlation function of the incident light.

Further perturbation theory expansion of (\ref{2.7}) can be done in the second-quantized formalism introduced for both the field and atomic subsystems. Then the first term in the interaction Hamiltonian (\ref{2.1}), expressed by atomic second quantized $\Psi$-operators, see \cite{LaLfIII}, and considered in the interaction representation transforms to the following form
\begin{eqnarray}
\hat{H}_{\mathrm{int}}^{(0)}(t)\!&=&\!-\sum_{n,m}\int d^3\!r\;%
d^{\mu}_{nm}\;\hat{E}_{\mu}^{(0)}(\mathbf{r},t)\;%
\hat{\Psi}_{n}^{(0)\dagger}(\mathbf{r},t)\;\hat{\Psi}_{m}^{(0)}(\mathbf{r},t)%
\ +\ \ H.c.,%
\label{2.9}%
\end{eqnarray}
where, for the sake of generality, we use covariant notation for the tensor indices in the dipole matrix element and for the electric field components.\footnote{We omit here and further (where it is not confusing with the detailed notations) the symbols of invariant sums over the tensor indices.} The $\Psi$-operators for the set of the ground states $m$ are given by
\begin{eqnarray}
\hat{\Psi}_{m}^{(0)}(\mathbf{r},t)&=&\sum_{\mathbf{p}}\frac{1}{\sqrt{\cal V}}\;%
\mathrm{e}^{\frac{i}{\hbar}\mathbf{p}\cdot\mathbf{r}}\, b_{\mathbf{p}m}(t)%
\nonumber\\%
\hat{\Psi}_{m}^{(0)\dagger}(\mathbf{r},t)&=&\sum_{\mathbf{p}}\frac{1}{\sqrt{\cal V}}\;%
\mathrm{e}^{-\frac{i}{\hbar}\mathbf{p}\cdot\mathbf{r}}\, b_{\mathbf{p}m}^{\dagger}(t)%
\label{2.10}%
\end{eqnarray}
Here $b_{\mathbf{p}m}$ and $b_{\mathbf{p}m}^{\dagger}$ are respectively the annihilation and creation operators of an atom with momentum $\mathbf{p}$ in the internal Zeeman quantum state $m$. The $\Psi$-operators in the excited states $n$ can be similarly defined by trivial substitution of $m\to n$. In the case of near resonant interaction one can smooth the time derivation and accept
\begin{eqnarray}
\hat{\mathbf{E}}^{(0,+)}(\mathbf{r},t)&\to&%
+\frac{i\omega_0}{c}\,\hat{\mathbf{A}}^{(0,+)}(\mathbf{r},t)%
\nonumber\\%
\hat{\mathbf{E}}^{(0,-)}(\mathbf{r},t)&\to&%
-\frac{i\omega_0}{c}\,\hat{\mathbf{A}}^{(0,-)}(\mathbf{r},t)%
\label{2.11}%
\end{eqnarray}
such that each frequency component of the electric field operator can be expressed by the similar component of the vector potential operator. The frequency $\omega_0$ should be associated here with an atomic resonance frequency, which can vary for each particular excitation transition. Also applying the standard assumptions of the rotating wave approximation (RWA) reduces the interaction Hamiltonian to the following form
\begin{eqnarray}
\hat{H}_{\mathrm{int}}^{(0)}(t)\!\!&\approx &\!\!-i\frac{\omega_0}{c}\!\sum_{n,m}\!\int\!\! d^3\!r\;%
d^{\mu}_{nm}\;\hat{A}_{\mu}^{(0,+)}\!(\mathbf{r},t)\;%
\hat{\Psi}_{n}^{(0)\dagger}\!(\mathbf{r},t)\;\hat{\Psi}_{m}^{(0)}\!(\mathbf{r},t)\ +\ \ H.c.,%
\label{2.12}%
\end{eqnarray}
which keeps only counter-rotating Heisenberg terms in the field and atomic subsystems.

Straightforward expansion of the evolution operators creates the operators' products, which can be always realigned to the N-ordered form, when the creation and annihilation operators are moved to the left and right sides of the averaging expression respectively. There are several types of Green's functions created in this procedure, which are associated with the different chronological ordering in the transposed operators. We will use the following convention, see \cite{Keldysh,LfPtX} for further details. If, as in the example of Eq.(\ref{2.7}), an operator contributes from the left brackets (generated by expansion of $S^{\dagger}$ and $\tilde{T}$-ordered), we associate it with the "$+$" sign and we alternatively associate it with the "$-$" sign if it contributes from the right brackets (generated by expansion of $S$ and $T$-ordered). Final transformation of any operator product to its $N$-ordered form is known as Wick's Theorem \cite{Wick,LfPtX} and, in an example of the field subsystem, one obtains the following set of Green's functions \begin{eqnarray}
iD_{\mu_1\mu_2}^{(\sigma_1\sigma_2)}(\mathbf{r}_1,t_1;\mathbf{r}_2,t_2)&=&T_{\sigma_1\sigma_2}%
\left[\hat{A}_{\mu_1}^{(0)}(\mathbf{r}_1,t_1)\,\hat{A}_{\mu_2}^{(0)}(\mathbf{r}_2,t_2)\right]%
\;-\;N\left[\hat{A}_{\mu_1}^{(0)}(\mathbf{r}_1,t_1)\,\hat{A}_{\mu_2}^{(0)}(\mathbf{r}_2,t_2)\right]%
\nonumber\\%
&&\nonumber\\%
&\Rightarrow&\begin{array}{c}\scalebox{1.0}{\includegraphics*{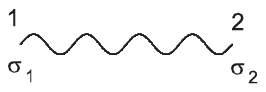}}\end{array}%
\label{2.13}%
\end{eqnarray}
where $\sigma_1,\sigma_2$ equals either $+$ or $-$ are the sign indicators showing the original location of the field operators and $T_{\sigma_1\sigma_2}$ is the chronological operator associated with this location. That stands for that $T_{--}=T$, $T_{++}=\tilde{T}$, $T_{+-}$ is the identical operator and $T_{-+}$ is the transposition operator. In the rotating wave approximation only positive frequency components of the Green's functions (\ref{2.13}) can appear for any chronological ordering, and these components can be specified in a diagram by the arrow on each wavy line directed from one end associated with the creation event to the other end associated with the annihilation event.

For the atomic subsystem we keep only the Gaussian type correlations for any product consisting of any number of the atomic $\Psi$-operators. This approximation is often fulfilled because of the macroscopic nature of the atomic subsystem. Then the atomic Green's functions are expressed by all possible chronological couplings between creation and annihilation operators
\begin{eqnarray}
iG_{m_1m_2}^{(\sigma_1\sigma_2)}(\mathbf{r}_1,t_1;\mathbf{r}_2,t_2)&=&%
\left\langle T_{\sigma_1\sigma_2}\hat{\Psi}_{m_1}^{(0)}\!(\mathbf{r}_1,t_1)\,%
\hat{\Psi}_{m_2}^{(0)\dagger}\!(\mathbf{r}_2,t_2)\right\rangle%
\nonumber\\%
&&\nonumber\\%
&\Rightarrow&\begin{array}{c}\scalebox{1.0}{\includegraphics*{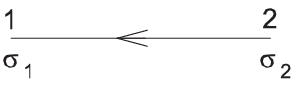}}\end{array}%
\label{2.14}%
\end{eqnarray}
It is important that any transposition of the $\Psi$-operators obeys the rule of quantum statistics, such that it should change the sign in the case of Fermionic operators. This permutation rule normally manifests itself for loop-type diagrams by an extra minus sign for each loop in the Fermionic case. Let us also point out that the applied Gaussian factorization is a relevant approximation only for a non-degenerate quantum gas and it fails in the case of quantum degeneracy, for example, in the case of the Bose-Einstein condensate.

The exact analytical expressions of the complete set of the original unperturbed Green's functions (\ref{2.13}) and (\ref{2.14}) for the field and atomic subsystems are given in \ref{Appendix.B}.

\subsection{Graphical images of the basic interactions and important diagram
blocks}\label{2.D}

\subsubsection{General analysis: The field subsystem}\label{2.D.1}
There are two types of interaction diagrams between light and atoms. If part of
the light is originally in a classical mode, being a superposition of the
eigenstates of the field annihilation operators (\ref{2.3}), then it generates
the following interaction diagrams
\begin{eqnarray}
\pm\frac{i}{\hbar}\,{d_{nm}^\mu}\,{\cal E}_{\mu}^{(+)}(\mathbf{r},t)\ \ %
&\Leftrightarrow\ \ &%
\raisebox{-0.5 cm}{\scalebox{1.0}{\includegraphics*{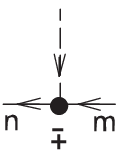}}}%
\nonumber\\%
\pm\frac{i}{\hbar}\,{d_{mn}^\mu}\,{\cal E}_{\mu}^{(-)}(\mathbf{r},t)\ \ %
&\Leftrightarrow\ \ &%
\raisebox{-0.5 cm}{\scalebox{1.0}{\includegraphics*{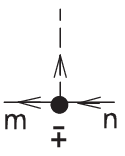}}}%
\label{2.15}%
\end{eqnarray}
The upper graph describes the excitation transition from the ground state $|m\rangle$ to the excited state $|n\rangle$ initiated by the positive frequency component of the classical field ${\cal E}_{\mu}^{(+)}(\mathbf{r},t)$. The lower graph respectively describes the induced transition from the excited to the ground state initiated by the negative frequency component ${\cal E}_{\mu}^{(-)}(\mathbf{r},t)$. In those cases when a classical field is applied on empty transitions, which can be involved in the excitation process with other fields, we will refer to it as a control mode.

Similar processes initiated by the quantized field are expressed by the diagrams
\begin{eqnarray}
\mp\frac{\omega_0}{\hbar c}\,{d_{nm}^\mu}\ \ %
&\Leftrightarrow&%
\raisebox{-0.5 cm}{\scalebox{1.0}{\includegraphics*{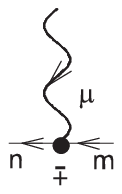}}}%
\nonumber\\%
\pm\frac{\omega_0}{\hbar c}\,{d_{mn}^\mu}\ \ %
&\Leftrightarrow&%
\raisebox{-0.5 cm}{\scalebox{1.0}{\includegraphics*{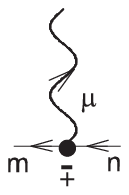}}}%
\label{2.16}%
\end{eqnarray}
In this diagram the short wavy line can be an edge fragment of the propagation function, given by (\ref{2.13}), but can be also associated with the correlation function of the external field $\emph{i.e.}$ with the light wave incident on the medium. In the latter case we will normally assume this field to be weak and in some examples even classical. Nevertheless, to distinguish it from the field (\ref{2.15}) which, in turn, will be mostly considered as a strong control mode unaffected by the medium, we will display such a probe as a short wavy line. The probe mode can be strongly modified by the medium and can be generated by the processes of spontaneous emission.

If the probe field originally performs a coherent pulse propagating through the atomic sample, then the dynamics of its coherent component is described by the following graphical equation
\begin{eqnarray}
\langle \hat{A}_{\mu}^{(+)}(\mathbf{r},t)\rangle&=&\left\langle S^{\dagger}%
T\left[S\hat{A}_{\mu}^{(0,+)}(\mathbf{r},t)\right]\right\rangle%
\nonumber\\%
&\Rightarrow&\raisebox{-2 cm}{\scalebox{1.0}{\includegraphics*{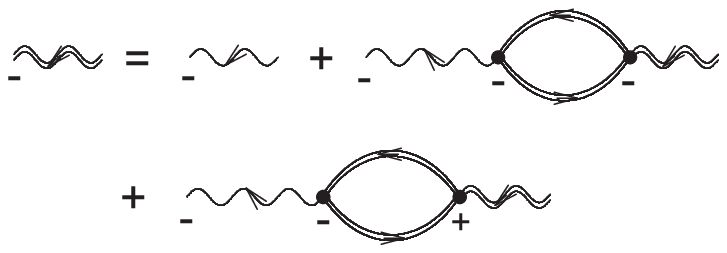}}}%
\label{2.17}%
\end{eqnarray}
Here, and further in the diagrams, we will omit evident specification of the atomic states and field polarizations. The self-energy part of this diagram performs the retarded-type polarization operator, and the double lines express the "dressed" atomic propagators (Green's functions), which accumulated all the interaction processes. Let us also point out that under the rotating wave approximation and in a low density limit normally there is no vertex-type correction in the structure of the polarization operator. One exception to this rule, associated with internal correlations in the coherent excitation process, will be discussed in Section \ref{3.C}. The fundamental solution of equation (\ref{2.17}) defines the positive frequency component of the retarded-type Green's function, which is given by
\begin{eqnarray}
i{\cal D}_{\mu_1\mu_2;+}^{(R)}(\mathbf{r}_1,t_1;\mathbf{r}_2,t_2)&=&\left\langle\left[\hat{A}_{\mu_1}^{(+)}(\mathbf{r}_1,t_1),\,\hat{A}_{\mu_2}^{(-)}(\mathbf{r}_2,t_2)\right]\right\rangle\,%
\theta(t_1-t_2)%
\nonumber\\%
&\Rightarrow&\begin{array}{c}\scalebox{1.0}{\includegraphics*{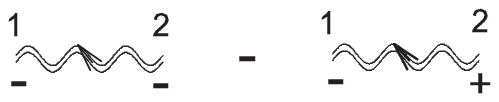}}\end{array}%
\nonumber\\%
&=&\begin{array}{c}\scalebox{1.0}{\includegraphics*{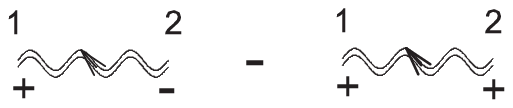}}\end{array}%
\label{2.18}%
\end{eqnarray}
where we put an index of the frequency component in the subscript line. This function performs the averaged commutator of the complete Heisenberg operators of the quantized field. Its graphical definition shows that this function can be expressed by the original vacuum lines (\ref{2.13}) subsequently "dressed" by the interactions generated in the expansion of the evolutionary operator. Any double wavy line can be interpreted as the relevantly ordered product of the dressed quantum field fluctuations. For example, the subtracted term in the second line of Eq.(\ref{2.18}) is the result of higher order interaction processes and it includes the contribution of incoherently scattered light as well as fluorescence associated with the processes of optical pumping, and with other non-linear processes initiated by the interacting fields. This term can be expressed analytically by the following product of the Heisenberg operators
\begin{eqnarray}
i{\cal D}_{\mu_1\mu_2;+}^{(-+)}(\mathbf{r}_1,t_1;\mathbf{r}_2,t_2)&=&%
\langle \hat{A}_{\mu_2}^{(-)}(\mathbf{r}_2,t_2)\,\hat{A}_{\mu_1}^{(+)}(\mathbf{r}_1,t_1)\rangle%
-\langle \hat{A}_{\mu_2}^{(-)}(\mathbf{r}_2,t_2)\rangle%
\langle\hat{A}_{\mu_1}^{(+)}(\mathbf{r}_1,t_1)\rangle%
\nonumber\\%
\nonumber\\%
&=&\left\langle\tilde{T}\!\left[\!S^{\dagger}\hat{A}_{\mu_2}^{(0,-)}(\mathbf{r}_2,t_2)\!\right]%
T\!\left[\!S\hat{A}_{\mu_1}^{(0,+)}(\mathbf{r}_1,t_1)\!\right]\right\rangle%
-\left\langle \tilde{T}\left[S^{\dagger}\hat{A}_{\mu_2}^{(0,-)}(\mathbf{r}_2,t_2)\right]S\right\rangle%
\left\langle S^{\dagger}T\!\left[\!S\hat{A}_{\mu_1}^{(0,+)}(\mathbf{r}_1,t_1)\!\right]\right\rangle%
\nonumber\\%
\nonumber\\%
&\Rightarrow&\raisebox{-0.5 cm}{\scalebox{1.0}{\includegraphics*{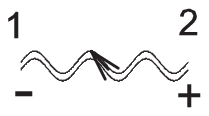}}}%
\label{2.19}%
\end{eqnarray}
With the acceptance of the relations (\ref{2.11}) this Green's function gives us the correlation function of the radiation mode diffusely propagating through the atomic medium. Other Green's functions, dressed by the interaction, can be similarly expressed accordingly to their ordering rules. The last line in Eq.(\ref{2.18}) indicates a useful symmetry relation for the retarded-type photon propagator.

If the probe field incident on the medium exists in a coherent state then equation (\ref{2.17}) is applicable in the most general case even if this field is strong. Then the last term of this equation legitimizes the situation when the upper energy levels of atoms can be significantly populated. Thus, this term becomes responsible for amplification of the propagating probe mode via induced emission and such an amplification manifests itself in the retarded propagator (\ref{2.18}) as well. In an optically dense gas the probe pulse will be immediately split into many scattered fragments, such that equation (\ref{2.17}) considered only for its forward propagation is actually insufficient. Nevertheless, further dynamics of each scattered fragment along a short free propagation segment between subsequent scattering events is also expressed by the retarded propagator (\ref{2.18}), $\emph{i.e.}$ by the fundamental solution of equation (\ref{2.17}). Amplification of the diffusely propagating light is a desirable scenario for the random lasing, which we will discuss in Section \ref{4.A} in more detail. We will further see from this and other examples that coherence is not completely lost in the light diffusion process.

\subsubsection{General analysis: The atomic subsystem}

We emphasize that the entire dynamics of the light subsystem should be considered together with the dynamics of the atomic subsystem and the atomic Green's functions (\ref{2.14}) are strongly modified by the interaction with the electromagnetic field. Let us define the following differential operator associated with free evolution of an atom
\begin{equation}
\hat G_{0j}^{-1}\ =\ i\frac{\partial}{\partial t_j} +\frac{\hbar}{2m_0}\triangle_j-\frac{1}{\hbar}E_j,%
\label{2.20}
\end{equation}
where $j$ denotes $\mathbf{r}_j,t_j$ as well as specifies the internal state $m_j,n_j\ldots$ with $j=1,2\ldots$; $m_0$ is the atomic mass, and $E_j$ denotes the energy of the specific atomic state in the definitions (\ref{2.10}), (\ref{2.14}). Then we can introduce the following graphical precursor of a kinetic equation for the atomic density matrix
\begin{eqnarray}
\lefteqn{\left(\hat G_{01}^{-1} - \hat G_{02}^{-1*}\right)\ %
\raisebox{-0.4 cm}{\scalebox{1.0}{\includegraphics*{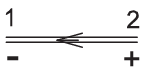}}}}%
\nonumber\\%
&\Rightarrow&\raisebox{-0.6 cm}{\scalebox{1.0}{\includegraphics*{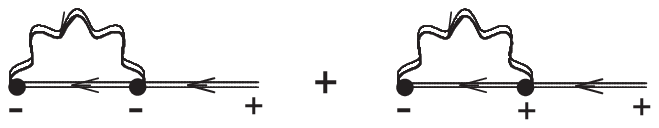}}}%
\nonumber\\%
&&\scalebox{1.0}{\includegraphics*{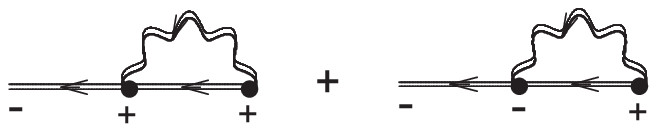}}%
\nonumber\\%
&+&\sum_{\mathrm{Coh}}\ \ \left(\raisebox{-0.5 cm}{\scalebox{1.0}{\includegraphics*{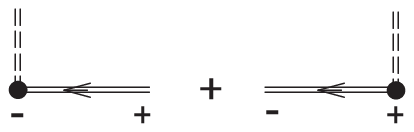}}}\right)%
\label{2.21}%
\end{eqnarray}
In this general diagrammatic form we accept any electromagnetic-type interactions coupling the atomic ground and excited states. The first line gives the free dynamics of the atoms, which is quantum mechanical for their internal degrees of freedom and assumed to be classical for their spatial motion. The next two lines describe the interaction with quantized modes and can include interactions with the scattered or trapped light. The doubly wavy lines express the correlation functions of the quantized modes dressed by interaction with the medium. The last line in Eq.(\ref{2.21}) symbolically indicates all possible interactions with external coherent modes, which can be optical and microwave and in some configurations these modes can be either depleted or modified by interaction with the medium. The inward or outward arrows on the dressed double dashed lines (not shown, but have to attribute them) depend on the concrete vertex type specified by the basic interaction diagrams (\ref{2.15}).

The single particle atomic density matrix $\rho_{m_1m_2}(\mathbf{p},\mathbf{r},t)$ is expressed by $G^{(-+)}$-type Green's functions, see (\ref{2.14}), dressed by the interaction processes
\begin{eqnarray}
i{\cal G}_{m_1m_2}^{(-+)}(\mathbf{r}_1,t_1;\mathbf{r}_2,t_2)&=& \pm\left\langle\hat{\Psi}_{m_2}^{\dagger}(\mathbf{r}_2,t_2)\;\hat{\Psi}_{m_1}(\mathbf{r}_1,t_1)\right\rangle%
\nonumber\\%
&=&\left\langle \tilde{T}\left[S^{\dagger}\hat{\Psi}_{m_2}^{(0)\dagger}(\mathbf{r}_2,t_2)\right]T\left[S\hat{\Psi}_{m_1}^{(0)}(\mathbf{r}_1,t_1)\right]\right\rangle%
\label{2.22}%
\end{eqnarray}
via the following Wigner-type transform
\begin{eqnarray}
i{\cal G}_{m_1m_2}^{(-+)}(\mathbf{r}_1,t_1;\mathbf{r}_2,t_2)&=&\pm\!\int\frac{d^3p}{(2\pi\hbar)^3}\,\exp\left\{\frac{i}{\hbar}\mathbf{p}(\mathbf{r}_1-\mathbf{r}_2)%
-\frac{i}{\hbar}\left[\frac{p^2}{2m}+\frac{E_{m_1}+E_{m_2}}{2}\right](t_1-t_2)\right\}%
\nonumber\\%
&&\times\;\rho_{m_1m_2}\left(\mathbf{p},\frac{\mathbf{r}_1+\mathbf{r}_2}{2},\frac{t_1+t_2}{2}\right),%
\label{2.23}
\end{eqnarray}
where $t_1-t_2\sim 0$ and the external sign, either $+$ or $-$, relates to either Bosonic or Fermionic statistics respectively. In semiclassical conditions, when the density matrix depends on the atomic momentum $\mathbf{p}$ and spatial coordinate $\mathbf{r}$ as classical variables, the difference in statistics vanishes in the final kinetic equation.

If the scattered light is weak and characterized by a small degeneracy parameter (small number of photons per coherence volume) it plays a negligible role in the atomic dynamics, which then is mostly affected by interaction with vacuum quantized modes. In such a situation the above graphical equation (\ref{2.21}) can be straightforwardly converted to a kinetic equation responsible for the dynamics of the single particle density matrix (\ref{2.23}). The density matrix describing the dynamics of an arbitrary atom driven by external fields obeys the following Lindblad-type master equation written in the operator form
\begin{eqnarray}
\left[\frac{\partial}{\partial t} + \frac{\mathbf{p}}{m_0}\nabla\right]{\hat{\rho}} &=&-\frac{i}{\hbar}\left[\hat{H}_0,\hat{\rho}\right]-\frac{i}{\hbar}\left[\hat{V}(t),\hat{\rho}\right]%
-\frac{i}{\hbar}\left[\hat{U}(t),\hat{\rho}\right]+\hat{{\cal R}}\hat{\rho}.
\label{2.24}%
\end{eqnarray}
Here $\hat{\rho}=\hat{\rho}(\mathbf{p},\mathbf{r},t)$ and the left-hand side performs a full time convective derivative of the density operator where $\mathbf{p}/m_0=\mathbf{v}$ is the atomic velocity and $\nabla$ is the gradient operator. The first term in the right-hand side is responsible for free dynamics of the atom driven by its internal Hamiltonian $\hat{H}_0$ and the second and third terms describe the interactions with external fields. We denote these interaction terms as $V(t)$ for optical modes and $U(t)$ for microwave modes.

For the monochromatic case, and in the rotating wave approximation, the interaction operators are given by
\begin{eqnarray}
\hat{V}(t)&=&-\hat{\mathbf{d}}^{(-)}\mathbf{E}_0\,\mathrm{e}^{-i\omega_c t+i\mathbf{k}_c\mathbf{r}}-\hat{\mathbf{d}}^{(+)}\mathbf{E}_0^*\,\mathrm{e}^{i\omega_c t-i\mathbf{k}_c\mathbf{r}},%
\nonumber\\%
\hat{U}(t)&=&-\hat{\mathbf{m}}^{(-)}\mathbf{H}_0\,\mathrm{e}^{-i\Omega t}-\hat{\mathbf{m}}^{(+)}\mathbf{H}_0^*\,\mathrm{e}^{i\Omega t},%
\label{2.25}
\end{eqnarray}
where $\hat{\mathbf{d}}^{(\mp)}$ and $\hat{\mathbf{m}}^{(\mp)}$ are the raising/lowering components of the electric dipole and magnetic moment operators; $\mathbf{E}_0$ and $\mathbf{H}_0$ are the complex amplitudes of the electric and magnetic fields respectively. In the case of the microwave field we can ignore its spatial profile since its wavelength is typically much larger than the size of the atomic cloud. In the case of linear polarizations for both fields in a direction along the $Z$-axis the interaction matrix elements are defined as follows,
\begin{eqnarray}
V_{n\tilde{m}}&=&\left(d_{z}\right)_{n\tilde{m}}E_0,%
\nonumber\\%
U_{m\tilde{m}}&=&\left(m_{z}\right)_{m\tilde{m}}H_0,%
\label{2.26}%
\end{eqnarray}
where we follow a convention where $n,n',\ldots$ specify any upper states and $m,m',\ldots$ any ground states. If, as in matrix elements (\ref{2.26}), we need to distinguish the ground states belonging to different hyperfine sublevels we additionally overscribe by tilde the states belonging to other hyperfine sublevels.

The spontaneous radiative decay of the upper states and of the optical coherences contributes in Eq.(\ref{2.24}) by the following relaxation terms
\begin{eqnarray}
\left(\hat{{\cal R}}\hat{\rho}\right)_{n'n}&=&-\gamma\; \rho_{n'n}(\mathbf{p},\mathbf{r},t),%
\nonumber\\%
\left(\hat{{\cal R}}\hat{\rho}\right)_{nm}&=&-\frac{\gamma}{2}\; \rho_{nm}(\mathbf{p},\mathbf{r},t),%
\label{2.27}%
\end{eqnarray}
where $\gamma$ is the natural radiative decay rate. In the important example of alkali-metal atoms the optical pumping repopulation process providing the atomic polarization transfer from the upper to the ground state via spontaneous decay is described by the following incoming-type term
\begin{eqnarray}
\left(\hat{{\cal R}}\hat{\rho}\right)_{m'm}&=&\gamma\,\sum_{n'n}\rho_{n'n}(\mathbf{p},\mathbf{r},t)\sum_{q}C_{F'_0M'_0\, 1q}^{F'M'}C_{F_0M_0\, 1q}^{FM}%
\nonumber\\%
&&\times\; (-)^{F_0-F'_0}\left[(2F'_0+1)(2F_0+1)\right]^{1/2}(2J+1)%
\;\left\{\begin{array}{ccc} S & I & F'_0\\ F' & 1 & J \end{array}\right\}%
\left\{\begin{array}{ccc} S & I & F_0\\ F & 1 & J \end{array}\right\}%
\label{2.28}%
\end{eqnarray}
where $m=F_0,M_0$, $m'=F'_0,M'_0$ and $n=F,M$, $n'=F',M'$. Other quantum numbers specifying the atomic states in (\ref{2.28}) are $S=1/2$, $I$ and $J=1/2,3/2$, the electronic spin, nuclear spin and total electronic angular momentum of the upper state respectively. Here and throughout we follow the definitions and notations for the Clebsch-Gordon coefficients and 6j-symbols as in \cite{LaLfIII,VMK} and some derivation details are clarified in \ref{Appendix.C}.

If the excitation of the atomic subsystem is driven by external coherent fields, undisturbed by the medium, the above equation becomes closed and can be solved. In some cases it transforms to the generalized optical pumping equation, where optical interaction channels could create quite non-trivial distribution of atomic amplitudes over the ground state Zeeman sublevels, see \cite{Happer}. Considered together with the light transport equation, this can make a closed calculation scheme for the entire dynamics of atomic and light subsystems.

Unfortunately, in the general case when the scattered light is not weak or for some reason cannot be ignored it is extremely hard to build the closed and solvable system of the self-consistent equations for the complete set of the field and atomic Green's functions; then additional approximations and simplifications are apparently needed. However in some important applications (random lasing, as an example) such an entire atom-field dynamics would be important to know and control in the calculation scheme. In this case one can generally describe the coupled atom-field dynamics by a set of generalized Maxwell-Bloch equations such that the field subsystem and atomic polarization exist in the spatially multimode and randomized configuration. In this case any local field inhomogeneity is strongly correlated with the local inhomogeneity of the atomic density matrix. An attempt to analyze these correlations in the three-dimensional configuration in application to the quantum information processing was made in \cite{Sorensen&Sorensen}.

\subsubsection{Weak probe, sample susceptibility and the scattering tensor}\label{2.D.2}

For the process of light diffusion the most typical situation is when the probe light only slightly disturbs the ground state sublevels, which are not affected by the control mode, such that the upper sublevels are unpopulated. If the scattered light is weak, such that we can safely ignore any small perturbation of the initial Zeeman distribution in the atomic subsystem, the last term in the graphical equation (\ref{2.17}) and second term in the first line of the graphical expansion (\ref{2.18}) can be neglected. Then the ground state Green's function can be expressed by the non-disturbed original coupling of the $\Psi$ operators (\ref{2.14}) and the polarization operator is visualized by the diagram
\begin{equation}
\begin{array}{c}{\scalebox{1.0}{\includegraphics*{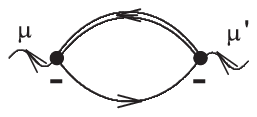}}}\end{array}%
\sim\ \chi_{\mu\mu'}(\mathbf{r},\omega)%
\label{2.29}%
\end{equation}
After analytical decryption this can be linked with the sample susceptibility $\chi_{\mu,\mu'}(\mathbf{r},\omega)$ at the specific frequency $\omega$, see Section \ref{2.E}. We ignore the effect of spatial dispersion but accept possible inhomogeneity of the atomic medium. It is an important advantage of the polarization operator (\ref{2.29}) that it can be calculated in the closed form under rather general assumptions for an atom with arbitrary energy structure.

Indeed, the upper state Green's function performs a single particle retarded propagator (which is a matrix function in the general case) dressed by interaction with the control mode and the vacuum environment
\begin{eqnarray}
i{\cal G}_{n_1n_2}^{(--)}(\mathbf{r}_1,t_1;\mathbf{r}_2,t_2)&=&%
\left\langle T\,\hat{\Psi}_{n_1}(\mathbf{r}_1,t_1)\,%
\hat{\Psi}_{n_2}^{\dagger}(\mathbf{r}_2,t_2)\right\rangle%
\nonumber\\%
&=&\left\langle S^{\dagger}T\left[S\hat{\Psi}_{n_1}^{(0)}(\mathbf{r}_1,t_1)\,%
\hat{\Psi}_{n_2}^{(0)\dagger}(\mathbf{r}_2,t_2)\right]\right\rangle%
\nonumber\\%
&\Rightarrow&\begin{array}{c}\scalebox{1.0}{\includegraphics*{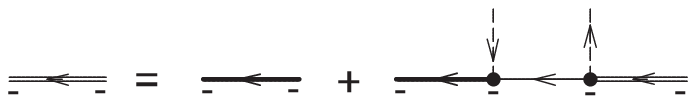}}\end{array}%
\label{2.30}%
\end{eqnarray}
where the solid line expresses the dynamics of the excited atomic state dressed by interaction only with the vacuum modes:
\begin{equation}
\scalebox{1.0}{\includegraphics*{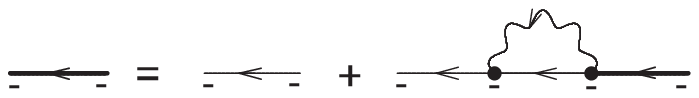}}%
\label{2.31}%
\end{equation}
If the control mode is coherent, stationary and monochromatic then the system of equations (\ref{2.30}), (\ref{2.31}) can be analytically solved and the sample susceptibility is given by
\begin{eqnarray}
\chi_{\mu\mu'}(\mathbf{r},\omega)\!&=&\!%
-\frac{1}{\hbar}\sum_{m,m'}\sum_{n,n'} \rho_{m'm}(\mathbf{r})\left(d_{\mu}\right)_{mn}%
\left(d_{\mu'}\right)_{n'm'}%
\ {\cal G}_{nn'}^{(--)}(\hbar\omega+E_m+i0)%
\label{2.32}%
\end{eqnarray}
where ${\cal G}_{nn'}^{(--)}(E)$ is the matrix function, performing the retarded-type propagator of the excited states specified by the quantum numbers $n,n'$, and given by solution of Eqs.(\ref{2.30}) and (\ref{2.31}). We give an example of such a solution below in sections \ref{4.B} and \ref{4.C}. This propagator contributes to the susceptibility tensor with its "on shell" energy argument with $E=\hbar\omega+E_m+i0$, where $E_m=E_{m'}$ are energies of the degenerate Zeeman sublevels $m$ and $m'$ and $\omega$ is the frequency of the probe photon.\footnote{Even possible action of a weak static magnetic field implies for considering the Zeeman sublevels of the ground state as degenerate in the optical domain.} The Green's function modifies the energy structure of only empty transitions such that the atomic density matrix $\rho_{m'm}(\mathbf{r})$ is assumed to be non-disturbed by the applied control and probe fields in the first approximation.
If, in diagram (\ref{2.29}), one breaks the atomic line, which is directed backward in time, it will transform to another important graph

\begin{equation}
\begin{array}{c}{\scalebox{1.0}{\includegraphics*{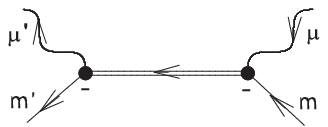}}}\end{array}%
\sim\ \alpha_{\mu'\mu}^{(m'm)}(\omega)\;|m'\rangle\langle m|%
\label{2.33}%
\end{equation}
which is known as the scattering tensor or scattering amplitude of the probe photon
on an atom
\begin{eqnarray}
\alpha_{\mu'\mu}^{(m'm)}(\omega)\;|m'\rangle\langle m|&=&%
-\frac{1}{\hbar}\sum_{n,n'}\left(d_{\mu'}\right)_{m'n'}%
\left(d_{\mu}\right)_{nm}\ %
{\cal G}_{n'n}^{(--)}(\hbar\omega+E_m+i0)\ |m'\rangle\langle m|%
\label{2.34}%
\end{eqnarray}
This tensor is responsible for either elastic or inelastic scattering of an incident photon accompanied by an atomic transition $|m\rangle\to |m'\rangle$. There is an evident analogy between the sample susceptibility and the scattering tensor, but there is an important difference between these quantities. The susceptibility allows only coherent coupling of the states $|m\rangle$ and $|m'\rangle$ for which $\rho_{m'm}(\mathbf{r})\neq 0$. This typically includes a few Zeeman sublevels in the ground state such that in a weakly disordered but homogeneous medium the macroscopic electromagnetic wave, obeying the graphical equation (\ref{2.17}), propagates in the same direction as the incident wave. The scattering process in the spontaneous regime, expressed by the scattering tensor (\ref{2.34}), is open for all possible outgoing spatial and spectral modes as well as for all available ground state transitions such that state $|m'\rangle$ belongs to any available energy level.

We introduced diagrammatic definitions (\ref{2.29}) and (\ref{2.33}) for a system of motionless atoms. This is a relevant approximation for an atomic system prepared or confined in a magneto optical trap at a temperature around 100 $\mu K$ or less. The atoms are considered as a weakly disordered system of scatterers, which preserves its configuration during the interaction time. However in some cases the random thermal motion of atoms can have an important influence on interference effects associated with the light scattering. Our formalism allows us to include the atomic motion via generalization of the atomic Green's function  ${\cal G}_{nn'}^{(--)}(E)\to {\cal G}_{nn'}^{(--)}(E,\mathbf{p})$, which should be now considered as a function of two arguments: energy $E$ and momentum $\mathbf{p}$. Then it will contribute to (\ref{2.32}) and (\ref{2.34}) with an "on shell" energy argument given by the initial energy of the atom and incident photon and with a momentum argument also given by the sum of their momenta. In a semiclassical description Eq.(\ref{2.32}) should be additionally averaged over the momentum distribution of atoms. We will mostly ignore the effects of atomic motion, which manifest themselves via the Doppler shifts and are often negligible for ultracold systems. However we shall point out that in an ensemble of warm atoms the presence of the control mode in combination with the Doppler effect makes the medium spatially dispersive, i. e. its susceptibility tensor becomes dependent on the wave vector of the probe mode: $\chi_{\mu\mu'}=\chi_{\mu\mu'}(\mathbf{r};\omega,\mathbf{k})$.

\subsubsection{Diagram expansion of the correlation function}

Summarizing the results of the previous sections, the correlation function of a weak probe propagating through the atomic medium can be expressed by the following series of the "bond"-type graph terms
\begin{eqnarray}
&&\ \ \scalebox{1.0}{\includegraphics*{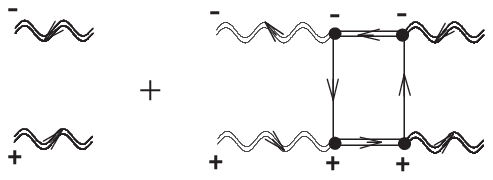}}%
\nonumber\\%
&&+\ \ \begin{array}{c}{\scalebox{1.0}{\includegraphics*{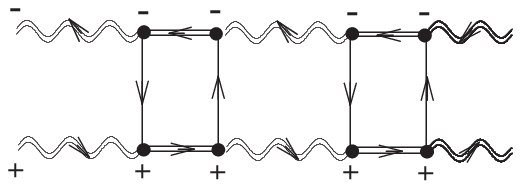}}}\end{array}%
\nonumber\\%
&&+\ \ \begin{array}{c}{\scalebox{1.0}{\includegraphics*{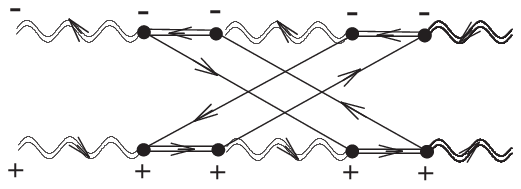}}}\end{array}%
\nonumber\\%
&&+\ \ \ldots
\label{2.35}%
\end{eqnarray}
where the first term is associated with the coherently transmitted part of the light, see Eq.(\ref{2.17}). The second and higher order terms express the contribution of the single and higher order scattering events and reveal a diagram expansion for the binding part of the Green's function (\ref{2.19}). For the sake of convenience the diagrams are rotated such that the upper and lower lines are respectively chained by the retarded-type ($-$) and advanced-type ($+$) propagators. \footnote{Strictly speaking these signs should be associated with causal or anti-causal time ordering, but for the positive frequency component and in a first non-vanishing order with respect to a weak probe there is no difference between these definitions.} The important point is that the expansion (\ref{2.35}) with the dressed atomic and photon propagators now is rapidly converging and can be numerically evaluated for a variety of applications.

Let us introduce the differential cross section as a standard representative characteristic of the scattering process. Then the diagram expansion (\ref{2.35}) leads to the natural expansion of the differential cross section for a macroscopic atomic ensemble
\begin{equation}
\frac{d\sigma}{ d\Omega}\ =\ \sum_{a=1}^{N} \frac{d\sigma_{a}}{ d\Omega}+%
\sum_{a\neq b}\frac{d\sigma_{ab}}{ d\Omega}+%
\sum_{a\neq b\neq c}\frac{d\sigma_{abc}}{d\Omega}+\ldots%
\label{2.36}
\end{equation}
which is expressed by subsequent contributions of the single, double, triple and higher orders of multiple scattering into the differential solid angle $d\Omega$. While making the sum over atomic scatterers contributing to this sequence we emphasize that all possible events of recurrent scattering are eliminated in such an expansion. That is justified by the fact that in a dilute medium the average distance between the scatterers $a,b,c,\ldots$ is around the mean free path for a photon, which is much longer than the radiation wavelength, such that any coincidence among $a,b,c,\ldots$ is a quite rare event. In the evaluation of the sum the actual point-like atomic distribution can be smoothed by the macroscopic density distribution $n_0=n_0(\mathbf{r})\equiv\sum\rho_{mm}(\mathbf{r})$. These approximations are generally constrained by a weak disorder regime in a configuration of cold atoms when $n_0\lambdabar^3\ll 1$ and the above expansion becomes nonapplicable in the alternative limit of strong disorder lying beyond the Ioffe-Regel boundary \cite{IoffeRegel}.

The diagram expansion (\ref{2.35}) introduces contributions of two types for each scattering order. For example, for subsequent double scattering on atom $"a"$ and atom $"b"$ one obtains the following representation of the partial cross-section
\begin{equation}
\frac{d\sigma _{ab}}{d\Omega}\ =\ \frac{d\sigma _{ab}^{(L)}}{d\Omega} + \frac{d\sigma _{ab}^{(C)}}{d\Omega}%
\label{2.37}%
\end{equation}
which are respectively given by the diagrams shown in the second and third lines of the graph series (\ref{2.35}). The first diagram defines so-called "ladder" terms expressed by the product of the scattering amplitude with its conjugated counterpart subsequently on $"a"$ and $"b"$. In the alternate "crossed" term the conjugated amplitude relates to the inverse order of scatterers $\emph{i.e.}$ on $"b"$ first and on $"a"$ second. In addition, one can transpose $a\leftrightarrow b$ such that there are four terms for each pair. The interference correction survives the configuration averaging and exists in all orders of the multiple scattering. It can be observed via enhancement of the light intensity scattered in the backward direction and is known as the coherent backscattering (CBS) effect.

We conclude this section by the following remarks. Even if one extended the diagram series (\ref{2.35}) up to the higher orders and incorporated all possible interference corrections (not only maximally crossed diagrams, but also all the combinatoric combinations of "ladder" and "crossed" terms) the entire sum still would not give us the exact result. What was evidently ignored in this expansion is the effective renormalization of the interaction coupling constant (atomic dipole moments) caused by either cooperative or anti-cooperative dynamics of proximal atomic scatterers. This effect is a part of the dependent or recurrent scattering effects, as pointed out above, but it also includes the difference between the real interaction Hamiltonian given by ({\ref{2.1}}), (\ref{2.2}), (\ref{2.9}) and its approximation (\ref{2.12}). Both these effects, which we will discuss in section \ref{Section V} typically give a correction of $O(n_0\lambdabar^3)\ll 1$ in order of magnitude and can be safely ignored for a dilute system.

There is another important constraint for the diagram expansion (\ref{2.35}). As follows from the beginning of our consideration, the built-up expansion is critically based on the assumption of Gaussian type decoupling of any product of atomic $\Psi$-operators. That allows only identical sequences of the scatterers in the retarded ($-$) and advanced ($+$) lines of each contributing graph. The non-Gaussian correlations were ignored as far as they are less statistically important than the Gaussian ones in the macroscopic limit. However such an assumption also cannot be relevant in the strong disorder limit $n_0\lambdabar^3\gtrsim 1$ when higher order correlations or Dicke-type entanglement of atomic states become ultimately important.

\subsection{The retarded-type Green's function}\label{2.E}

For the stationary conditions, the retarded-type Green's function (\ref{2.18}) depends only on the difference of its time arguments $\tau=t_1-t_2>0$. Then it can be equivalently expressed by its Fourier transform
\begin{equation}
{\cal D}_{\mu_1\mu_2}^{(R)}(\mathbf{r}_1,\mathbf{r}_2;\omega)\ =\ %
\int_0^\infty\!\! d\tau\ \mathrm{e}^{i\omega\tau}\ %
{\cal D}_{\mu_1\mu_2;+}^{(R)}(\mathbf{r}_1,\mathbf{r}_2;\tau),%
\label{2.38}%
\end{equation}
which is considered for $\omega>0$, $\emph{i.e.}$ for its positive frequency component. For a weak probe this function performs the fundamental solution of Eq.(\ref{2.17}).  That is, it obeys the following wave equation
\begin{eqnarray}
\left[\triangle_1+\frac{\omega^2}{c^2}\right]%
{\cal D}_{\mu_1\mu_2}^{(R)}(\mathbf{r}_1,\mathbf{r}_2;\omega)\;%
+\;\sum_{\mu}\int\!\!d^3r\ {\cal P}_{\mu_1\mu}^{(R)}(\mathbf{r}_1,\mathbf{r};\omega)\,%
{\cal D}_{\mu\mu_2}^{(R)}(\mathbf{r},\mathbf{r}_2;\omega)%
\!\!&=&\!\!4\pi\hbar\;\delta_{\mu_1\mu_2}^{\perp}(\mathbf{r}_1-\mathbf{r}_2),%
\label{2.39}%
\end{eqnarray}
where the polarization operator, given by the diagram definition (\ref{2.29}), can be expressed via the dielectric susceptibility of the medium, as indicated in the graph. The latter quantity is rigorously introduced by the Kubo formula, see \cite{LfPtIX}.  However because of the transverse structure of the field operators in the Coulomb and dipole gauges ($\mathrm{div}\hat{\mathbf{A}}=0$, see (\ref{2.3})) the diagram (\ref{2.29}) actually gives the following integral relation between the polarization operator and the susceptibility
\begin{eqnarray}
{\cal P}_{\mu_1\mu}^{(R)}(\mathbf{r}_1,\mathbf{r};\omega)\!\!&=&\!\!%
4\pi\frac{\omega^2}{c^2}\sum_{\mu'\mu''}\int\!\!d^3r'\ \delta_{\mu_1\mu'}^{\perp}(\mathbf{r}_1-\mathbf{r}')%
\;\chi_{\mu'\mu''}\!(\mathbf{r}',\omega)\;\delta_{\mu''\mu}^{\perp}(\mathbf{r}'-\mathbf{r}),%
\label{2.40}%
\end{eqnarray}
where
\begin{eqnarray}
\delta_{\mu\mu'}^{\perp}(\mathbf{r}-\mathbf{r}')\!\!&=&\!\!\int\!\! \frac{d^3k}{(2\pi)^3}%
\left[\delta_{\mu\mu'}-\frac{k_\mu k_{\mu'}}{k^2}\right]\,\mathrm{e}^{i\mathbf{k}\cdot(\mathbf{r}-\mathbf{r}')}%
\ =\ \delta_{\mu\mu'}\delta(\mathbf{r}-\mathbf{r}')+%
\frac{1}{4\pi}\frac{\partial^2}{\partial x_{\mu}\partial x_{\mu'}}\frac{1}{|\mathbf{r}-\mathbf{r}'|}%
\label{2.41}%
\end{eqnarray}
and the same transverse delta function contributes in the right-hand side of the basic equation (\ref{2.39}).

The solution of equation (\ref{2.39}) describes the radiation response generated by a point-like atomic dipole source. Inside the medium such a response is normally associated with a secondary, tertiary, etc. wave generated in the scattering process. Attenuation of the scattered spherical wave is scaled by an extinction length $l_0$ (same in order of magnitude as the transport length $l_{\mathrm{tr}}$) and much longer than the wavelength $\lambdabar$. Thus for practical applications it is enough to find a long distance asymptote for ${\cal D}_{\mu_1\mu_2}^{(R)}(\mathbf{r}_1,\mathbf{r}_2;\omega)$ when $|\mathbf{r}_1-\mathbf{r}_2|\sim l_0\gg\lambdabar$. This can be done with the phase integral technique.

The relevant distribution volume in the integral over $\mathbf{r}$ in Eq.(\ref{2.39}) is a small zone around the point $\mathbf{r}_1$, which can be bounded by a box of a few $\lambda$. The atomic density distribution as well as the sample susceptibility can be considered as homogeneous functions inside this volume. That allows the Green's function to be factorized in the product
\begin{equation}
{\cal
D}_{\mu_1\mu_2}^{(R)}(\mathbf{r}_1,\mathbf{r}_2;\omega)\ \approx\ %
-\hbar\, X_{\mu_1\mu_2}(\mathbf{r}_1,\mathbf{r}_2;\omega)\,%
\frac{\exp\left[ik|\mathbf{r}_1-\mathbf{r}_2|\right]}{|\mathbf{r}_1-\mathbf{r}_2|},%
\label{2.42}%
\end{equation}
where the slowly varying amplitude $X_{\mu_1\mu_2}(\mathbf{r}_1,\mathbf{r}_2;\omega)$ satisfies the following differential equation
\begin{eqnarray}
\frac{\partial}{\partial z_1}X_{\mu_1\mu_2}\!(\mathbf{r}_1,\mathbf{r}_2;\omega)&=&%
\frac{2\pi i\omega}{c}\sum_{\mu}\,\chi_{\mu_1\mu}(\mathbf{r}_1,\omega)\;%
X_{\mu\mu_2}\!(\mathbf{r}_1,\mathbf{r}_2;\omega)%
\nonumber\\%
X_{\mu_1\mu_2}(\mathbf{r}_1,\mathbf{r}_2;\omega)&\to&\delta_{\mu_1\mu_2}^{\perp}%
\ \ \ \mathrm{at}\ \ \mathbf{r}_1\to\mathbf{r}_2%
\label{2.43}%
\end{eqnarray}
This equation is written in a special reference frame where the $z$-axis is directed along the ray from point $\mathbf{r}_2$ to point $\mathbf{r}_1$. The ray is associated with the local plane wave with the wave vector $\mathbf{k}=k(\mathbf{r}_1-\mathbf{r}_2)/|\mathbf{r}_1-\mathbf{r}_2|$. In turn, the polarization components $\mu_1,\mu_2,\mu$ belong to the plane $x,y$, which is transverse to the propagation direction. The Kronecker symbol $\delta_{\mu_1\mu_2}^{\perp}$ is now transverse in the real (not reciprocal) space. The equation (\ref{2.43}) describes the transformation of the complex amplitude of a light wave propagating in space from point $\mathbf{r}_2$ to point $\mathbf{r}_1$.

The dielectric susceptibility tensor contributes to equation (\ref{2.43}) by its components in the $x,y$-plane. Because of the specific symmetry of atomic transitions, which is clarified in \ref{Appendix.C}, it is more natural and convenient to define these components in the angular momentum basis set. It can be defined by the following basis expansion $\mathbf{e}_{0}=\mathbf{e}^{0}=\mathbf{e}_{z}$, $\mathbf{e}_{\pm 1}=-\mathbf{e}^{\mp 1}=\mp (\mathbf{e}_{x}\pm i\mathbf{e}_{y})/\sqrt{2}$ with respect to a Cartesian frame and it requires co/contravariant notation in writing the tensors indices, see  \cite{VMK}. Let us denote the transverse components of the susceptibility tensor in the angular momentum representation as $\tilde{\chi}_{q_1}{}^{q_2}(\ldots)$, where $q_1,q_2=\pm 1$. Then $\tilde{\chi}_{q_1}{}^{q_2}(\ldots)$ performs a $2\times 2$ matrix which can be expanded in the set of Pauli matrices $\overrightarrow{\sigma}=(\sigma_{\mathrm{x}},\sigma_{\mathrm{y}},\sigma_{\mathrm{z}})$ as follows
\begin{equation}
\tilde{\chi}_{q_1}{}^{q_2}(\mathbf{r},\omega )\ =\ \chi_0(\mathbf{r},\omega)\,%
\delta _{q_1}{}^{q_2}+%
\left( \overrightarrow{\chi}(\mathbf{r},\omega)\cdot\overrightarrow{\sigma }\right)_{q_1}{}^{q_2}%
\label{2.44}%
\end{equation}%
We subscript the Pauli matrices by "$\mathrm{x},\,\mathrm{y},\,\mathrm{z}$" indices (in Roman style) and this symbolic vector notation should not be confused with the Cartesian frame notation. In this expansion the upper row and left column of Pauli matrices are associated with the $+1$ index and the respective lower row and right column with the $-1$ index.

The expansion coefficients $\chi_{0}(\mathbf{r},\omega)$ and $\overrightarrow{\chi }(\mathbf{r},\omega )$ are parameterized by components of a particular tensor of the dielectric susceptibility. Once this tensor is known, then the expansion coefficients can be found via its rotational transformation from laboratory to the local reference frame defined above. Such transformation is most natural to do from the major reference frame where the susceptibility tensor has a diagonal form.

The solution of equation (\ref{2.43}) can be expressed by the parameters of the expansion (\ref{2.44}). Let us enumerate the polarization components $\mu_1,\mu_2$ by $1$ ($x$-axis) and $2$ ($y$-axis), then the solution is given by
\begin{eqnarray}
X_{11}(\mathbf{r}_{2},\mathbf{r}_{1};\omega )\!\!&=&\!\!%
e^{i\phi _{0}(\mathbf{r}_{2},\mathbf{r}_{1})}%
\left[ \cos \phi (\mathbf{r}_{2},\mathbf{r}_{1})-i\sin \phi (\mathbf{r}_{2},\mathbf{r}_{1})n_{\mathrm{x}}\right]%
\nonumber \\%
X_{22}(\mathbf{r}_{2},\mathbf{r}_{1};\omega )\!\!&=&\!\!%
e^{i\phi _{0}(\mathbf{r}_{2},\mathbf{r}_{1})}%
\left[ \cos \phi (\mathbf{r}_{2},\mathbf{r}_{1})+i\sin \phi (\mathbf{r}_{2},\mathbf{r}_{1})n_{\mathrm{x}}\right]%
\nonumber \\%
X_{12}(\mathbf{r}_{2},\mathbf{r}_{1};\omega )\!\!&=&\!\!%
e^{i\phi_{0}(\mathbf{r}_{2},\mathbf{r}_{1})}\,i\sin \phi (\mathbf{r}_{2},\mathbf{r}_{1})%
\left(n_{\mathrm{y}}+i\,n_{\mathrm{z}}\right)%
\nonumber\\%
X_{21}(\mathbf{r}_{2},\mathbf{r}_{1};\omega )\!\!&=&\!\!%
e^{i\phi_{0}(\mathbf{r}_{2},\mathbf{r}_{1})}\,i\sin \phi (\mathbf{r}_{2},\mathbf{r}_{1})%
\left(n_{\mathrm{y}}-i\,n_{\mathrm{z}}\right)%
\label{2.45}%
\end{eqnarray}%
where
\begin{eqnarray}
\phi _{0}(\mathbf{r}_{2},\mathbf{r}_{1})&=&\frac{2\pi \omega }{c}%
\int_{\mathbf{r}_{1}}^{\mathbf{r}_{2}}\chi_0(\mathbf{r},\omega )ds;%
\nonumber\\%
\phi (\mathbf{r}_{2},\mathbf{r}_{1})&=&\frac{2\pi \omega }{c}%
\int_{\mathbf{r}_{1}}^{\mathbf{r}_{2}}\chi (\mathbf{r},\omega )ds%
\label{2.46}%
\end{eqnarray}%
perform the phase integrals along the path $s$ linking the points $\mathbf{r}_1$ and $\mathbf{r}_2$ such that $\mathbf{r}=\mathbf{r}(s)$ in the integrand.

In these integrals $\chi(\mathbf{r},\omega)$ is the complex "length" of the symbolic vector $\overrightarrow{\chi}(\mathbf{r},\omega)$ and $\overrightarrow{n}(\omega )$ is its complex "director", which are given by
\begin{eqnarray}
\chi^2(\mathbf{r},\omega)&=&%
\chi_\mathrm{x}^2(\mathbf{r},\omega)+\chi_\mathrm{y}^2(\mathbf{r},\omega)+\chi_\mathrm{z}^2(\mathbf{r},\omega)%
\nonumber\\%
\overrightarrow{n}&=&\overrightarrow{n}(\omega)=%
\overrightarrow{\chi }(\mathbf{r},\omega)/\chi (\mathbf{r},\omega )%
\label{2.47}%
\end{eqnarray}
We additionally assume (and it is a crucial point in the derivation) that the atomic polarization is homogeneous along the atomic sample such that the "director" $\overrightarrow{n}(\omega )$ is conservative (does not depend on $\mathbf{r}$) along the path in the phase integrals representation of the Green's function (\ref{2.42}), (\ref{2.45}). That is a relevant approximation for many applications and it mainly assumes that the principle directions of the dielectric susceptibility tensor are conserved on a spatial distance scaled by the extinction length $l_0$.

The long range asymptote of the positive frequency component for the advanced-type Green's function ${\cal D}_{\mu_1\mu_2}^{(A)}(\mathbf{r}_1,\mathbf{r}_2;\omega)$ is given by complex conjugation and matrix transposition of expression (\ref{2.42}) for the retarded propagator. Thus we have expressed the retarded and advanced Green's functions as well as the scattering tensor (\ref{2.32}) via the atomic propagator given by diagrams (\ref{2.30}) and (\ref{2.31}). By this we have determined all the ingredients for further numerical evaluation, which implies a Monte-Carlo iteration scheme with high order approaching the diagram expansion (\ref{2.35}) for correlation functions of the transmitted and scattered light.

As a basic result of the performed diagram analysis, the entire process of light scattering by a dilute atomic sample can be approached by the Green's function formalism and Monte-Carlo simulations. It is important that each diagram term, expressing a particular order of the multiple scattering in (\ref{2.35}), requires the knowledge of all the contributions for prior orders. However such calculations are quite cumbersome and often hide from us the qualitative picture of the process. Many features of the multiple scattering can be more easily, but still correctly, understood in the framework of a light transport or a diffusion equation, which we shall briefly discuss below.

\subsection{The light transport equation and the diffusion approximation}\label{2.F}

\subsubsection{The light transport equation}\label{2.F.1}

If, in the expansion (\ref{2.35}), we ignore any interference and keep only the main ladder-type sequence, the sum can be converted to the following graph equation
\begin{equation}
\scalebox{1.0}{\includegraphics*{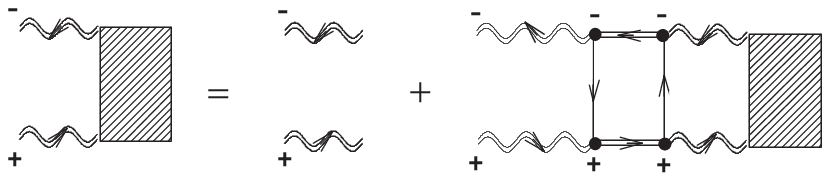}}%
\label{2.48}%
\end{equation}
where the dashed block denotes the complete sum. The sum approximately expresses the averaged product of the field operators
\begin{eqnarray}
\begin{array}{c}\scalebox{1.0}{\includegraphics*{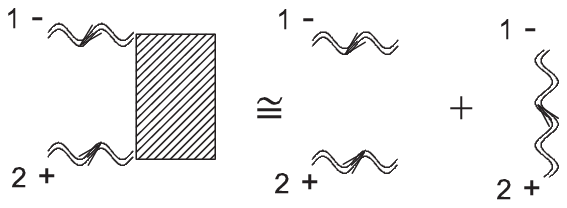}}\end{array}%
&\Rightarrow&%
\langle \hat{A}_{\mu_2}^{(-)}(\mathbf{r}_2,t_2)\,\hat{A}_{\mu_1}^{(+)}(\mathbf{r}_1,t_1)\rangle%
\label{2.49}%
\end{eqnarray}
in accordance with definitions given above by Eqs. (\ref{2.17}) and
(\ref{2.19}).

Graph equation (\ref{2.48}) performs a Bethe-Salpeter type equation \cite{LfPtIX} written for the single particle Green's function equivalent to the light correlation function.\footnote{The introduced diagram equation, which has a ladder structure  given by a sequence of the retarded and advanced photon propagators, has a certain analogy with the equation originally introduced by E.E. Salpeter and H.A. Bethe for two interacting Fermi-Dirac particles in \cite{Bethe}. Such an equation is often considered and similarly named in theories of wave transport in random media \cite{Rossum}.}. In an optically dense sample the first term in the right-hand side of (\ref{2.49}) performs a source contribution responsible for the coherent penetration of the probe inside the medium to a depth of around $l_0$. Equation (\ref{2.48}) becomes closed if the retarded/advanced -type photon's propagator as well as the scattering tensor, forming the kernel of the integral operator in its right-hand side, are known in analytical form. This justifies the Monte-Carlo iteration scheme for its further solution as we pointed out in the end of the previous section. But with certain simplifications and assumptions the Bethe-Salpeter equation can be transformed to a more simple and convenient form of light transport equation tracking the light energy transfer.

Let us assume that the correlation function (\ref{2.49}) can be expressed by the following integral expansion
\begin{eqnarray}
\langle\hat{A}_{\mu_2}^{(-)}(\mathbf{r}_2,t_2)\,\hat{A}_{\mu_1}^{(+)}(\mathbf{r}_1,t_1)\rangle&=&%
\frac{2\pi c}{\bar{\omega}^2}\int d\Omega\int_0^{\infty}\frac{d\omega}{2\pi}%
\ \exp\left[i\frac{\omega}{v_p(\omega)}\mathbf{n}\,\delta\mathbf{r}-i\omega\tau\right]\ %
I_{\mu_1\mu_2}(\mathbf{n},\omega,\mathbf{r},t)%
\label{2.50}%
\end{eqnarray}
where $\delta\mathbf{r}=\mathbf{r}_1-\mathbf{r}_2$, $\mathbf{r}=(\mathbf{r}_1+\mathbf{r}_2)/2$ and $\tau=t_1-t_2$, $t=(t_1+t_2)/2$. We will assume that the integrand  $I_{\mu_1\mu_2}(\mathbf{n},\omega,\mathbf{r},t)$ considered as function of $\omega$ and director $\mathbf{n}\equiv \{\theta,\phi\}\equiv \Omega$ provides the integral convergency in such a way that the left-hand side, considered as a function of $\delta\mathbf{r}$ and $\tau$, is rapidly vanishing on mesoscopic scales i.e. respectively much shorter and faster than the radiation transport length and time. Then definitions of $\mathbf{r}$ and $t$ in the left-hand side would be insensitive to any relevant variations of the fast variables $\delta\mathbf{r}$ and $\tau$, further associated with the radiation coherence time and lengths (longitudinal and transverse), and the integrand $I_{\mu_1\mu_2}(\mathbf{n},\omega,\mathbf{r},t)$ could be treated as a slowly varying function of its arguments $\mathbf{r}$ and $t$.  For radiation trapping in a dilute gas the above requirements are easy to fulfil for spectrally broad radiation located near carrier frequency $\bar{\omega}$ as the phase speed of the light $v_p(\omega)$ is approximately equal to the speed of light in vacuum $c$. The expansion (\ref{2.50}), when it is acceptable, generalizes definitions (\ref{2.5}) and (\ref{2.6}) for local and instantaneous amount of spectral energy distributed with light in a particular polarization and has a clear analogy with the Wigner representation of a density matrix in kinetic theory. The tensor function $I_{\mu_1\mu_2}(\mathbf{n},\omega,\mathbf{r},t)$ is transverse to the director $\mathbf{n}$ such that
\begin{equation}
\int_0^{\infty}\frac{d\omega}{2\pi}\;I_{\mu\mu}(\mathbf{n},\omega,\mathbf{r},t)\ =\ %
I(\mathbf{n},\mathbf{r},t)%
\label{2.51}%
\end{equation}
coincides with the Poynting vector for the local plane wave propagating in
direction $\mathbf{n}$, which was defined earlier by Eq.(\ref{2.6}).

Let us introduce the wave operator in the medium, which can be accepted from equation (\ref{2.39})
\begin{equation}
\hat{D}^{-1}_j\ =\ \triangle_j-\frac{1}{c^2}\frac{\partial^2}{\partial t_j^2}\,%
+\,\sum_{\mu}\int\!\!d^3r\ {\cal P}_{\mu_j\mu}^{(R)}(\mathbf{r}_j,\mathbf{r};t_j-t)\ldots%
\label{2.52}%
\end{equation}
Then by applying this operator to the graph equation (\ref{2.48}) it can be transformed to
\begin{eqnarray}
\lefteqn{\left(\hat D_{1}^{-1} - \hat D_{2}^{-1*}\right)\ %
\begin{array}{c}{\scalebox{1.0}{\includegraphics*{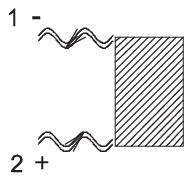}}}\end{array}}%
\nonumber\\%
&=&\begin{array}{c}{\scalebox{1.0}{\includegraphics*{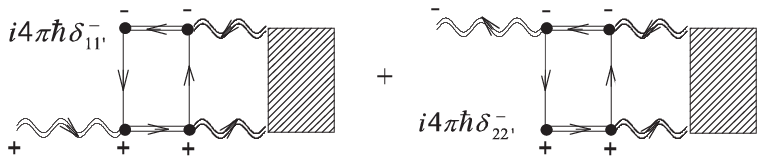}}}\end{array}%
\label{2.53}%
\end{eqnarray}
The action of the wave operators (\ref{2.52}) on the retarded and advanced photon propagators is indicated by the presence of transverse delta-functions in this graph, see definition (\ref{2.41}). The built up graph equation can be decoded and in its analytical form can straightforwardly generate the sought equation for the polarization sensitive spectral distribution of light energy $I_{\mu_1\mu_2}(\mathbf{n},\omega,\mathbf{r},t)$. Such an equation can be performed in closed form if the sample is described by the isotropic susceptibility
\begin{equation}
\chi(\mathbf{r},\omega)\ =\ \chi'(\mathbf{r},\omega)+i\chi''(\mathbf{r},\omega)
\label{2.54}%
\end{equation}
which should also be uniform on a spatial scale of a few $l_0$.

Finally the radiative transport equation has the following form
\begin{eqnarray}
\lefteqn{\hspace{-1cm}\frac{1}{v_g(\mathbf{r},\omega)}\frac{\partial}{\partial t}%
I_{\mu_1\mu_2}(\mathbf{n},\omega,\mathbf{r},t)+%
\mathbf{n}\nabla I_{\mu_1\mu_2}(\mathbf{n},\omega,\mathbf{r},t)}%
\nonumber\\%
&&\hspace{-1cm}=\;\sum_{i}\!\int d\Omega'\!\sum_{\mu'_1,\mu'_2}%
I_{\mu'_1\mu'_2}^{(i)}(\mathbf{n}',\omega',\mathbf{r},t)%
\; n_0(\mathbf{r})\;\frac{\omega}{\omega'}\;%
{}^{(i)}\!Q_{\mu_1\mu_2}^{\mu'_1\mu'_2}(\mathbf{n}',\omega'\to \mathbf{n},\omega)%
\ -\ n_0(\mathbf{r})\,\sigma_{\mathrm{tot}}(\omega)\,I_{\mu_1\mu_2}(\mathbf{n},\omega,\mathbf{r},t)%
\label{2.55}%
\end{eqnarray}
The time and convective derivative terms in the left-hand side of this equation are scaled by the local group velocity of light in the medium $v_g(\mathbf{r},\omega)$, which is given by
\begin{equation}
\frac{1}{v_g(\mathbf{r},\omega)}=\frac{1}{c}\,+\,\frac{2\pi\bar{\omega}}{c}\,%
\frac{d}{d\omega}\chi'(\mathbf{r},\omega)%
\label{2.56}
\end{equation}
and it can significantly differ from the light phase speed. The equation (\ref{2.55}) can track the dynamics of light energy distributed in several spectral domains, specified by the "$i$"-index and coupled via Rayleigh and Raman scattering mechanisms. The generalized polarization-dependent differential cross-section for the single atom scattering process is given by\footnote{For a sake of notation convenience in this \ref{2.F} section we change notation for the differential scattering cross section $\frac{d\sigma}{d\Omega}\to {}^{(i)}Q...$ with specifying it by all accessible spectral and polarization input and output scattering channels.}
\begin{eqnarray}
{}^{(i)}\!Q_{\mu_1\mu_2}^{\mu'_1\mu'_2}(\mathbf{n}',\omega'\to \mathbf{n},\omega)&=&%
\frac{\omega^3\omega'}{c^4}\sum_{m,m'}\rho_{m'm'}\;%
\alpha_{\bar{\mu}_1\bar{\mu}'_1}^{(mm')}(\omega')\ \alpha_{\bar{\mu}_2\bar{\mu}'_2}^{(mm')*}(\omega')%
\nonumber\\%
&&\times\ \delta_{\mu_1\bar{\mu}_1}^{\perp}(\mathbf{n})\,\delta_{\mu'_1\bar{\mu}'_1}^{\perp}(\mathbf{n'})\,%
\delta_{\mu_2\bar{\mu}_2}^{\perp}(\mathbf{n})\,\delta_{\mu'_2\bar{\mu}'_2}^{\perp}(\mathbf{n'}),%
\label{2.57}%
\end{eqnarray}
and parameterized by the quantum numbers for input and output scattering channels, and
\begin{equation}
\delta_{\mu\bar{\mu}}^{\perp}(\mathbf{n})\ =\ \delta_{\mu\bar{\mu}}-n_\mu n_{\bar{\mu}}%
\label{2.58}%
\end{equation}
is the transverse delta-function (\ref{2.41}) in the reciprocal space. The atomic density matrix is factorized $\rho_{m'm'}(\mathbf{r})=n_0(\mathbf{r})\,\rho_{m'm'}$, such that $n_0(\mathbf{r})$ is the density distribution and $\rho_{m'm'}$ performs the distribution of atoms over their Zeeman sublevels. The frequencies $\omega'$ and $\omega$ are linked by the energy conservation law in the scattering process: $\omega=\omega'-\omega_{mm'}$, where $\omega_{mm'}$ is the transition frequency for the specific $i$-th channel. The total scattering cross section $\sigma_{\mathrm{tot}}(\omega)$, responsible for the overall emission into elastic as well as into inelastic channels, is expressed by the imaginary part of the sample susceptibility
\begin{equation}
4\pi\,\chi''(\mathbf{r},\omega)\ =\ \frac{c}{\bar{\omega}}\,n_0(\mathbf{r})\,\sigma_{\mathrm{tot}}(\omega).%
\label{2.59}%
\end{equation}
Then the extinction length can be defined as $l_0^{-1}\equiv l_0^{-1}(\mathbf{r},\omega)=n_0(\mathbf{r})\sigma_{\mathrm{tot}}(\omega)$. The kinetic-type radiative transport equation similar to (\ref{2.55}) for the first time was introduced in astrophysics, see \cite{Chandrasekhar}, where it is perfectly applicable. In case of ultracold atomic systems it gives a good initial approximation for the description of diffuse light transport in randomly scattering media.

\subsubsection{The diffusion approximation}\label{2.F.2}

If the medium is dense for Rayleigh and transparent for the Raman channel and if the scattering process is isotropic, such that the trapped radiation is approximately unpolarized, then equation (\ref{2.55}) can be integrated over the light spectrum and rewritten in the following form
\begin{eqnarray}
\frac{1}{\bar{v}}\frac{\partial}{\partial t}%
I(\mathbf{n},\mathbf{r},t)+%
\mathbf{n}\nabla I(\mathbf{n},\mathbf{r},t)%
\!\!&=&\!\!\int d\Omega'\,%
I(\mathbf{n}',\mathbf{r},t)\, n_0\, \bar{Q}(\mathbf{n}'\to \mathbf{n})\;%
-\;\frac{1}{\bar{l}_0}\,I(\mathbf{n},\mathbf{r},t)%
\label{2.60}%
\end{eqnarray}
It is additionally assumed that the medium is homogeneous and all the spectral dependencies can be smooth and averaged over the light spectrum. Then overbar indicates that the group velocity $\bar{v}$, the differential cross-section $\bar{Q}(\mathbf{n}'\to \mathbf{n})$ and the extinction length $\bar{l}_0$ are considered now as averaged parameters of the process.

The integration of the first term in Eq.(\ref{2.60}) over all $\mathbf{n}$ directions leads to
\begin{equation}
\int d\Omega\,\frac{1}{\bar{v}}\;I(\mathbf{n},\mathbf{r},t)\ \equiv\ W(\mathbf{r},t)%
\label{2.61}%
\end{equation}
where $W(\mathbf{r},t)$ can be associated with the local instantaneous density of radiant energy accumulated in the medium. Similar integration of the second term defines the local energy flux
\begin{equation}
\int d\Omega\;\mathbf{n}\;I(\mathbf{n},\mathbf{r},t)\ \equiv\ \mathbf{J}(\mathbf{r},t)%
\label{2.62}%
\end{equation}
Equation (\ref{2.60}) yields the following continuity relation
\begin{equation}
\frac{\partial}{\partial t}W(\mathbf{r},t)+\nabla\mathbf{J}(\mathbf{r},t)\ =\ %
-\bar{v}\,\frac{1-a}{\bar{l}_0}\,W(\mathbf{r},t)%
\label{2.63}%
\end{equation}
where the existence of inelastic scattering channels is included into the albedo ratio denoted by $a$. In the case of an ideally closed transition without any Raman losses $a=1$.

The continuity relation (\ref{2.63}) can be transformed to the form of a diffusion equation after the following assumption
\begin{equation}
I(\mathbf{n},\mathbf{r},t)\ \approx\ \frac{\bar{v}}{4\pi}W(\mathbf{r},t)+\ldots%
\label{2.64}%
\end{equation}
which is consistent with the hypothesis of the isotropic nature of the trapped radiation. After straightforward mathematical manipulations with Eq.(\ref{2.60}) one arrives at the following diffusion form of the light transport equation
\begin{equation}
\frac{\partial}{\partial t}W(\mathbf{r},t)-D\,\Delta W(\mathbf{r},t)\ =\ %
-\bar{v}\,\frac{1-a}{\bar{l}_0}\,W(\mathbf{r},t)%
\label{2.65}%
\end{equation}
where the diffusion constant $D$ is given by
\begin{equation}
D\ =\ \frac{1}{3}\frac{\bar{l}_0}{1-\langle\cos\theta\rangle}\bar{v}\ \equiv\ %
\frac{1}{3}l_{\mathrm{tr}}\,\bar{v}%
\label{2.66}%
\end{equation}
Here $\langle\cos\theta\rangle$ denotes the averaged polar angle between directors $\mathbf{n}'$ and $\mathbf{n}$. The diffusion approximation to the radiative transport process in a random medium implies as a spatial scale the radiation transport length $l_{\mathrm{tr}}$ instead of the extinction length $\bar{l}_0$.

The presence of the spectrally weighted group velocity $\bar{v}$ in the above relations indicates that any fraction of the reradiated light propagates through an atomic sample slower (and sometimes much slower) than in vacuum. The observable delay time has two physical contributions: the ballistic time to propagate from one scatterer to the next and the additional "dwell" time associated with delay of the scattering process for itself, see \cite{AlbTigLag91}. These two contributions naturally appeared in Eq.(\ref{2.56}) after its averaging and multiplying by $l_0$ with observation that the second term can be many times bigger than the first one. For a near resonant radiation the group velocity is sharply varying from slow light up to superluminal regime, but in average the effect of "dwell" time dominates the process. Such a "dwell" time delay was observed in experiment \cite{LVMDMWK03} where the diffusion speed was demonstrated as almost five orders of magnitude smaller than the vacuum speed.

The description of light propagation in disordered media via the radiative transport equation and its diffusion approximation is an often used and popular approach in the literature, see review \cite{Rossum} showing a number of examples. The diffusion approach has been applied for description of such subtle phenomena as coherent backscattering and weak/strong light localization, see \cite{AkkWolfMayn,BerkKav}. In this sense let us make the following remark concerning its validity. As we see from the above derivation, the main requirement to convert the ladder-type diagram series to the transport equation (\ref{2.55}) consists in the applicability of expansion (\ref{2.50}). This expansion is sufficiently justified if the intensity distribution, satisfying either equation (\ref{2.55}) or (\ref{2.60}), (\ref{2.65}), can only slightly vary on a spatial/time scale associated with coherence length/time of the radiation. As follows from the above derivation this is fulfilled if the light has a coherence time shorter than a typical time for the diffusion process. In the case of atomic systems that means that the diffusion approach is surely valid for the transport of radiation with the spectrum comparable or broader than the spontaneous decay rate $\gamma$. However such a basic requirement cannot be fulfilled for monochromatic light and for the long pulses with a smoothed profile. The direct addressing to the graph series (\ref{2.35}) gives much better approximation in this case.

\section{Coherent backscattering of light}\label{SecIII}
\setcounter{equation}{0}
\setcounter{figure}{0}

\subsection{Observations specific to atomic systems}

Considering a dilute system of light scatterers randomly distributed in space one meets an interesting phenomenon called coherent backscattering (CBS) of light. The second (cross) term in the overall multiple cross-section (\ref{2.37}) survives the disorder configuration self-averaging and gives a contribution comparable with the first (ladder) term for a particular narrow range centered around the backward scattering direction. As can be easily verified, the configuration sensitive phase difference $\Delta\phi$ for direct and reciprocal paths vanishes for each scattering loop in the diagram series (\ref{2.35}) and for motionless system of scatterers is given by
\begin{equation}
\Delta\phi\ \sim\ (\mathbf{k}_{\mathrm{in}}+\mathbf{k}_{\mathrm{out}})%
(\mathbf{r}_{\mathrm{out}}-\mathbf{r}_{\mathrm{in}}).%
\label{3.1}%
\end{equation}
In the equation, $\mathbf{k}_{\mathrm{in}}$ and $\mathbf{k}_{\mathrm{out}}$ are the wave vectors of the incident and scattered photons, while $\mathbf{r}_{\mathrm{out}}-\mathbf{r}_{\mathrm{in}}$ is the spatial separation between the first and last scatterer. In the backscattering direction $\mathbf{k}_{\mathrm{out}}\sim -\mathbf{k}_{\mathrm{in}}$, so the accumulated phase difference tends to zero. For classical Rayleigh type scatterers, when each direct and reciprocal scattering paths are time reversible, the constructive interference can enhance the output intensity by as much as a factor of two. In the general case the deviation of the backward scattered intensity from the simplest prediction based on an incoherent description of the process can be expressed in terms of the so-called enhancement factor, which is given by
\begin{equation}
\eta\ =\ \frac{\left.\frac{d\sigma}{d\Omega}\right|_{\Omega\to 0}}%
{\left.\frac{d\sigma^{(L)}}{d\Omega}\right|_{\Omega\to 0}}%
\label{3.2}%
\end{equation}
where in the numerator we keep all the contributions of single and multiple scattering from both the ladder and interference terms, but in the denominator keep only the contributions of the ladder type and single scattering processes, see definitions (\ref{2.36}),(\ref{2.37}). In practice the CBS effect is observed as an enhancement of the light intensity scattered into a normally small solid angle near the backward direction.

First observation of CBS of light from a disordered scattering sample was made in 1984 \cite{TIsh,IshKg}. These reports were quickly followed by experimental and theoretical work \cite{AlbLag}, \cite{WolfMaret} that discussed and explained CBS in the context of classical optics for electromagnetic wave scattering by a disordered medium. Currently many important aspects of the CBS phenomenon of light scattered from liquid and condensed mater phases and for a system consisting of classical scatterers are understood, explained and discussed in a number of reviews \cite{Sheng,KuzRom,KSSH06,AegrtMaret}. It would seem naively expectable that this intrinsically classical effect should be naturally observed in an ensemble of atomic scatterers as well.

Given this background, why is it that the CBS effect had not been observed in atomic systems until the later 1990's? The problem lies in the relatively fast atomic motion, which induces uncontrollable frequency shifts in each event of Rayleigh scattering and makes this process not precisely elastic. Indeed, because of the Doppler effect the scattering of a photon with incoming wave vector $\mathbf{k}$ to an outgoing mode with wave vector $\mathbf{k}'$ is accompanied by the following frequency change
\begin{equation}
\omega'\ =\ \omega+(\mathbf{k}'-\mathbf{k})\mathbf{v},%
\label{3.3}%
\end{equation}
where $\mathbf{v}$ is the velocity of the atomic scatterer. If atoms are in a thermal equilibrium then at any segment of the scattering path this frequency change accumulates a random phase shift, which can be estimated in its order of magnitude as
\begin{equation}
\delta\phi\ \sim\ 2\pi\,k\frac{d\chi'(\omega)}{d\omega}k\bar{\mathrm{v}}\,l_{\mathrm{tr}}\sim%
\frac{n_0\lambdabar^2}{\gamma}k\bar{\mathrm{v}}\,l_{\mathrm{tr}}\sim%
\frac{k\bar{\mathrm{v}}}{\gamma}%
\label{3.4}%
\end{equation}
where $\chi'(\omega)$ is the dispersive part of the local dielectric susceptibility of the medium, see Eq.(\ref{2.54}), and $\bar{\mathrm{v}}$ is the average atomic velocity. To make this thermal shift negligible it is necessary that the atomic scatterers would not move on a distance of $\lambdabar$ during the time $\gamma^{-1}$ and the relevant Doppler shift should be smaller than $\gamma$. This crucial requirement cannot be fulfilled for a warm atomic ensemble prepared at room temperature, but it is easy attainable for cold atoms samples prepared, for instance, in a magneto optical trap. Then in the performed estimate we can have in mind an ideal situation of an isolated dilute atomic ensemble, cooled below the "Doppler limit" with temperature $T< \hbar\gamma$, such that Eq. (\ref{3.4}) gives a necessary condition for the CBS observation. The subtler estimate showing how the residual thermal motion of mobile atoms affect the CBS interference has been given in \cite{LDKM06} and demonstrated the enforcing of a random phase as function of scattering orders. But let us also point out that in a more general situation of arbitrary disordered medium the internal dynamics of atomic scatterers can be quite complicated because of anisotropic structure and environment disturbing the susceptibility tensor (\ref{2.32}) and scattering tensor (\ref{2.34}), which, as a consequence, strongly affects the interference of the partial scattering waves in the output channel see \cite{KSSH06,Rossetto09}.

\begin{figure}[t]\center
{$\scalebox{0.5}{\includegraphics{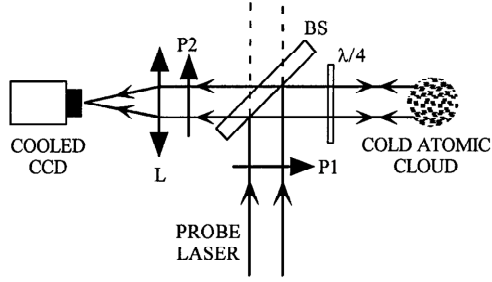}}$ }%
\caption{The CBS detection scheme. P1, P2: polarizers; $\lambda/4$: quarter-wave plate; BS:
beamsplitter (transmittance T = 90\%); L: analysis lens (f= 500 mm), copyright 1999 by the American
Physical Society}
\label{fig31}
\end{figure}

The first experiment on observation of the CBS effect in a cold atomic ensemble consisting of ${}^{85}$Rb atoms, confined with a magneto-optical trap, was performed in Nice \cite{LTBMMK99} in 1999. A schematic diagram of the coherent backscattering detection scheme used in \cite{LTBMMK99} and in the series of later experiments is shown in Fig.~\ref{fig31}. A weak and well-collimated probe laser beam was passed through a polarizer and then scattered on a cold atomic cloud. The portion of light scattered in the nearly backward direction was detected in a selected polarization channel. In the experiment the following four polarization channels were observed. For linearly polarized input radiation, the backscattered light was detected in two output polarization modes either parallel ($\mathrm{Lin}\|\mathrm{Lin}$) or orthogonal ($\mathrm{Lin}\bot\mathrm{Lin}$) to the input polarization. The other two polarization schemes were related to the CBS observation for the circularly polarized probe light in the helicity preserving ($\mathrm{Hel}\|\mathrm{Hel}$) and helicity orthogonal ($\mathrm{Hel}\bot\mathrm{Hel}$) polarization channels.

\begin{figure}[t]\center
{$\scalebox{0.6}{\includegraphics{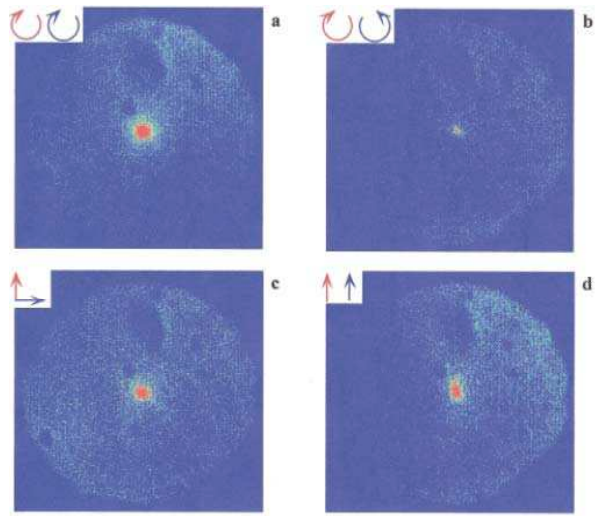}}$ }%
\caption{(Color online) Atomic CBS images in the circular and linear polarization channels. For each
channel, the inset shows the incident (red) and detected (blue) polarization.
Data (a) and (b) correspond to the helicity non-preserving and preserving
channels (circular polarization). Data (c) and (d) correspond to the orthogonal
and parallel channels (linear polarization),  copyright 1999 by the American
Physical Society}%
\label{fig32}%
\end{figure}

In Fig.~\ref{fig32} we reproduce the experimental images from Refs.\cite{LTBMMK99,LMWMK00} demonstrating the spatial distribution of backscattered light in the above four polarization channels. There is clear evidence of enhancement in the light intensity backscattered into a small solid angle, which is visualized as a bright spot near the central point of each image. These spots accumulate the light propagating in the coherence angle of radiation scattered in the backward direction and focused by the lens shown in Fig.~\ref{fig31}. By scanning the observation angle the enhancement factor (\ref{3.3}) can be measured as a ratio of light intensity scattered at a particular angle to its angularly insensitive background level.

\begin{figure}[t] \center
{$\scalebox{0.8}{\includegraphics{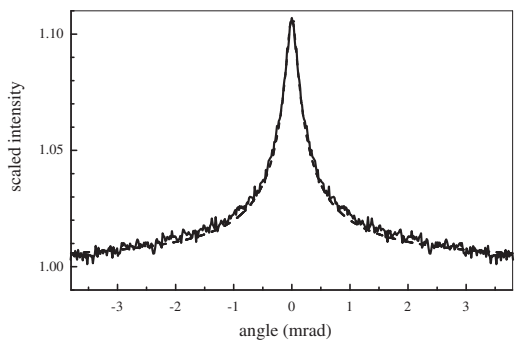}}$ }
\caption{ A CBS cone in the helicity non-preserving channel associated with
the $F_0 = 3\rightarrow F = 4$ transition in cold atomic ${}^{85}$Rb, reprinted from \cite{LTBMMK99};
copyright 1999 by the American Physical Society}
\label{fig33}%
\end{figure}

The typical dependence on detection angle $\theta$ is shown in Fig.~\ref{fig33} and reveals the cone-type angular distribution of the enhancement factor. The most critical characteristics of the cone shape are its amplitude and angular width. Both the parameters are very sensitive to polarization, geometry of the observation channel, distribution of atoms in the cloud, and to residual effects of atomic motion. The observed amplitude of the enhancement factor in atomic system is smaller than in colloidal suspensions and powders. This is because of the presence of elastic Raman scattering in the photon transport via a diffusion mechanism. The existing Zeeman degeneracy of the hyperfine sublevels of an alkali-metal atom opens the Raman scattering channels, which break the optimal time-reversal symmetry, relations between scattering amplitudes for direct and reciprocal paths of any scattering loop. As a consequence the cross terms are reduced and the ladder terms dominate in the overall cross section such that the interference effect is manifested more weakly. However for a "two-level" atom such as strontium, which has zero electronic angular momentum in its ground state, only the Rayleigh scattering channel is allowed and the classical limit of "two" for the enhancement factor becomes experimentally attainable \cite{BKBDLMWK02}. There is one intriguing property of elastic Raman process in the multiple light scattering, and in the CBS in particular, it can lead to destructive interference behavior, which is known as an anti-localization phenomenon. This effect was predicted in \cite{KHS04} and has been a rather challenging effect for experimental verification. We shall briefly discuss it in the next section.

The first reliable observations of the CBS effect in light scattering on atomic systems have stimulated a wide range of experimental and theoretical studies. The modification of the CBS cone with variation of probe detuning and its spectrum, the influence of spatial and velocity distribution of atoms in the cloud on the scattering process, were considered and reported in \cite{LMWMK00,BKBDLMWK02,KSBHKS03,KSLKSBH04}. Serious theoretical efforts were addressed to analysis of possible extension of the CBS phenomenon in the context of the Mollow problem \cite{Mollow,ChTnDRGr}, when the saturation effects of the driving field manifest themselves in cooperative fluorescence \cite{SMB05,GWDM06,KBS14}. The experimental observations of the saturation and nonlinear effects in CBS were reported in \cite{CWBKM04,BKSHKS05,Labeyrie08}. The interaction with static electric or magnetic field can also affect the CBS observables. As was demonstrated in \cite{LMMSDK02,SLJDKM04} the presence of an external static magnetic field can eliminate the single scattering contribution in the entire multiple scattering process and can increase the enhancement factor up to its classical Rayleigh limit of "two".

Below we discuss some of the above features of the CBS effect for light scattering on a cold atomic ensemble. The relevant theoretical tool for describing the experimental situation is the Monte-Carlo simulation of the entire scattering process based on the Green's function formalism, which calculation details were clarified in Section \ref{2.D}. As a realistic approximation for an atomic sample, which we will further assume throughout our discussion in this section, we will describe the atomic cloud by a spherically symmetrical Gaussian-type distribution of atoms given by
\begin{equation}
n(\mathbf{r})\ =\ n_0\exp\left(-\frac{\mathbf{r}^2}{2r_0^2}\right),%
\label{3.5}
\end{equation}
where $n_0$ is the peak density in the middle of the cloud and $r_0$ is the radius (root of square variance) of the cloud. For a closed transition the optical depth on resonance for a light ray crossing the sample through its central part is given by $b_0=\sqrt{2\pi}n_0\sigma_0r_0$, where $\sigma_0$ is the resonance scattering cross-section. Most of the experiments, which we discuss in the section below, are performed for alkali-metal atoms. As is customary, these atoms are specified by the ground and excited state angular momentum and its projection respectively defined as $F_0,M_0$ and $F,M$. For the case of closed transition the resonance scattering cross section for atoms, equally populating Zeeman sublevels of the ground state, is given by
\begin{equation}
\sigma_0=\frac{2F+1}{2F_0+1}\frac{2\pi}{k_0^2}=\frac{2F+1}{2F_0+1}\frac{\lambda_0^2}{2\pi}%
\label{3.6}
\end{equation}
where $k_0$ and $\lambda_0$ are the resonance wave number and wavelength respectively.

\subsection{Polarization and anisotropy effects, anti-localization}\label{SecIIIB}

First demonstrations of the CBS effect for light scattered from ultracold atomic ensembles \cite{LTBMMK99}, \cite{LMWMK00} revealed a very small value of enhancement factor, which for ${}^{85}$Rb did not exceed $1.2$. That is, it was essentially smaller than typical value corresponded with classical scatterers. The small CBS enhancement observed in atoms has been confirmed in a series of subsequent experiments with ${}^{85}$Rb and ${}^{87}$Rb in \cite{KSBHKS03,KSLKSBH04,LDMMK04,WBCDJKLMMSK04} showing the intrinsic quantum nature of the atomic scatterers. The ground states of rubidium atoms are degenerate and the Zeeman sublevels are populated in cold atom samples with approximately equal probabilities. Any specific combination of Rayleigh and Raman transitions contributing into a given multiple scattering chain results in imperfection of constructive interference between direct and reciprocal paths. As was pointed out in the previous section, under equilibrium conditions the most effective interference enhancement up to the classical limit of "two" was demonstrated in strontium, which has a non-degenerate ground state \cite{BKBDLMWK02}.

\begin{figure}[tp] \center
{$\scalebox{1.00}{\includegraphics*{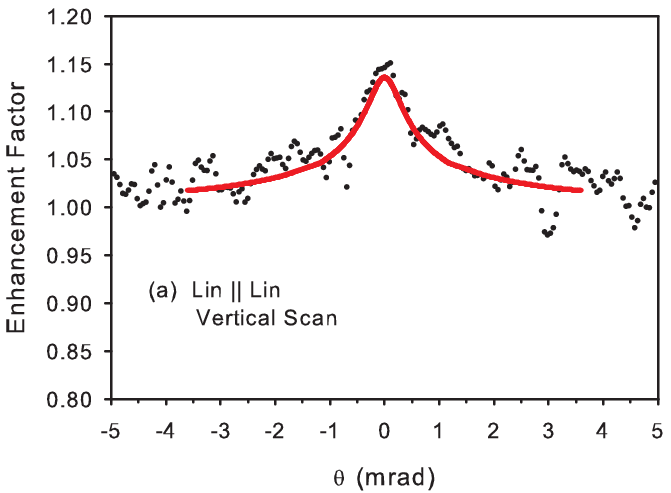}}$ }
{$\scalebox{1.0}{\includegraphics*{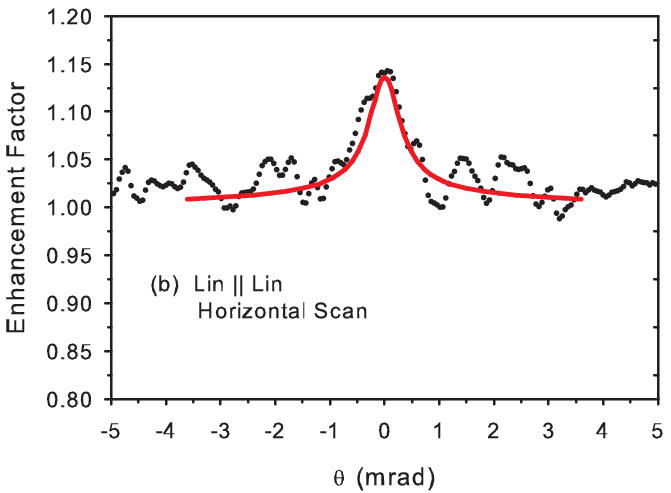}}$ }
{$\scalebox{1.0}{\includegraphics*{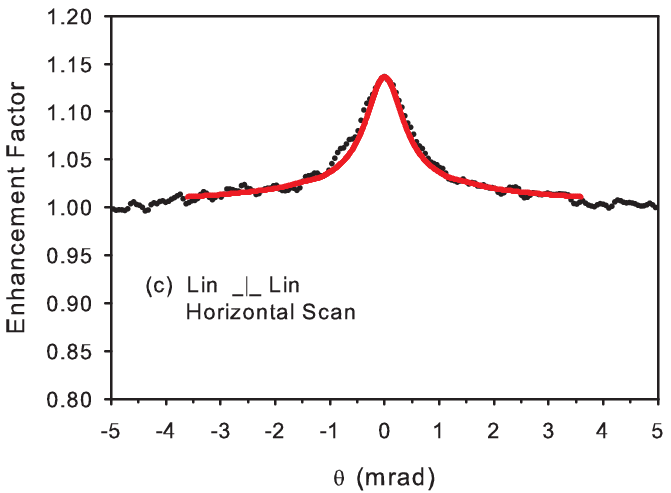}}$ }
\caption{ (Color online) Line scans through the CBS cone for the parallel linearly polarized
output and input modes: (a) - vertical scan, (b) - horizontal scan,
specified by geometry of figure \ref{fig32}(d). (c) -
horizontal scan for the orthogonal linearly polarized
output and input modes. In each case, results of quantum Monte Carlo simulations
are indicated by solid lines, reprinted from \cite{KSBHKS03};
copyright 2003 by the American Physical Society.}
\label{fig34}%
\end{figure}

However, the complicated internal energy structure of atomic scatterers makes a background for various polarization effects accompanying the CBS of light on atoms. As visualized on the CBS images in Fig.~\ref{fig32}, reproducing the experimental results from Refs.\cite{LTBMMK99,LMWMK00}, the CBS enhancement is quite sensitive to the polarization channel under observation. As was later shown in \cite{KSBHKS03} not only the absolute value of the enhancement factor differs for the above four selected polarization channels but the shape of the CBS cone itself is changed with polarization. Anisotropy of the CBS cone was observed in experiments \cite{LMWMK00,KSBHKS03} in the basis of parallel linearly polarized output and input modes. The observed cone had an elliptical shape in its cross-section in the transverse plane, which was narrowed in the direction orthogonal to the light polarization. A fragment of the experimental data for a line scan through the CBS cone in different directions illustrating the variation of the cone profile is reproduced in Fig.~\ref{fig34}.

\begin{figure}[tp] \center
{$\scalebox{0.8}{\includegraphics{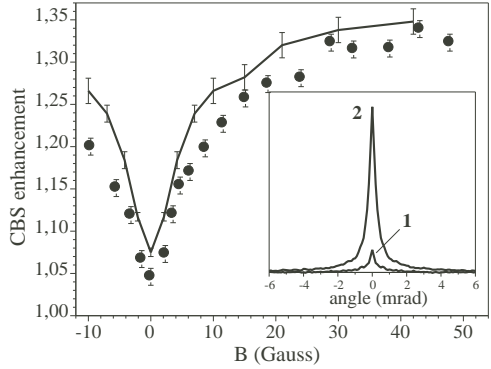}}$} \caption{CBS enhancement factor of light for a cold Rb atomic cloud, measured in the helicity preserving channel, as a function of the transverse magnetic field $B$; The solid line is the result of a Monte-Carlo simulation. The inset shows profiles of the CBS cones when $B = 0$ (1) and $B = 43$ G (2), reprinted from Ref.\cite{SLJDKM04}; copyright 2004 by the American Physical Society.}
\label{fig35}
\end{figure}

Among different aspects of weak localization in multilevel systems it was found that one can effectively manipulate the observed interferences by reinforcing those scattering channels that lead to interference and suppressing those that do not. Such effects can be realized for example by polarization of the atomic angular momenta \cite{KSLKSBH04}. If the optically oriented atomic ensemble is probed by light circularly polarized in the direction of the collective atomic spin the single scattering contribution can vanish and the optimal interference conditions can be fulfilled. If such a medium becomes transparent such that only the double Rayleigh scattering in the helicity-preserving channel can contribute to the scattering process that makes ideal conditions for observation of maximal enhancement in the CBS process. An equivalent effect can be achieved by applying an auxiliary static magnetic field and by tuning the frequency of the light in such a way that light would interact only with the specific Zeeman sublevel, see \cite{SLJDKM04}. In Fig.~ \ref{fig35} we reproduced the basic graph of \cite{SLJDKM04}, showing the experimental verification of this effect. This demonstrates a quite unusual behavior of the enhancement factor in the CBS process for the spectrally selected Zeeman hyperfine transition $F_0=3,M_0=3\to F=4,M=4$ in $D_2$-line of ${}^{85}$Rb in the presence of a strong magnetic field. For the considered excitation scheme there is domination of the double scattering contribution in the helicity preserving Rayleigh-type scattering channel which is accompanied by maximal (by a factor of "two") enhancement of the scattered intensity.

\begin{figure}[tp]\center
{$\scalebox{0.9}{\includegraphics*{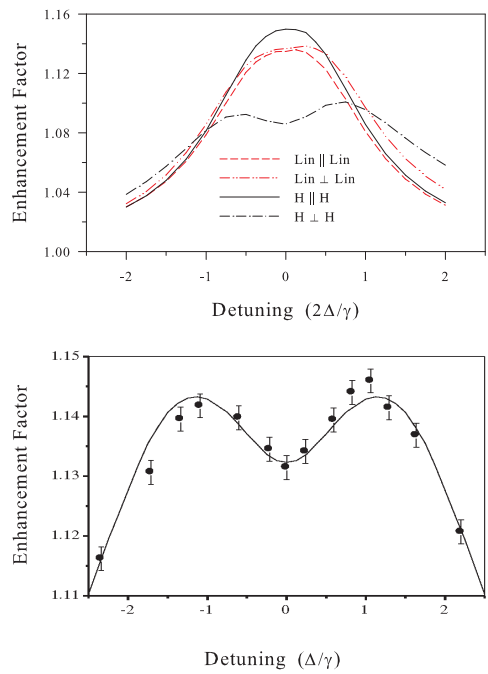}}$ } \caption{(Color online) (Upper panel) The spectral behavior of the enhancement factor, for the $F_{0}=3\to F=4$ hyperfine transition in ${}^{85}\mathrm{Rb}$ for different polarization channels, calculated for a spherical Gaussian-type atomic cloud with radius $r_{0}\,=\,1\,\mathrm{mm}$ and peak density $n_{0}\,=\,16\cdot 10^{9}\,\mathrm{cm}^{-3}$. The frequency detuning $\Delta $ is scaled in units of the natural half-linewidth $\gamma /2$. (Lower panel) Spectral asymmetry of the CBS enhancement observed for helicity orthogonal channel in \cite{LDMMK04}.}
\label{fig36}
\end{figure}

The hyperfine interaction in an alkali-metal atom also makes the enhancement factor quite sensitive to the variation of the spectral detuning of the probe field, which was experimentally verified in Refs.\cite{KSBHKS03,KSLKSBH04,WBCDJKLMMSK04}. The hyperfine interaction is very important for realistic description of the CBS effect in an ensemble consisting of alkali-metal atoms. In \cite{KSKSH03} an asymmetric behavior of the CBS enhancement was predicted for the spectral variation of the probe mode near the resonance of $F_{0}=3\to F=4$ transition in $D_2$-line of ${}^{85}$Rb, see Fig.~\ref{fig36}. The asymmetry of the spectral profile is more manifestable in the case of circular polarization, which has a dip near the resonance point for the helicity orthogonal polarization channel. This indicates a useful spectroscopic tool relating with non-trivial behavior and for Raman-type interference between the ladder and crossed terms near the resonance. The spectral asymmetry can be explained by interference among all the hyperfine transitions including the off-resonant ones. This circumstance was unambiguously established in \cite{KSKSH03} by formally keeping only the $F_{0}=3\to F=4$ transition, in which case a symmetric spectral profile results how it is shown in the lower panel of Fig.~\ref{fig36}.  Many important features of CBS associated with the hyperfine interaction were studied in experimental papers \cite{KSLKSBH04,LDMMK04}, the spectral asymmetry of the enhancement factor and the spectral dip at the resonance point were experimentally observed in \cite{LDMMK04}.

\begin{figure}[tp]\center
{$\scalebox{1}{\includegraphics*{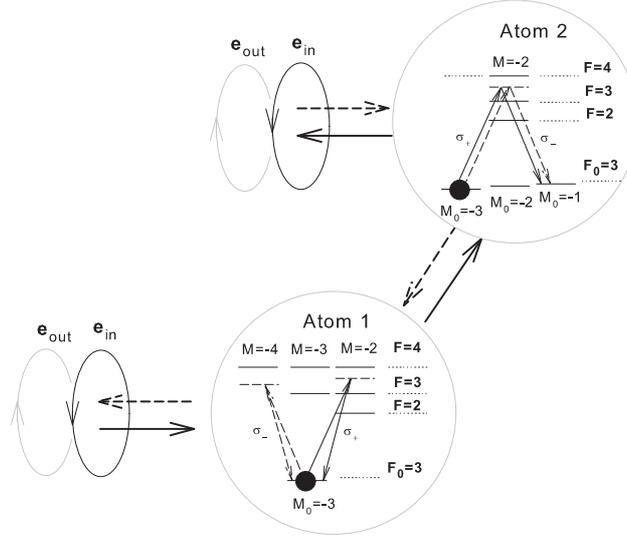}}$ } \caption{Diagram explaining the anti-localization phenomenon in example of the helicity preserving scattering channel for double scattering of circular polarized light from an ensemble of ${}^{85}$Rb atoms oriented opposite to the helicity vector of the probe beam. The destructively interfering amplitudes are a combination of Rayleigh- and Raman-type transitions. The solid and dashed lines, in case of approximately collinear location of the atoms, indicate the interfering direct and reciprocal scattering paths for probing between $F_0 = 3 \to F = 4$ and $F_0 = 3 \to F = 3$ hyperfine transitions.}
\label{fig37}%
\end{figure}

General analysis of the CBS enhancement in a wide spectral interval covering all hyperfine structure of atomic excited states was performed in \cite{KHS04,KLSH05} and it revealed an interesting qualitatively new phenomenon called anti-localization. It occurs that in some spectral regions the enhancement factor can be expected to be less than unity, $\emph{i.e}$ the interference can have destructive character for some observation channels. The nature of the destructive interference in coherent backscattering can be understood by considering the transition diagrams displayed in Fig.~\ref{fig37}. This figure shows an example of double scattering of right-handed helicity light by a system composed of two ${}^{85}$Rb atoms; the output channel also detects right-handed helicity light. A particular case is considered where both atoms initially occupy the same Zeeman sublevel $|F_0, M_0=-F_0 \rangle$ and have a spin orientation opposite to polarization of the incident light. For the sake of simplicity, let us further assume that the atoms are approximately collinearly located with respect to the probe beam. Then both the interfering channels, forward and backward, represent a sequence of scattering events of the Raman and Rayleigh types. There is a fundamental difference in the Rayleigh scattering amplitudes for the two interfering channels. For the forward channel, the $\sigma_+$-polarized light induces in the first atom transitions through the Zeeman sublevels $M = -2$ of the hyperfine excited states $F$ = 2, 3, and 4. For the backward channel, the same atom interacts with the $\sigma_-$polarized light and undergoes transitions only through the ($F = 4, M = -4$) upper sublevel. Varying the light frequency, one can make the corresponding amplitudes almost equal in magnitude but their phases will differ by $\pi$. For this reason, the interference of the two processes shown in Fig.~\ref{fig37} will be destructive and coherent decay rather than coherent amplification will be observed for the backscattered light. The influence of constructively interference transitions essentially decreases anti-localization, however this does not suppress it completely. This effect can be observed in an atomic medium with equilibrium population of the Zeeman sublevels, but, as was shown in \cite{KHS04}, spin orientation of an atomic ensemble can enhance this effect.

\begin{figure}[tph]\center
{$\scalebox{0.9}{\includegraphics*{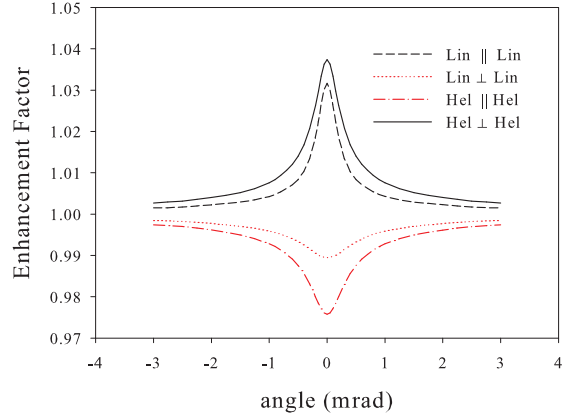}}$ } \caption{(Color online) Angular dependence of the CBS enhancement factor for different polarization channels. The calculated data are presented for a spherically symmetric Gaussian-type cloud of ${}^{85}$Rb atoms with $r_0 = 1 mm$ and with the resonance optical depth $b_0 = 5$, and for a specific detuning $\Delta = -13.25\Gamma$ of the probe mode from the frequency of the $F_0= 3 \rightarrow F = 4$ transition in $D_2$-line.}
\label{fig38}%
\end{figure}%

In Fig.~\ref{fig38} we demonstrate the anti-localization phenomenon and reproduce the results of our calculations of the CBS cone for different polarization channels and at specific frequency detuning $\Delta = -13.25\Gamma$ of the probe mode from the frequency of the $F_0= 3 \rightarrow F = 4$ transition of $D_2$-line of ${}^{85}$Rb. The anti-localization is evident in $\mathrm{Lin}\bot\mathrm{Lin}$ and $\mathrm{Hel}\|\mathrm{Hel}$ observation channels. It is an interesting peculiarity in the formation of the CBS cone in the anti-localization regime that the odd and even orders of multiple scattering makes competitive either destructive or constructive contribution to the overall outgoing intensity. As result the observed profile, given by the sum of all partial contributions, has broader width and smaller amplitude.

\subsection{Saturation and non-linear effects in coherent backscattering}\label{3.C}

The physical problem of interaction of stronger light fields with an atomic system has been driven by milestone or seminal papers \cite{RautianSobelman,Mollow} and now is commonly referred as the Mollow problem. Interaction of a single two-level atom with a monochromatic driving field of arbitrary amplitude manifests itself in the correlation and spectral properties of its radiation emission. The coherent component of the emission can be expressed by the following diagram
\begin{equation}
\scalebox{0.8}{\includegraphics*{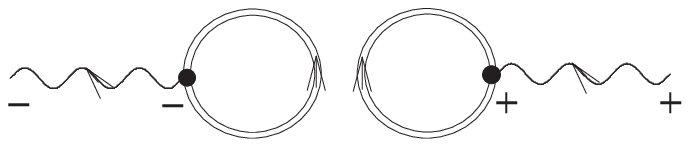}}%
\label{3.7}%
\end{equation}
where the loop atomic lines perform the optical coherency of the atomic density matrix driven by the field.\footnote{In the diagram analysis we assume here Gaussian factorization of the atomic operators even in a single atom case. This can be justified by the Markovian nature of the stochastic dynamics of atomic variables and by quantum non-degeneracy of a macroscopic atomic system where such a single atom can be selected.} The external thin wavy lines are free propagators responsible for propagation of the emitted photons outward and away from the system.  All the density matrix components can be found as a solution of the Bloch equations (\ref{2.24}). As a direct consequence of that solution the optical coherency is proportional to, and has a phase dependent on, the field amplitude. Diagram (\ref{3.7}) describes the light coherently scattered by an atomic oscillator over all directions and the total intensity of this part of the radiation emission is given by
\begin{equation}
 I_{\mathrm{coh}}\ \propto\ \frac{1}{2}\frac{s}{(1+s)^2},%
 \label{3.8}%
 \end{equation}
 where
 \begin{equation}
 s\ =\ \frac{\Omega_c^2/2}{\Delta_c^2+\gamma^2/4}%
 \label{3.9}%
 \end{equation}
 is the saturation parameter defined by the Rabi frequency of the driving field $\Omega_c$ and by its detuning $\Delta_c$ from the atomic resonance. As follows from Eq.(\ref{3.8}) the coherent part of the radiation, associated with the classical vision of the fluorescence emission, should vanish in the saturation limit $s\gg 1$.

In the saturation limit the nonlinear fluorescence transforms to the incoherent part of the radiation emission, which can be expressed by the following diagram sequence
\begin{equation}
\scalebox{0.8}{\includegraphics*{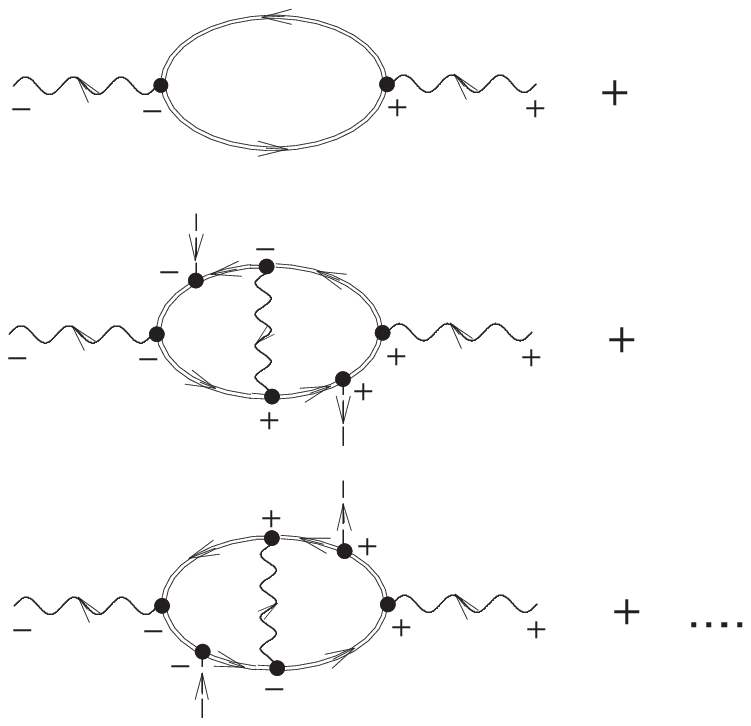}}%
\label{3.10}%
\end{equation}
and the total intensity of incoherent emission is given by
\begin{equation}
 I_{\mathrm{incoh}}\ \propto\ \frac{1}{2}\frac{s^2}{(1+s)^2}%
 \label{3.11}%
\end{equation}
In this expansion the first graph reproduces the incoherent emission, if it was initiated only by direct spontaneous decay from the upper to the ground atomic states, both "dressed" by interaction with the driving field and reservoir modes, and ignoring any correlations between emission events. However such a naive description misses internal correlations induced by the coherent excitation channel and is insufficient for a correct reproduction of the fluorescence spectrum. The next two diagrams open the series responsible for the steady state dynamics of an atomic dipole driven by both the coherent excitation and spontaneous processes. The crucial peculiarity of this dynamics is that each subsequent event of spontaneous emission is delayed and prevented by an act of coherent pumping from the ground to the upper state. This effect, known as the anti-bunching phenomenon, induces correlations between different radiated modes and essentially modifies the entire fluorescence spectrum, see \cite{ChTnDRGr} for more details.

Considered together, both the contributions (\ref{3.7}) and (\ref{3.10}) reproduce the complete dynamics of the autocorrelation function of the atomic dipole moment
\begin{eqnarray}
\left\langle \hat{d}^{(-)}_{\mu'}(t')\hat{d}^{(+)}_{\mu}(t)\right\rangle&=&\bar{d}^{(-)}_{\mu'}(t')\bar{d}^{(+)}_{\mu}(t)\ %
+\ \left\langle \Delta\hat{d}^{(-)}_{\mu'}(t')\Delta\hat{d}^{(+)}_{\mu}(t)\right\rangle,%
\label{3.12}
\end{eqnarray}
where $\hat{d}^{(+)}_{\mu}(t)$ and $\hat{d}^{(-)}_{\mu'}(t')$ are respectively the Heisenberg operators for positive and negative frequency components of the atomic dipole. The dynamics of these operators can be tracked by the Heisenberg-Langevin formalism and the spectral profile of the Mollow triplet can be calculated in closed form.

\begin{figure}[tp]\center
{$\scalebox{0.8}{\includegraphics*{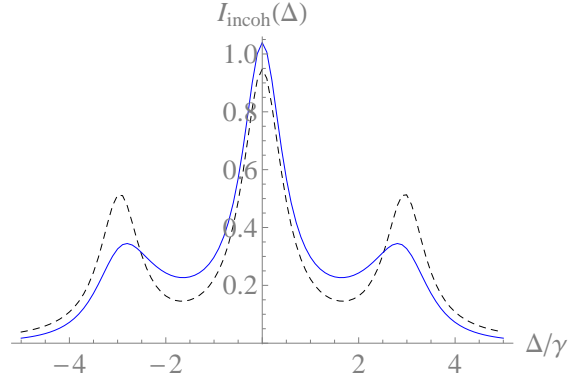}}$ }
\caption{(Color online) The spectrum of the fluorescence emitted by a single two-level atom driven by the resonant coherent mode with $\Delta_c=0$ and $\Omega_c=3\gamma$ ($s=18$). To emphasize the importance of anti-bunching and correlations between different emission channels we have additionally plotted (dashed) the contribution given by the first term in expansion (\ref{3.10}) responsible only for the "dressing" effects. The integral over the fluorescence spectrum is normalized as $\int I_{\mathrm{incoh}}(\Delta)d\Delta/2\pi=I_{\mathrm{incoh}}=s^2/2(1+s)^2\sim1/2$, see (\ref{3.11}).}%
\label{fig39}
\end{figure}%

Then the first term in the right hand side indicates the average value and the contribution of (\ref{3.7}) and the second term performs the contribution of the dipole's stochastic dynamics, $\emph{i.e.}$, the incoherent part (\ref{3.10}). The former leads to a monochromatic response similar to light scattering by a harmonic oscillator in a classical visualization of the process. The incoherent contribution has a spectral distribution with a profile which strongly correlates with the driving field intensity and its frequency detuning. If the driving field is weak ($s\ll 1$) then this contribution is practically negligible and gives a small correction to the overall fluorescence revealing a preliminary signature of the quantum nature of the process. In the alternative limit of the saturation regime ($s\gg 1$) when the driving field is strong the coherent term vanishes, see (\ref{3.8}), and the incoherent part dominates in the fluorescence spectrum. The population of the upper and low atomic states are equalized and become independent of the field intensity. In the general case the correlation function (\ref{3.12}) has a nonlinear dependence on the driving field intensity and the fluorescence spectrum, considered as function of $\Delta=\omega-\omega_c$ (where $\omega$ is the emitted photon frequency and $\omega_c$ is the driving field frequency), develops a triplet resonance structure, see \cite{RautianSobelman,Mollow,ChTnDRGr}. In Fig.~\ref{fig39} we show an example of the Mollow triplet calculated for $\Delta_c=0$ and $\Omega_c=3\gamma$ and compare it with the uncorrelated contribution given by the first term of expansion (\ref{3.10}).

Consider now the configuration of two atoms well separated by a radiation zone. The coherent contributions (\ref{3.7}) can be generalized by a sum of such products where the mean dipole components can be associated with the same as well as with the different atoms. The coherent radiation outgoing from a collection of coherently excited dipole sources has an evident tendency for interference. If we extended the situation up to a macroscopic number of atoms and introduced the configuration averaging then the sum of all the coherent terms could be converted to the diagram expansion (\ref{2.35}) and would give us the mesoscopic visualization for the propagation of a weak coherent pulse in a weakly disordered and diffusely scattering medium. The coupled ladder terms in the graph expansion (\ref{2.35}) are normally referred to as incoherent scattered events. But in the context of the present section they describe completely coherent radiation, with spectral properties not modified by the interaction process, and the definition of coherency/incoherency associated with the spatial inhomogeneity of the medium should not be confused with the different definition associated with the stochastic temporal dynamics of the atomic dipoles in the entire Mollow problem.

\begin{figure}[tp]\center
{$\scalebox{0.9}{\includegraphics*{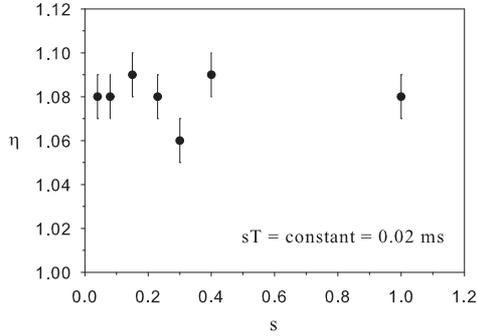}}$ }
\caption{Dependence of the CBS enhancement factor on the saturation parameter s, for observation in the helicity preserving polarization channel of $F_0 = 3 \rightarrow F = 4$ hyperfine resonance in ${}^{85}$Rb, from \cite{BKSHKS05}. Each point corresponds to the varying storage time $T=0.02/s\ \mathrm{ms}$.}
\label{fig310}%
\end{figure}

\begin{figure}[t]\center
{$\scalebox{0.8}{\includegraphics*{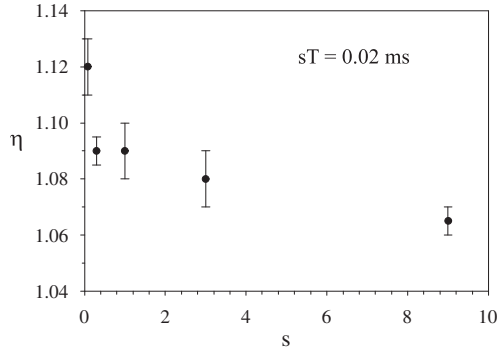}}$ }
\caption{Same as in figure \ref{fig310} but for observation in the channel of orthogonal linear polarizations, from \cite{BKSHKS05}}
\label{fig311}%
\end{figure}

It seems intuitively reasonable that only coherent outgoing fields could interfere and the incoherent part of the fluorescence emitted from different atoms would not. In particular, if accepting this point, the CBS phenomenon should vanish in the saturation limit. The reduction of the enhancement factor as a function of the saturation parameter was verified in two independent experiments \cite{CWBKM04,BKSHKS05}. Typical experimental behavior of the enhancement factor in the saturation regime in the example of the  $F_0=3\to F=4$ closed transition of the $D_2$ line of ${}^{85}$Rb is shown in Figs.~\ref{fig310} and \ref{fig311}. However it has been the subject of many discussions in the literature that the above statement is not exactly correct and the incoherent parts of the radiation emitted in the backward direction can interfere as well \cite{WGDM04,WGDM05,SMB05,GWDM06,WGDM06,KBS14}.

To explain why this is possible let us turn again to the radiation emitted from two atoms and consider it in the high saturation limit and also ignoring any coherent contributions associated with the average dipole polarization. Then the correction of the diagram series (\ref{3.10}) associated with the presence of the second atom would appear in two ways. First, the outward wavy line should be modified by the contribution to the polarization operators and would build up the retarded and advanced Green's functions for a photon propagating in the medium, see Eqs.(\ref{2.17}) and (\ref{2.18}). That would be the standard precursor of the macroscopic behavior in the system. Second, there would be corrections to the atomic Green's functions associated with the presence of another atom. Part of such corrections could be incorporated into the dressed atomic lines where incoherent radiation from each of the atoms contributes to the excitation process of another atom. But among these there can be specified the following diagram
\begin{equation}
\scalebox{0.8}{\includegraphics*{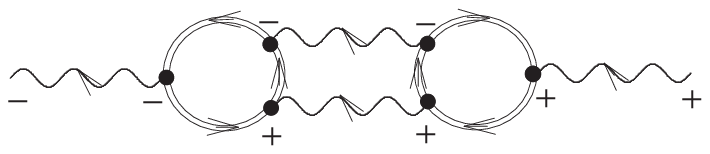}}%
\label{3.13}
\end{equation}
which contributes to the correlation function of the dipole operators of different atoms
\begin{equation}
\left\langle \hat{d}^{(2,-)}_{\mu'}(t')\,\hat{d}^{(1,+)}_{\mu}(t)\right\rangle%
\label{3.14}
\end{equation}
and is phase sensitive to the location of the atoms. In this diagram we kept only the low order vertex-type coupling and the diagram for itself is proportional to a small parameter $\lambdabar^2/r_{12}^2$, where $r_{12}$ is the interatomic separation. It is a crucial property of this graph that it describes the cooperative part of incoherent radiation emitted by two atoms as a complex system. Physically it is based on original interfering contributions to the coherent emission by each of atom, see diagram (\ref{3.7}), but the process is disturbed by interchange of a virtual photon between the atoms with making it incoherent. Considering contribution (\ref{3.13}) together with a similar diagram, with replaced atoms, we have no knowledge which atom has emitted the photon. The diagram overall phase factor is given by Eq.(\ref{3.1}) and if we voluntarily extend our estimate up to a multiatom ensemble and accumulate the contribution of many such terms then we get a perceptible enhancement of the fluorescence in the backward direction. Furthermore, more complex diagrams can be performed for an atomic chain consisting of three and more atoms. Evaluation of these diagrammatic terms in the environment of other contributions is not so straightforward a task since it requires the solution of the Heisenberg-Langevin equations for all the dyadic-type operators of a multiatom system. But as claimed by the set of independent calculations presented in \cite{WGDM04,WGDM05,SMB05,GWDM06,WGDM06,KBS14} such an interference enhancement could survive and be observable in experiment.

However as was pointed in \cite{Labeyrie08} the real experimental conditions are much more complicated and non-linear interaction modifies the macroscopic properties of the medium as well. That requires the complete analysis of light dynamics in the regime of non-linear diffusion. The simple model of the cooperative fluorescence, which we described above, is apparently insufficient. The imperfection of an experimental setup and any mechanisms of decoherence can wash out any presence of the CBS-type interference in the nonlinear excitation regime.

\section{Diffuse transport in the gained and Lambda-configured systems}\label{Section IV}
\setcounter{equation}{0}
\setcounter{figure}{0}

\subsection{Random lasing}\label{4.A}

\subsubsection{Light diffusion in an amplifying medium and the random lasing effect}

Consider the situation when the atomic scatterers are randomly distributed in a homogeneous medium which has an excess in population of its upper energy states and can emit extra photons into the propagating light. Then such an active medium contributes to the retarded-type polarization operator by the sum of the graphs
\begin{equation}
\raisebox{-0.4 cm}{\scalebox{1.0}{\includegraphics*{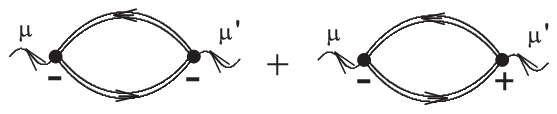}}}%
\ \;\sim \chi_{\mu\mu'}^{(\mathrm{A})}(\omega)%
\label{4.1}%
\end{equation}
see the general definition (\ref{2.17}) and related discussion in section \ref{2.D.1}. The internal lines in these diagrams belong to the atoms associated with the active medium, which are affected by a non-depleting pumping process. The crucial requirement for the medium susceptibility is $\mathrm{Im}[\chi_{\mu\mu}^{(\mathrm{A})}(\omega)]<0$ such that any freely propagating wave is amplified by it. For the sake of simplicity let us additionally assume that the active medium for itself is ideally homogeneous and does not contribute to the scattering process.

Then the derivation of the light transport equation, presented in Section \ref{2.F}, can be partly reconsidered and one finally arrives at the following modified diffusion equation
\begin{equation}
\frac{\partial}{\partial t}W(\mathbf{r},t)-D\,\Delta W(\mathbf{r},t)=%
+\frac{\bar{v}}{\bar{l}_g}\,W(\mathbf{r},t)%
\label{4.2}%
\end{equation}
The right-hand side is responsible for the amplification process and in the case of an isotropic medium, when $\chi_{\mu\mu'}^{(\mathrm{A})}(\omega)=\delta_{\mu\mu'}\chi^{(\mathrm{A})}(\omega)$ the relevant gain length $\bar{l}_g$ is given by
\begin{equation}
\bar{l}_g^{-1}\ =\ -4\pi \frac{\bar{\omega}}{c}\,\mathrm{Im}\left[\chi^{(\mathrm{A})}(\bar{\omega})\right]
\label{4.3}%
\end{equation}
This equation assumes an ideal lossless scattering medium consisting of atoms randomly but more or less uniformly distributed inside the amplifying medium. Thus in accordance with Eq.(\ref{2.63}) its albedo $a=1$ and the scattering medium does not contribute in the right-hand side of Eq.(\ref{4.2}).

Let us apply equation (\ref{4.2}) to a spherical sample of radius $r_0$. The needed boundary conditions can be built up from the physical assumption that any portion of energy density distributed with light via a diffusion process freely passes through the medium boundary and emerges from the sample. Then one obtains
\begin{equation}
-\left.D\frac{d}{d r}W(\mathbf{r},t)\right|_{r=r_0}\ =\ \left.\frac{1}{2}\bar{v}\,W(\mathbf{r},t)\right|_{r=r_0}%
\label{4.4}%
\end{equation}
If the sample is optically thick and the spatial inhomogeneity associated with the macroscopic scale $r_0$ is large enough such that $r_0\gg l_{\mathrm{tr}}$ the condition (\ref{4.4}) can be simplified to $W(\mathbf{r},t)\sim 0$ at the bounding surface. Then the standard analysis of the differential equation (\ref{4.2}) leads to the following important criterion: if
\begin{equation}
r_0>\pi\left(\frac{l_{\mathrm{tr}}l_g}{3}\right)^{1/2}%
\label{4.5}%
\end{equation}
the solution becomes unstable and generates infinite amplification of the light emitted by the active medium.

This criterion was introduced first by Letokhov in \cite{Letokhov} and it determines a threshold requirement for lasing in a disordered system. Of course, in a realistic scenario there cannot be infinite amplification. If the sample volume is large enough such that this threshold is overcome then the amplification initiates the saturating feedback of radiation with the active medium and finally the radiation of the entire system approaches a steady state regime. The role of randomly distributed elastic scatterers in this process consists of effective trapping of the amplified radiation such that they play an equivalent role as the cavity in a conventional laser scheme.

There have been many demonstrations of the random lasing phenomenon in condensed matter systems. For instance, a narrow spectral peculiarity was observed in the emission of metal ions distributed in paint powder \cite{Markushev}. Another example is the radiation of an active dye medium, which can be diffusely trapped by TiO${}_2$ microparticles distributed in the dye solution \cite{Lawandy}. We address the reader to a review \cite{HuiCao} for the representative references related to this important subject. The physical mechanism of lasing from the disordered system is probably also observable in astrophysics as stimulated emission from certain stellar objects, see \cite{LetokhovJohansson}. In laboratory realizations it has a special interest to the systems with a one dimensional configuration where the lasing in the disordered medium can be accompanied by the localization phenomenon, see \cite{MilnerGenack,Matos}.

However in a gas phase the random lasing is still a challenging problem for experimental verification. There is a proposal of the Nice group claiming that this effect could be observed in the system of ultracold alkali-metal atom sample prepared in a large optical depth MOT, see \cite{KaiserRandLaser1}. The unique statement formulated in that paper is that the radiation can be emitted and trapped by the same medium, which is active and scattering simultaneously.  In this case the sample is considered to consists of billions of atoms at a temperature of $\mu$K with extremely high optical thickness. The authors made their estimates based on a two-level model of an optical transition pumped by a strong coherent field $\emph{i.e.}$ for the Mollow-type system and they predicted the optical depth of around several hundred to approach the threshold condition.

But as was recognized later the complicated multilevel structure of an alkali-metal atom would be even more helpful in realization of the random lasing conditions. Indeed in the alkali-metal system the atoms can be repopulated among different hyperfine sublevels and then can be considered as distinguishable fractions of the matter. In the next section we discuss how the amplification and light trapping can be organized in such system.  The progress in the architecture of ultracold atomic systems is rapidly developing and the required conditions can be attainable in the laboratory in the near future. Recently preliminary experimental results, showing that the threshold conditions can in fact be overcome in a system of ultracold atoms, have been reported in \cite{KaiserRandLaser2,KaiserRandLaser3}.

\subsubsection{Raman amplification of the trapped radiation}\label{SecIVA2}

\begin{figure}[tp]\center
{$\scalebox{0.75}{\includegraphics*{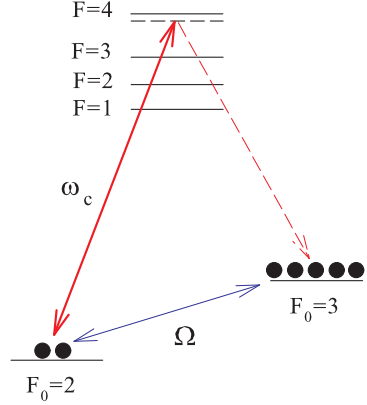}}$ }
\caption{(Color online) Energy levels and the excitation diagram for the Raman process initiated in an ensemble of ${}^{85}$Rb via $F_0=3 \to F_0=2 \to F=3,2$ transitions. The Raman emission is a result of the simultaneous action of microwave $\Omega$ and optical $\omega_c$ excitations, both are linearly polarized along the quantization direction. The emitted light is trapped on the closed $F_0= 3 \to F =4$ transition in the optically thick atomic ensemble. While propagating through the atomic sample this light stimulates additional Raman emission and may increase the output fluorescence emerging the sample.}
\label{fig41}%
\end{figure}

In this section we consider the mechanism of Raman-type amplification for radiation trapped in a disordered atomic medium. Such a mechanism can be organized around the $D_2$-line hyperfine manifold of most alkali-metal isotopes and in Fig.~\ref{fig41} we show ${}^{85}$Rb as an example. Let the rubidium atoms initially populate the upper hyperfine sublevel in their ground state. Then a part of them can be driven to the lower sublevel with a microwave field. With applying the strong coherent field resonant with the forbidden $F_0=2\to F=4$ transition one can initiate the off resonant Raman process via the $F=3,2$ upper states with transporting atoms back to sublevel $F_0=3$. The crucial peculiarity of such an excitation channel is that the light emitted in the spontaneous Raman process will be strongly trapped by elastic scattering on the practically closed $F_0=3\to F=4$ transition if the optical depth of the sample is large. Then the portion of light, originally created in the spontaneous Raman process, and then diffusely propagating though the disordered atomic ensemble, can stimulate additional Raman emission. If we unbalance the trapping conditions in such way that the amplification of the transporting light overcomes its losses, then we would arrive at the situation of a random lasing mechanism as discussed in the previous section. This mechanism has been recently considered in \cite{GerasimovRandLaser}.

Let us specify the physical requirements for the process shown in Fig.~\ref{fig41}. The dipole-type interaction $\hat{V}(t)$ (defined by Eqs.(\ref{2.15}) and (\ref{2.25})) of atoms with the optical mode initiating the Raman process will be assumed to have linear polarization along a $z$-axis of the laboratory frame. The interaction $\hat{U}(t)$ of the magnetic dipole moment with the microwave radiation, see Eq.(\ref{2.25}), will be considered in linear polarization as well. If in order of magnitude the matrix elements of interaction with optical and microwave modes are estimated as $\bar{V}$ and $\bar{U}$ respectively, then we will assume the following relation $\gamma\gg \bar{V}^2\gamma/\Delta_{\mathrm{hpf}}^2\gg \bar{U}$, where $\Delta_{\mathrm{hpf}}$ is the hyperfine splitting in the upper state.\footnote{To simplify notation in these and below estimates of this \ref{4.A} subsection we take $\hbar=1$} This inequality sets that the rate of spontaneous decay is greater than the rate of spontaneous Raman scattering but the latter is greater than interaction with the microwave field. In other word, the interaction with the microwave field is the weakest process in the system.

The above conditions justify that only a small portion of the atoms are repopulated from the upper hyperfine sublevel of the ground state and the stimulated amplification is expected to take place at a longer path length than the diffusion transport length. Then the amplification process can be subsequently considered in perturbation theory and conveniently described in the diagram framework beyond any simplifications associated with light transport or the diffusion equation, see comment at the end of section \ref{2.F.2}. The dynamics of the field obeys the following Bethe-Salpeter type equation for the correlation function of light generated in the Raman process
\begin{equation}
\scalebox{1}{\includegraphics*{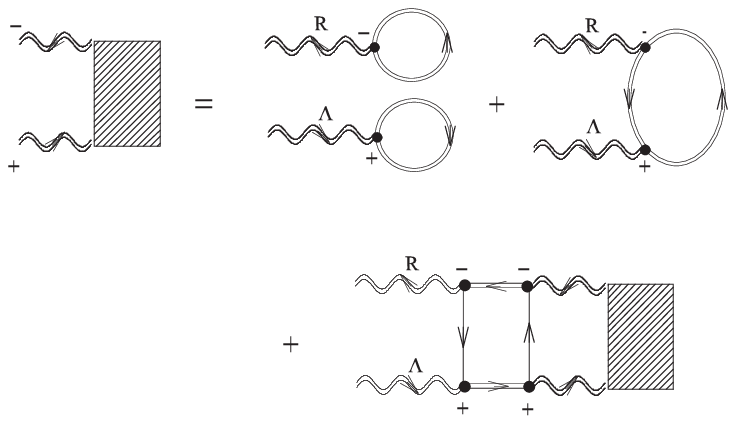}}%
\label{4.6}
\end{equation}
The outward doubly wavy line in this equation respectively performs the exact retarded $(R)$ and advanced $(A)$ propagators of the photon propagating in the nonlinear medium, see definition (\ref{2.18}). The crucial difference between the Bethe-Salpeter equation in the form (\ref{4.6}) with (\ref{2.48}), is just in the presence of these propagators, which make a resource for amplification of the light fragment propagating in any direction.

The seeding terms in the right-hand side of Eq.(\ref{4.6}) are the light sources created by the Raman excitation, shown in Fig.~\ref{fig41} and distributed in the atomic sample. The reader can recognize here the analog of coherent and incoherent excitations, introduced before by Eqs. (\ref{3.7}) and (\ref{3.10}) in the Mollow problem, but now created by the Raman mechanism in the $\Lambda$-type configured system. In reality this observation is not so straightforward and let us clarify this point. Indeed, the source terms in equation (4.6) consist of two contributions. The first term, which has a factorized form, performs the coherent generation of the Raman mode in the three-wave mixing process: $\omega = \omega_c -\Omega$. Since the input and output optical modes are associated with electric dipole type transitions (polar vectors) but the microwave mode with magnetic dipole type transitions (axial vector), and all three waves have the same polarization, this coherent process is forbidden if the atoms symmetrically populate the Zeeman states of different signs. So the first contribution vanishes in the considered excitation geometry. Such an excitation channel would be allowed only if the atomic medium had a spin angular momentum polarization in its ground state. Thus, only the second term corresponds to the process of spontaneous Raman scattering on the atoms, populating the sublevel $F_0 = 2$ and dressed by interaction with the driving modes as shown in Fig.~\ref{fig41}, and providing the actual contributing emission source. The random emission events are uncorrelated in this case.

The last term in Eq.(\ref{4.6}) is responsible for multiple re-scattering of the emitted light, $\emph{i.e.}$ for the diffusion process, which would further develop in the optically dense medium in the nonlinear regime. The basic requirements to consider the diagram equation (\ref{4.6}) as a mathematically closed integral equation claim that the retarded/advanced Green's function and the scattering tensor should be known in analytic form. Then evaluation of the retarded propagator repeats the calculation scheme described in section \ref{2.E}. We only have to take into account the complete structure of the retarded type polarization operator given by the sum of diagrams (\ref{4.1}). The details of the calculations are presented in \cite{GerasimovRandLaser} and we only discuss here the final results.

The diffusive propagation of light mostly depends on the following kinetic parameters responsible for the light transport. The imaginary part of the sample susceptibility determines the extinction length for the plane wave entering the sample at the $e^{-1}$ level of losses. The inverse extinction length is given by
\begin{equation}
l_{\mathrm{ex}}^{-1}(\omega)\ =\ n_0\,\sigma_{\mathrm{ex}}(\omega)=4\pi k\, \mathrm{Im}\left[\chi(\omega)\right],%
\label{4.7}
\end{equation}
where $n_0$ is a typical local atomic density and for the sake of simplicity we omitted in this estimate its spatial dependence. In an anisotropic sample this quantity can be defined only for a plane wave propagating along a specific direction associated with the main reference frame, such that $\chi(\omega)$ in Eq. (\ref{4.7}) can be any of the major components of the susceptibility tensor. Otherwise the definition has to depend on the propagation direction and track the polarization properties of light. It is important to point out that in the case of amplification the extinction length accumulates not only losses but also gain associated with the stimulated Raman scattering and we distinguish its notation with the previous definition (\ref{2.59}). This parameter can even be negative in the case of population inversion and then it indicates the length scale for $e^{+1}$ amplification.

\begin{figure}[tp]\center
{$\scalebox{0.8}{\includegraphics*{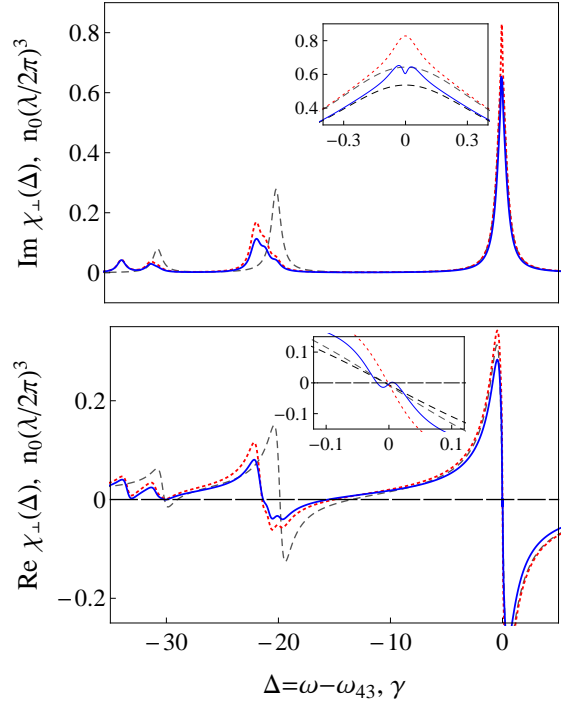}}$ }
\caption{(Color online) Spectral dependence of the imaginary part (upper panel) and real part (lower panel) of the $\chi_{\bot}(\omega)$ component of the susceptibility tensor in units of the dimensionless local density of atoms $n_0\lambdabar^3$. The gray dashed curve reproduces the original undisturbed multiplet of the D$_2$ line of $^{85}$Rb probed from the $F_0=3$ ground state hyperfine sublevel [Fig. \ref{fig41}]. The red dashed curve indicates how this multiplet is modified by the interaction with the optical control mode and the blue solid line shows the spectral profile affected by both the optical and microwave modes. The Rabi frequency for the optical transition (specified by the reduced matrix element for $D_2$-line) has a value of $2\bar{V}= 25\gamma$, for the microwave mode it was selected for the ``clock'' transition $F_0=2,M_0=0\to F_0=3,M_0=0$ and chosen as $2\bar{U}= 0.05\gamma$. The details of spectral behavior near the $F_0=3\to F=4$ resonance are magnified in insets, where the black dashed curve shows how the isotropic part (gray dashed) is reduced by depopulation of the $F_0=3$ state. }
\label{fig42}%
\end{figure}

As an example, in Fig.~\ref{fig42} we reproduce the susceptibility component $\chi_{\bot}(\omega)$ for the probe propagation in a direction orthogonal to the polarization vector of the optical control mode. For the presented calculations the Rabi frequency of the optical mode was specified by the reduced matrix element for the $D_2$-line as $2\bar{V}=25\gamma$ and the Rabi frequency for the microwave mode was defined by the matrix element for the "clock" transition $F_0=2,M_0 =0\to F_0=3,M_0=0$ and given by $2\bar{U}=0.05\gamma$. The details of evaluation of the matrix elements for the electric dipole and magnetic moment operators, contributing in the interaction terms (\ref{2.25}) and (\ref{2.26}), are specified in \ref{Appendix.C}. The graph illustrates how the original hyperfine energy spectrum is distorted by the Autler-Townes effect, see \cite{AutlerTownes,LethChebt}. As clarified in the  inset that creates additional quasi-energy resonance near the $F_0=3\to F=4$ transition, indicated by a dashed bar in Fig.~\ref{fig41}, which overlaps with the original Lorenzian shaped resonance and plays a crucial rule in diffusion and trapping of the emitted light.

Another kinetic parameter responsible for the scattering process, and in particular for light trapping, is the scattering length, which is given by
\begin{eqnarray}
l_{\mathrm{sc}}^{-1}(\omega)&=&n_0\,\sigma_{\mathrm{sc}}(\omega)%
\nonumber\\%
\sigma_{\mathrm{sc}}(\omega)&=&\frac{\omega\omega'^3}{c^4}\frac{1}{2F_0+1}\sum_{m'\!,m,\mathbf{e}'}\int\left|\alpha_{\mu'\mu}^{(m'm)}(\omega)e'_{\mu'}e_{\mu}\right|^2d\Omega%
\label{4.8}
\end{eqnarray}
where we introduced the scattering cross section $\sigma_{\mathrm{sc}}(\omega)$ with the scattering tensor earlier defined by Eq.(\ref{2.33}). Here $\omega,\,\mathbf{e}$ and $\omega',\,\mathbf{e}'$ are the frequencies and polarization vectors of the input and output photons respectively and the integral is over the full scattering angle. Similarly to the extinction length, this quantity critically depends on the polarization direction of the incident photon.

Light diffusion mostly depends on elastic contributions when $\omega=\omega'$ and both $m$ and $m'$ belong to the same $F_0=3$ level. With keeping only elastic contribution we can define the characteristic length associated with losses,
\begin{equation}
l_{\mathrm{ls}}^{-1}(\omega)\ \equiv\  -l_{\mathrm{g}}^{-1}(\omega)\ =\ l_{\mathrm{ex}}^{-1}(\omega)-l_{\mathrm{sc}}^{-1}(\omega) ,%
\label{4.9}
\end{equation}
which evaluates the averaged distance that a photon travels before being lost by an inelastic scattering event. The first equality indicates that in an amplifying medium it is possible to have this quantity negative and then redefine it as the gain length $l_{\mathrm{g}}(\omega)$. In such a gain medium the radiation trapping and amplification mechanisms can overcome the instability point, entering the regime of random Raman laser generation.

\begin{figure}[tp]\center
{$\scalebox{0.8}{\includegraphics*{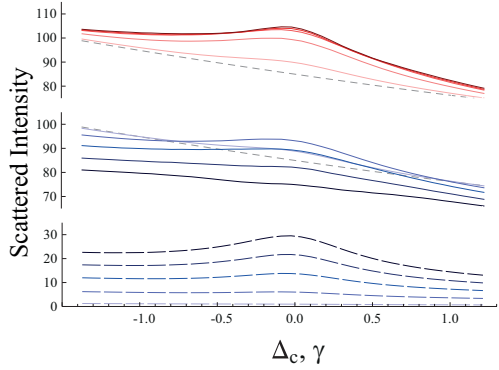}}$ }
\caption{(Color online) Intensity of the light (number of photons per time unit and per spectral unit) scattered at the angle $\theta=45^0$ to the polarization direction of the driving fields, see Fig.~\ref{fig41}. The frequency of the optical mode scans vicinity of the $F_0=2\to F=4$ forbidden transition and $\Delta_c$ is the respective detuning. The lower panel shows the contribution of the inverse anti-Stokes scattering, the middle panel is the contribution of elastic scattering, and the upper panel gives the sum for both the channels. The curve thickness in the plotted graphs is associated with optical thickness changed from lower to higher values $b_0=1,5,10,15,20$.}
\label{fig43}%
\end{figure}

In Fig.~\ref{fig43} we show one example of the Monte-Carlo simulations for the spectral variation of the intensity of the light originally emitted by the Raman process and further scattered by the atomic ensemble under the trapping conditions. The detection channel is chosen at the angle $\theta=45^0$ to the polarization direction of the driving fields. In the presented data the frequency of the optical driving mode was scanned near the forbidden $F_0=2\to F=4$ transition, the Rabi frequencies for both the driving modes are the same as in Fig.~\ref{fig42}, and the optical thickness of the atomic sample on $F_0=3\to F=4$ was varied from low to high values $b_0=1,5,10,15,20$. For the sake of simplicity we omit in our discussion some specific details concerning the coordination of the microwave and light induced energy shifts and the problem with the spectral distribution of the emitted light, see \cite{GerasimovRandLaser}. We just point out that the frequency detuning of the optical control mode $\Delta_c=\omega_c-\omega_{42}$ in the spectra shown in Fig.~\ref{fig42} incorporates the distortion of the energy structure associated with these effects.

The graphs of Fig.~\ref{fig43} are non-trivially explainable via the broad spectral domain of the emitted radiation which undergoes a sequence of scattering on rather complicated Autler-Townes resonance structures created by the driving fields, see Fig.~\ref{fig42}. The susceptibility spectrum, modified by both the control optical and microwave modes, was calculated in \cite{GerasimovRandLaser}. As was pointed out above, the $\chi^{(3)}$-type nonlinearity (with respect to the optical mode) in particular manifests itself in creation of additional quasi-energy resonances located in the atomic spectrum near the frequency of the optical mode as indicated by the dashed bar in Fig.~\ref{fig41} and visualized in the inset of Fig.~\ref{fig42}. The quasi-energy level always interacts with the modes emitted in the Raman process and traps them even if the optical mode is tuned far off-resonance from the $F =4$ reference level $\emph{i.e.}$ in the normally transparent spectral domain. But such a trapping mechanism is not closed because the interaction via the quasi-energy level opens inelastic inverse anti-Stokes scattering channels as well. In terms of a diffusion model, see Eq.(\ref{4.2}) and surrounding comments, the albedo $a<1$ in the considered case. The anti-Stokes scattering process redistributes the Raman emission into those spectral modes, which further escape the sample without any absorption, and in the context of the random lasing effect this process should be treated as responsible for additional losses. When the optical mode scans the vicinity of the upper level $F=4$ the situation noticeably changes and the trapping effect, also initiated by elastic scattering on the closed $F_0=3\to F=4$ transition, leads to additional amplification, which is visualized as a bump-shaped enhancement of the spectra. However as follows from the presented data a significant part of the light still escapes the sample via a inelastic inverse anti-Stokes scattering channel.  Although this type of scattering essentially reduces the light intensity transporting via the elastic channel, when the optical mode scans the vicinity of the $F_0=2\to F=4$ transition it nevertheless stimulates Raman amplification of the trapped light, which can be seen in both the elastic and inelastic channels.

\subsubsection{Random lasing in a inhomogeneous and disordered system of cold atoms}\label{SecIVA3}

As was commented on in \cite{GerasimovRandLaser} the problem with the extra losses, related to the control mode nonlinearity, can be solved for in a spatially inhomogeneous configuration. The crucial requirement is to separate the amplification and trapping areas. In the amplification area the atoms have to populate only the  pumping level i.e. $F_0=2$ in the considered case. That eliminates any inverse scattering processes. If this active area was surrounded by the atoms trapping the emitted light, that would induce a certain soft cavity feedback and can create instability in the system. Such option has been recently demonstrated by a round of Monte-Carlo simulations in \cite{GerasimovRandLaserI}

\begin{figure}[tp]\center
{$\scalebox{0.6}{\includegraphics*{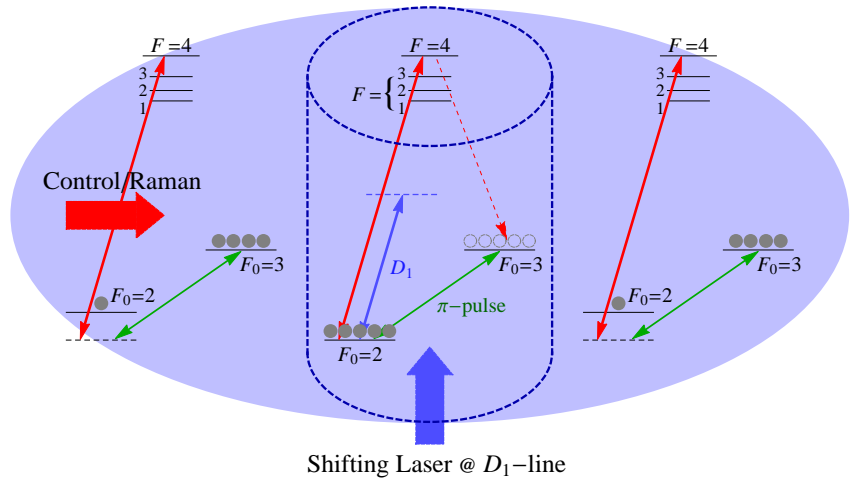}}$ }
\caption{(Color online) The excitation geometry, energy structure and transition diagram for the spatially inhomogeneous atomic system. The ${}^{85}$Rb atoms located in the cylindrical volume crossing the middle part of the cloud have the lower energy level shifted by the light shifting laser operating near resonance with a hyperfine component of the $D_1$-transition. After applying a $\pi$-type microwave pulse these atoms populate the hyperfine transition $F_0=2$. The control mode is tuned in resonance with these atoms and initiates the initial Raman-type spontaneous emission. The emitted light is trapped by such a soft cavity and at certain conditions can enter the random lasing regime.}
\label{fig44}%
\end{figure}

In Fig.~\ref{fig44} we show the proposed experimental architecture, which implies preparation of a spatially inhomogeneous energy structure and population distribution of the hyperfine sublevels in the atomic ensemble. This can be performed via a controllable light shift of only one hyperfine energy level for the atoms located in a spatially selected volume of cylindrical symmetry inside the atomic cloud. If we assumed the level $F_0=2$ as light shifted and organized the population inversion for all the atoms in the selected volume onto this level (for example, with $\pi$-type microwave pulse) then these atoms could form an active medium for the photon emission. Indeed, as follows from the transition scheme, shown in Fig.~\ref{fig41}, and explaining diagrams in Fig.~\ref{fig44}, the control mode would create the photon emission only for the atoms inside the selected volume and the interaction would be off-resonant with the control field for the other atoms of the ensemble. Then the atoms outside of the active volume would trap the emitted light and play the role of a soft cavity redirecting the light in a quasi-one-dimensional propagation channel associated with that volume. That could lead to instability if each spontaneously emitted photon would have a diffusion path long enough for stimulation of extra photon emission while it propagates through the channel.

\begin{figure}[tp]\center
{$\scalebox{0.8}{\includegraphics*{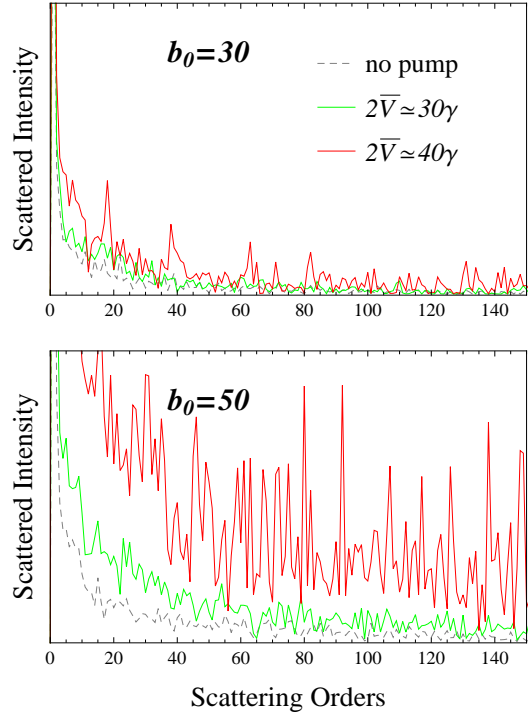}}$ }
\caption{(Color online) Intensity distribution over the scattering orders for a pointlike dipole-type light source emitting light from the center of an atomic sample. The light emerges from the trapping system along the channel shown in Fig.~\ref{fig44} and is detected at infinity at the angle of the channel direction. The gray dashed curve on both the panels refers to the intensity distribution in different scattering orders without the optical driving field, i.e., without amplification. Other curves show the Raman-type amplification induced by the control field (with the reduced transition matrix element $\bar{V}$) of different pump intensities with the Rabi frequencies $2\bar{V}=30\gamma$ (green curve) and $2\bar{V}=40\gamma$ (red curve). In this round of the Monte-Carlo simulations only the incoherent scattering mechanism were taken into consideration and were ignored additional contributions to the trapping process due to coherent scattering from the channel boundaries. Nevertheless the upper and lower panels, plotted for slightly different values of the sample optical depth $b_0=30$ and $b_0=50$, clearly indicate that the amplification process approaches unstable behavior for a critical optical depth. This instability can be associated with transforming the
emission process towards the regime of random laser generation.}
\label{fig45}%
\end{figure}

In Fig.~\ref{fig45} we show an example of our Monte-Carlo simulation performed for such a system under conditions demonstrating the instability behavior. These graphs track how the intensity of a single external light source located in a central point inside such a soft cavity is expanded in different orders of light scattering and depends on the amplitude of the pumping initiated by the control mode. At low pump intensity we observe an amplification of the outgoing light intensity but converging sum over a limited number of the scattering orders. At high intensity of the control mode the process becomes diverging and unstable. This effect depends on the optical density of the surrounding atoms $b_0$ and is more evident as far as the optical depth is higher. We can associate this instability with transforming the emission process towards the regime of random laser generation.

Preparation of a controlled realization of a soft cavity for this and other potential applications is a challenging experimental enterprize. But some basic elements of the soft cavity configuration have been already demonstrated in the laboratory \cite{OWKRBHSK,RKHSK}. One important constraint is that the density of the atomic sample be relatively high, so that radiation following optical excitation within the soft cavity has a tendency to be confined within the cavity. This suggests that one route towards experimental implementation would be to use an optical dipole trap to achieve the necessary high density. A second constraint would be that the soft cavity should be transversely small, on the order of a few microns or less, in order to limit the number of transverse optical quasimodes within the approximately cylindrical channel geometry. Further discussion of experimental feasibility of the scheme is performed in \cite{GerasimovRandLaserI}.

As a certain alternative to cylindrical channel geometry, which seems also feasible for experimental verification, the active medium can be fabricated with the atoms arrayed along a dielectric nanofiber passed through a magneto optical trapped atomic sample. The idea implicates experimental architecture originally proposed in \cite{BHKLM04,KBH04} and later successfully demonstrated in \cite{GCADLPTSK12} as a nanofiber trap designed for locking cesium atoms. The locked atoms are located in the wells formed by two far off-resonant red and blue detuned laser standing waves confined with the fiber.  In the random lasing configuration the locked atoms could be involved in the Raman pumping as active medium and their spontaneous emission can be further trapped and fed back by the soft cavity environment similar to the situation shown in Fig.~\ref{fig44}. The trapped radiation would stimulate spontaneous emission such that above a critical point the Raman emission would transform towards the lasing regime. The outgoing light would emerge the system via evanescent modes propagating along the fiber. Strictly speaking we would deal here not with a fully random laser configuration but perhaps with a combination of conventional cavity, associated with the dielectric fiber, and trapping feedback associated with elastic scatterers forming the soft cavity and incoherent internal reflection. Nevertheless such a light source could be interesting for applications as a possible source of non-Gaussian states of light, which could contain a few and well defined number of photons in a certain time interval.

\subsubsection{Coherence properties of the random laser radiation}

At present there is no microscopic quantum theory of the random laser above threshold and in the saturation regime. The performed analysis partly explains that such a theory would be not so easy to develop. There are mostly phenomenological approaches presented in the literature: some extension of a diffusion model is done in \cite{WiersLagnd}, and for more information and representative references we address the reader to the important review \cite{HuiCao}.

As known from the standard theory of a conventional cavity laser, see \cite{Haken}, the laser radiation below threshold performs an amplified spontaneous emission of the atoms of the active medium considered as a random Langevin source. The spectral properties of the initial spontaneous emission are further modified by the amplification gain and by decay dynamics of the cavity mode. In this sense we see a certain analogy with equation (\ref{4.6}) and the standard single mode subthreshold laser equations in a cavity. The source term performs the spontaneous emission, which would exist in any type of the pumping mechanisms in any laser. We emphasize that just amplification of this weak radiation source should be associated with any precursor of further laser generation. The main difference between cavity trapping and diffusion trapping is that the latter does not select any specific spatial spectral mode and the spectral properties of the emerging light are mostly controllable by the gain spectral profile, see (\ref{4.7})-(\ref{4.9}).

Let us briefly discuss the coherent properties of the radiation emerging from a disordered medium in the lasing regime. A common vision is that the diffusion approach and the radiative transport equation are not consistent to describe the wave nature of light. Such kinetic-type equations can only track the amplification process and be applicable to identifying a steady state energy balance between the active medium and the emitted radiation. The precise theory should turn us back to the complete diagram expansion of the correlation function and to the entire Green's function formalism discussed in the previous section but substantially extended up to the saturation regime. In addition, we should introduce a self-consistent master equation for the density matrix of the atomic subsystem, or more strictly for the atomic correlation functions, in the complete form performed by the diagram equation (\ref{2.21}).

In the considered example of the Raman laser the source generation originally occurs only in a narrow spectral domain in the radiation spectrum associated with the rate of spontaneous Raman emission: $\Delta\omega\sim\bar{V}^2\gamma/\Delta_{\mathrm{hpf}}^2$. If the amplification process, described by the Bethe-Salpeter equation (\ref{4.6}), involves high scattering orders then it would become more effective for those modes, which are closer to the Raman resonance conditions. That would make the output radiation more monochromatic and therefore more coherent. We also expect that spatial inhomogeneity is an important requirement, see our comment in the end of previous section, that could lead to the spatial self-organization of the emission in the saturation regime. So we point out the difference between the standard radiation trapping phenomenon and the random lasing effect, as well as between a conventional laser and a random laser. The quantum nature of the state of light also seems a challenging problem to consider. But again, as we pointed in the beginning of this subsection, better theory is required for describing the random laser generation and its statistical peculiarities in the saturation regime.

\subsection{Electromagnetically induced transparency with scattered light}\label{4.B}

In this section we are concerned with some aspects of coherent control of the scattering and propagation properties of light in cold atomic gas samples. This is a very large field of study, and has found many applications in spectroscopy \cite{Lukin1,CMJLKK05,MatzKuz09,DJDBBSMBG05,HFK92,BBYH03,BGKRYW02,BKRY99}, in quantum information, single photon manipulation \cite{Lukin1,CMJLKK05,MatzKuz09}, and high precision magnetic field measurements \cite{BGKRYW02,BKRY99}. Many of these approaches have employed two photon configurations typical of investigations of electromagnetically induced transparency (EIT) or absorption (EIA). A remarkable early paper \cite{HHDB99} for instance, demonstrated the strong effect that an external dressing electromagnetic field can have on the propagation of a weak probe in such a medium.

Many aspects of coherent control of optical properties of cold atomic gases have been studied in detail \cite{Lukin1,Fleischhauer}. However nearly all studies of electromagnetically induced coherent effects deal with the properties of the transmitted probe light, or, in terms of scattering theory, with properties of the coherently forward scattered radiation. At the same time, it is obvious that auxiliary external quasi-resonant electromagnetic fields influence also incoherent scattering of the probe light, $\emph{i.e.}$ its scattering into other modes. Incoherent scattering is important from different points of view. It is well known that the scattered field in optically dense media strongly influences such effects as optical pumping and optical orientation \cite{Happer,PetAnd91,SSWCL96,ASMSL93,Fleischhauer99}. In \cite{MNSW01} the influence of incoherent scattering on electromagnetically induced transparency in gas cells was experimentally studied. It was found there that the effective decay rate of Zeeman coherence, generated in the atomic vapor by laser light, increases significantly with the atomic density.  This effect was explained as a result of radiation trapping. It gave the authors of \cite{MNSW01} the opportunity to make a conclusion that to fully understand many electromagnetically induced transparency experiments with optically thick media one has to take into account multiple light scattering. In cold atomic ensembles the incoherent scattering is even more important because it influences additionally the optical cooling process. It restricts the minimal temperature of atomic clouds in traps \cite{MetStra99}.

In all these studies, the effect of radiation trapping was described as an external incoherent pumping of the atomic transitions. The influence of the control field on the properties of secondary radiation was not analyzed. Such influence on subsequent dynamics of the light scattered out of the coherent beam, that is to say the propagating diffusive flux normally present in an optically deep atomic medium, was studied theoretically in \cite{SKH11,DSKH06,DSKH08}.

The key problem in the description of the scattered light properties lies in accounting for anisotropic polarization properties of the atomic gas caused by the control mode and with different propagation directions of this mode and the scattered light.  The dependence of an electromagnetically induced transparency resonance on the angle between control and probe beams in an atomic cesium vapor in a gas  cell was measured in experiment \cite{CLT04}. The strong influence of the non-collinearity of the beams on the broadening of the EIT resonance was observed. The authors explained this effect as a result of different Doppler shifts for different atomic transitions. Note that a similar effect was noted earlier in  \cite{YeZib02}, though in experiment \cite{ACV91} no influence of  the nonparallelism was found.

\begin{figure}[tp]\center
{$\scalebox{1.05}{\includegraphics{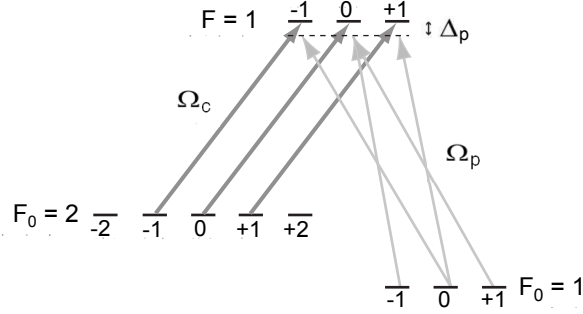}}$}
\caption{Relevant ${}^{87}$Rb hyperfine lambda scheme.  The nearly-resonant strong control laser is linearly polarized along $z$-direction and has Rabi frequency $\Omega_c$ (defined in respect to $F_0,M_0=2,0\to F,M=1,0$ transition). The probe mode has Rabi frequency $\Omega_p$, and detuning from the bare atomic resonance of $\Delta_p$. The transitions are shown for orthogonal mutual polarizations of the probe and control fields as in experimental geometry shown in Figs.~\ref{fig48} and \ref{fig49}.}
\label{fig46}
\end{figure}

The anisotropy induced by the control field in the case of very cold ensembles, where the Doppler effect plays a negligible role, was investigated in \cite{DSKH08}. The anisotropic properties of the gas were described with a susceptibility tensor given by Eq.(\ref{2.32}) where the retarded-type atomic propagator could be expressed by the diagram equations (\ref{2.30}) and (\ref{2.31}). For the experimental configuration in the example of ${}^{87}$Rb, and keeping only one hyperfine sublevel of the excited states, see Fig.~\ref{fig46}, we obtain the following analytical expression for the atomic propagator
\begin{eqnarray}
{\cal G}_{nn'}^{(--)}(E)&=&\delta_{nn'}\,{\cal G}_{n}(E)%
\nonumber\\%
{\cal G}_{n}(E)&=&\hbar\left[E - E_n + i\hbar\frac{\gamma}{2} - \frac{|V_{nm'}|^2}{E\!-\!\hbar\omega_c\!-\!E_{m'}}\right]^{-1}%
\label{4.10}
\end{eqnarray}
The control mode with frequency $\omega_c$ is linearly polarized along the quantization direction $z$ and, as evident from the transition scheme, the propagator has only diagonal components such that each ground state Zeeman sublevel $m'=m'(n)$ is coupled with only one excited state sublevel $n$.

For a probe beam, which experiences scattering and changes its propagation direction, its local reference frame is arbitrarily oriented with respect to the laboratory frame. The susceptibility tensor components in the local frame can be built up via the rotational transformation, which is more natural to perform in the angular momentum (spherical) basis, see our comment prefacing Eq. (\ref{2.44}). Then the transverse components (with $q_1,q_2=\pm 1$) contributing in the left-hand side of Eq.(\ref{2.44}) are expressed by the susceptibility tensor components (\ref{2.32}) in a laboratory frame as follows
\begin{eqnarray}
\tilde{\chi}_{q_1}{}^{q_2}(\mathbf{r},\omega )&=&\sum_{q_1'q_2'}D^{1}_{q_1'q_1}(\alpha,\beta,\gamma)D^{1\ast}_{q_2'q_2}(\alpha,\beta,\gamma)%
\;\chi_{q_1'}{}^{q_2'}(\mathbf{r},\omega )%
\label{4.11}
\end{eqnarray}
Here $\alpha,\beta,\gamma$ is a set of Euler angles for transforming from the laboratory frame to the local frame and $D^{k}_{qq'}(\alpha,\beta,\gamma)$ (with $k=1$ in our case) are the Wigner $D$-functions responsible for this transformation for $k$-irreducible covariant components, see \cite{LaLfIII,VMK}. The contravariant components are transformed via the complex conjugated $D$- functions.

\begin{figure}[tp]\center
$\scalebox{1.0}{\includegraphics*{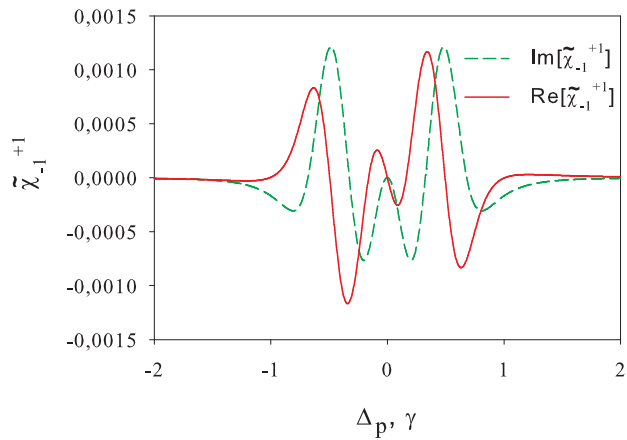}}$
\caption{(Color online) Real and imaginary part of a non-diagonal component $\tilde{\chi}_{-1}{}^{1}=\tilde{\chi}_{-1}{}^{1}(\Delta_p)$ of the susceptibility tensor scaled in units of $n_0\lambdabar^3$. The calculations are performed for a probe beam orthogonal to the control laser polarization direction for $\beta=\pi/2$ and for the Rabi frequency $\Omega_c=\gamma$, see. \ref{fig46}.}
\label{fig47}%
\end{figure}

In the considered multilevel configuration optical anisotropy appears even if the atoms equally populate the Zeeman sublevels. As a consequence, the EIT effect becomes strongly modified by its dependence on the angle between the probe and control fields wave vectors. If the probe propagates in an arbitrary direction the medium has a signature of birefringence and mutually orthogonal circular polarization components interfere in the probe transmitted through an optically deep sample such that the light becomes elliptically polarized. The generated orthogonal component is very sensitive to varying the experimental conditions. In Fig.~\ref{fig47} we show the spectrum of the real and imaginary parts of the off diagonal element of the susceptibility tensor for $\beta=\pi/2$, which indicate rather complicated spectral behavior.

For description of the scattering process in disordered media we have to introduce the scattering tensor (\ref{2.34}) in the presence of the control field.  This scattering process includes both elastic Rayleigh and elastic and inelastic Raman scattering channels. The entire dynamics of the probe pulse in the atomic sample can be realistically simulated by a relevant Monte Carlo scheme for the subsequent multiple scattering process as similarly described in the previous section for random lasing. Detailed theoretical description of the dynamics of pulse scattering under EIT conditions were performed in \cite{DSKH06,DSKH08}. It is shown that in realistic conditions with non-ideal transparency a significant portion of the incident probe pulse could be transferred into Rayleigh and Raman incoherent scattering channels. The light scattered into the Rayleigh channel emerges from the sample with an EIT time delay. This effect was checked experimentally in \cite{OWKRBHSK}. In that research, a particular hyperfine lambda scheme in a spatial geometry where the probe and control fields were collinearly incident on a sample was studied. Light emission from a cold atom sample was detected in a direction orthogonal to the propagation axis and the respective results of numerical simulations for specific experimental conditions were compared with experimental data.

Here we provide some illustrative theoretical and experimental results from \cite{OWKRBHSK}. In general, emission of optically polarized fluorescence from dressed atoms can show complex behavior reflecting the atom-light dynamics of the system.  This is particularly true when we consider the combined time and polarization evolution of the emitted radiation. These results show some of the dynamical complexity that can result through representative time-resolved results of optically polarized probe and control field induced sample fluorescence in a $90^0$ geometry. The illustrative results pertain to the transition diagram of Fig.~\ref{fig46}.

\begin{figure}[tp]\center
{$\scalebox{1.00}{\includegraphics*{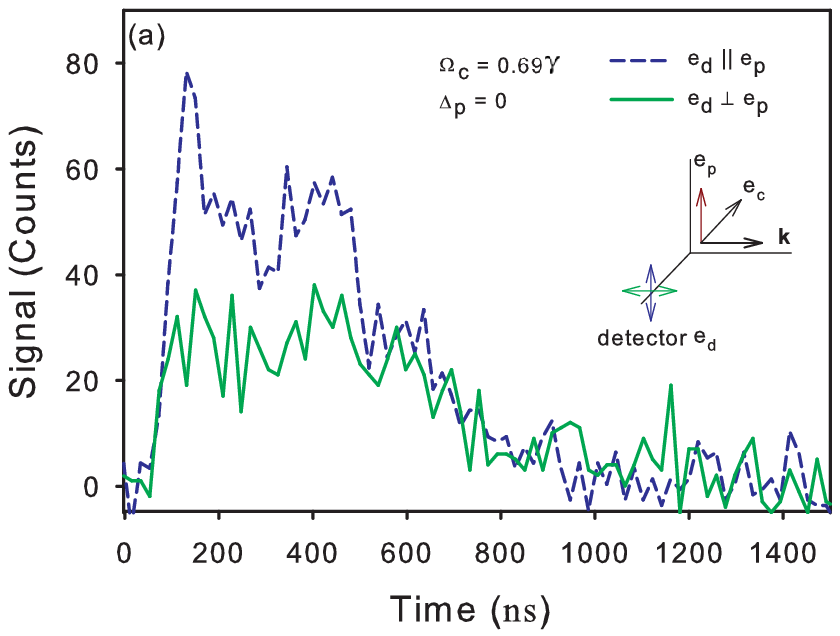}}$ }
{$\scalebox{1.0}{\includegraphics*{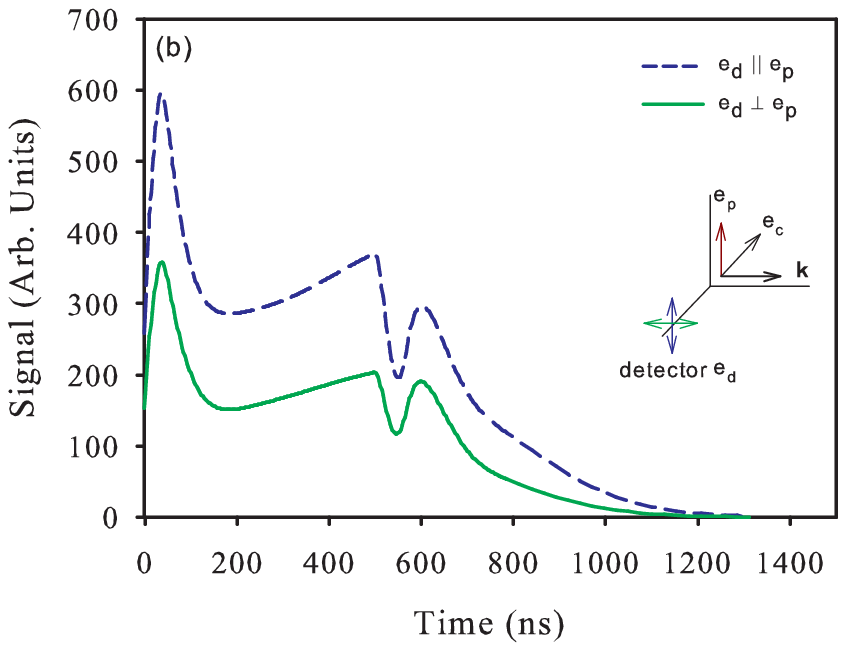}}$ }
\caption{(Color online) Comparison of (a) experimental results and (b) theoretical calculations of the time-dependent polarized fluorescence under similar conditions.  For these results the probe linear polarization is vertical while the control field linear polarization is horizontal. For both (a) and (b) two results are shown; in one case the linear polarization of the detected light is in the same direction as the probe linear polarization, while in the other they are orthogonal.}
\label{fig48}
\end{figure}

\begin{figure}[tp]\center
{$\scalebox{1.02}{\includegraphics*{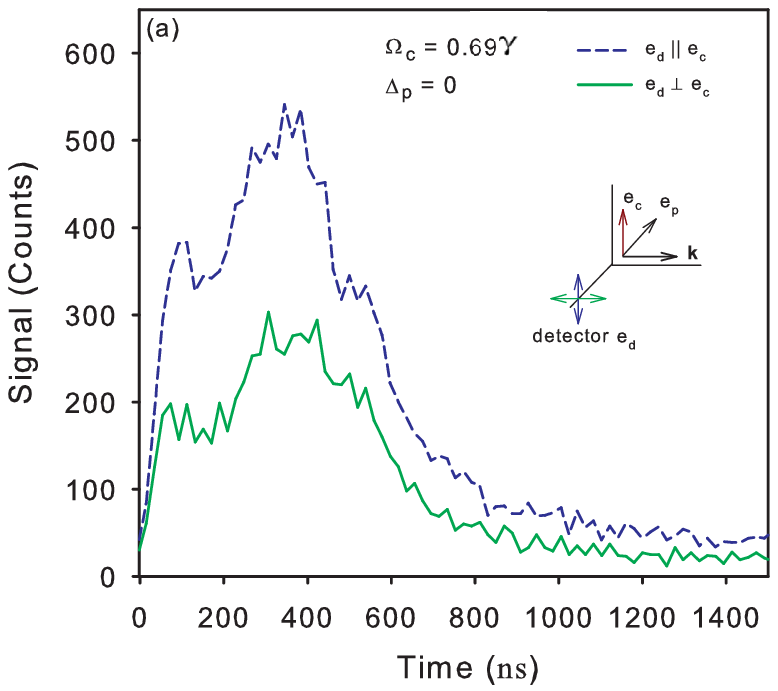}}$ }
{$\scalebox{0.95}{\includegraphics*{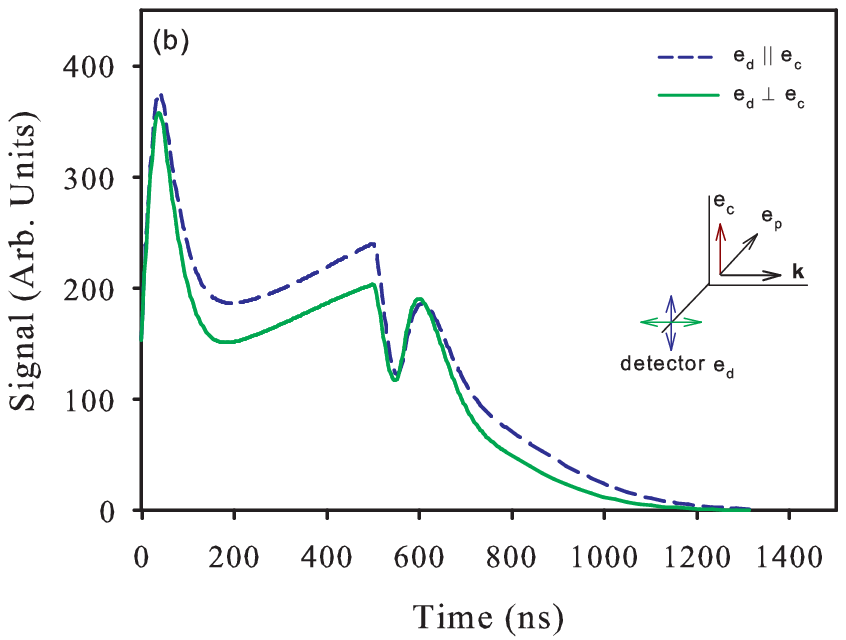}}$ }
\caption{(Color online) Comparison of (a) experimental results and (b) theoretical calculations of the time-dependent polarized fluorescence under similar conditions.  For these results the control field linear polarization is vertical while the probe field linear polarization is horizontal. For both (a) and (b) two results are shown; in one case the linear polarization of the detected light is in the same direction as the control field linear polarization, while in the other they are orthogonal.}
\label{fig49}
\end{figure}

To show particular possibilities, we consider four cases for which the experimental data and calculation results are compared in Figs.~\ref{fig48}, \ref{fig49}. Note that these transient responses result from an approximate rectangular probe pulse which has a slight overshoot on the rising edge.  This overshoot is partly responsible for the rather sharply peaked initial transient seen in all four spectra.  The rectangular shape of the pulse was taken into account in the theoretical calculations as well. The presented results differ by their excitation and detection geometry. In plots of Fig.~\ref{fig48} the probe field is vertical, the control field polarization vector is in the detection plane, and results are obtained for the fluorescence signals with the detector polarization analyzer either collinear with the probe field direction or perpendicular to it.  In the other two cases, plotted in Fig.~\ref{fig49}, the control field polarization is vertical, and the probe field vector lies in the detection plane.   Again, measurements are made of the fluorescence signals alternately with the detector polarization analyzer collinear with the control field direction and perpendicular to it.  These experimental polarization configurations are illustrated by insets to Figs.~\ref{fig48}(a),(b) and Figs.~\ref{fig49}(a),(b).

Experimental time-resolved fluorescence signals (from \cite{OWKRBHSK}) with the $\emph{probe field}$ vertically polarized, and for two orthogonal states of detected light linear polarization, are shown in Fig.~\ref{fig48}(a).  In the figure, the overall evolution of the intensity is a rise to an approximate steady state followed by decay of the signal after the probe is extinguished.   The fluorescence also has a nonzero degree of linear polarization, with an average linear polarization degree in the range $30 - 35 \%$, while the probe and control field are both on. If this excitation were purely from the probe $F_0=1\rightarrow F=1 \rightarrow F _0=1$ transition, then the linear polarization degree would be about $33 \%$, well within the uncertainty of the measured polarization as in Fig.~\ref{fig48}(a).  On the other hand, addition of contributions from the $F_0=1 \rightarrow F=1 \rightarrow F_0=2$ Raman transition, a linear polarization of 3/11 ($\sim 27$ \%) would result.  Theoretical calculations for similar conditions are shown in Fig.~\ref{fig48}(b).  It is evident that there is good qualitative agreement between theoretical and experimental results during the optically driven phase of the process.  During the transient shut-off phase of the dynamics, however, the theoretical results show a clear linear polarization dependence, whereas the measurements are essentially unpolarized. The fluorescence in the actual experiments must then emerge from spectral parts of the probe field that are outside the transparency window.

As another illustration through experimental results, consider time-resolved fluorescence signals (from \cite{OWKRBHSK}) with the $\emph{control field}$ vertically polarized.  Again there are two orthogonal linear polarization directions relevant to the fluorescence signals.  In Fig.~\ref{fig49}(a), the overall evolution of the measured intensity is more structured than in the previous case.   In this case the average linear polarization degree is about $35 \%$. Analogous theoretical results are shown in Fig.~\ref{fig49}(b).  As before, we see qualitative agreement between the theoretical and experimental results.  However, in this case the linear polarization degree as determined from theory and experiment are similar, even in the transient decay after the probe and control beams are extinguished.

\begin{figure}[tp]\center
{$\scalebox{1}{\includegraphics*{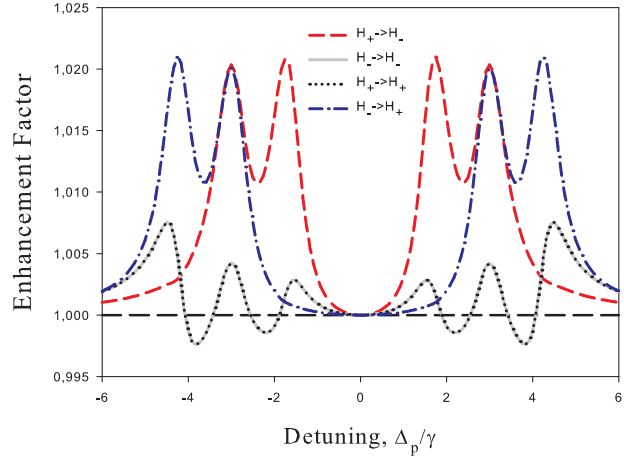}}$ }
\caption{(Color online) Spectrum of the CBS enhancement factor for different helicity channels, $\Omega_{c}=3\gamma$, see Fig.~\ref{fig46}.}
\label{fig410}
\end{figure}

To close this subsection let us turn back to our discussion in Section \ref{SecIII} and consider a coupled area of research, which is evidently connected with propagation in optically dressed media, and is concerned with multiple coherent scattering of quasi-monochromatic and near-resonant radiation in cold atomic gases.  In this case, multiple scattering of the field leads to interferences normally associated with the so-called weak localization or coherent backscattering (CBS) regime \cite{Sheng,AkkMont07,KSSH06,Labeyrie08}. As far as the coherent scattering and propagation of radiative flux in a sample is also influenced by the dressing field, this implies that the interferences responsible for CBS can be manipulated by an external control field \cite{DSKH06,DSKH08}. The influence of the dressing field on interference under multiple scattering is not just quantitative; it can cause qualitative changes in weak localization observables. For instance, spectral analysis of the enhancement factor of CBS in \cite{SKH11} revealed that for some polarization channels (follow definitions of Section \ref{SecIII}) the expected constructive interference can change to a destructive one (see Fig.~\ref{fig410}). In some spectral regions, for helicity channels $H_- \rightarrow H_-$ (left-handed to left-handed)  and $H_+ \rightarrow H_+$ (right-handed to right-handed) the enhancement factor becomes less than unity. Instead of a CBS cone we have a spectral CBS gap and meet here a signature of an anti-localization effect earlier discussed in subsection \ref{SecIIIB}. Fig.~\ref{fig410} proves that the interferences responsible for weak light localization can be manipulated by an external control field. In fact, this and recent experiments \cite{OWKRBHSK}  have lent support to this idea, and provide important guidance for further investigation.

\subsection{Quantum memory assisted by light trapping mechanism}\label{4.C}

\subsubsection{Importance of the problem}
One important aspect of light transport implies it as a natural, broadly implemented and highly developed information carrier. Moreover the information processing via an optical channel at the level of quantum uncertainty is now attainable but makes a significant challenge for ultimate quantum information technologies. As far as the losses, thermal and quantum noise strongly affect any quantum logic gates and data transmission, which should be performed by optical unitary operations in an ideal case, it becomes an extremely important implementation and developing of the protocols of quantum repeater and swapping. Such protocols are physically based on quantum mapping of a fragile non-classical light state onto the conservative and long-lived matter state and are entirely addressed as a problem of quantum memories and quantum interface, see \cite{PolSorHam,Simon,LeGouetMoiseev12}. Simplifying its complicated physics and considering any quantum network or processor in a low temperature environment with eliminated thermal noise, one could say that the quantum memory protocol would crucially help to overcome the problem of quantum noise and channel losses. Apparently, for fabrication of a quantum interface, the reliable sources of non-classical light, such as a single photon, entangled photon pairs, quadrature squeezed and entangled light are necessary.

There is a certain progress toward elaboration of non-classical light sources during the past decade. For example, generation of non-classical light in the regime of photon pairs in spontaneous down conversion developing in periodically poled nonlinear crystals has recently demonstrated high-quality entanglement reliably confirmed by the quantum test with violation of the Bell's inequality \cite{HYTPSS13,Zeilinger16}. The whispering gallery technique suggests a source of single photons and photon pairs with variously manipulable spectral properties and located in a narrow spectral domain \cite{Chekhova13}. There is a substantial progress in developing the various schemes of conditional preparation protocols towards engineering the non-Gaussian quantum states of light such as single photon, Schr\"{o}dinger cat and $NOON$ states, see \cite{Grangier14,GrangierLvovsky16} and reference therein. Further integration of such non-classical light sources with highly efficient and long-lived quantum memory makes a physically accessible platform for realization of the quantum repeater scheme and future long distance quantum communication.

Apart from the problem of the data processing, there is a direct physical and technological interest in preparation of macroscopic or mesoscopic multimode quantum correlated states in a matter subsystem. Nonlinear optics and diffractive scattering gives a tool for preparation of highly dimensional quantum states of light \cite{MolinaTerriza07,Grafe14}, which can be spatially separated by a macroscopic scale. Substantial progress in preparation of quadrature squeezed and macroscopically entangled states of bright light \cite{Chekhova15,Harder16} and multimode entangled Gaussian states of light \cite{TrepsFabre15,TrepsFabre16} also extends experimental capabilities for observation of quantum correlations in macroscopic conditions. With mapping such macroscopic quantum states on atomic subsystems via a swapping protocol the multimode quantum correlations can be converted to a spatial quantum hologram of the light and suggests various potentially interesting interferometric manipulations and measurements as well as quantum sensors operating beyond the standard quantum limit.

The standard schemes of the quantum interface and memories are based on principles of photon echoes \cite{MoiseevKroll01}, Raman conversion \cite{KozhMolmPolzk00}, and EIT delay \cite{FleischhLukin02}. The progress as well as difficulties in developing quantum interface systems have been discussed in reviews \cite{PolSorHam,Simon,LeGouetMoiseev12}. Recent theoretical studies have examined the role of disorder in the process of electromagnetically induced amplification or absorption (EIA) in the context of the quantum memory problem \cite{GSKOH,GSKH}. These investigations have prompted initial examination of the possibility that coherent multiple scattering and light trapping, mediated by a disordered assemblage of atoms, could lead to the development of a spin wave coherence distributed among a disordered collection of cold atoms. This, in turn, stimulated the possibility that the much longer scattering path lengths possible in disordered systems (in comparison with the path length associated with propagation of the coherent forward scattered beam in traditional one dimensional Raman based systems) might lead to improvements in efficiency and to more general types of EIA-based quantum memories for light \cite{GSKOH,GSKH}.

\subsubsection{General description of the memory protocol based on light diffusion}

\begin{figure}[tp]\center
{$\scalebox{1}{\includegraphics*{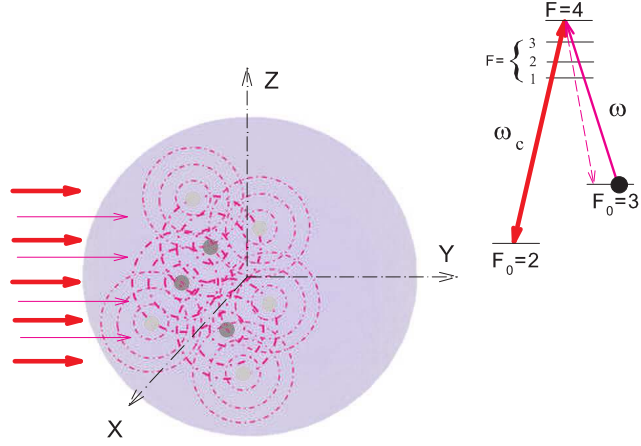}}$ }
\caption{(Color online) The mechanism of diffuse storage of light for the example of light trapping on the $F_0 = 3 \to F = 4$ closed transition in ${}^{85}$Rb. The diffuse propagation of a signal mode of frequency $\omega$ is indicated by incoming pink thin arrows. The diffusion process is affected by a strong control mode of frequency $\omega_c$ indicated by red and thick arrows. This converts a signal pulse into a long-lived and disordered spin coherence in the atomic subsystem.}
\label{fig411}%
\end{figure}

The considered light storage mechanism is based on stimulated Raman conversion of a signal pulse into a long-lived spin coherence. The main difference with traditional approaches, see \cite{PolSorHam,Simon}, requires that the pulse diffusely propagates through an atomic sample and is trapped in it for quite a long natural delay. The hyperfine energy structure of heavy alkali-metal atoms, such as rubidium or cesium, allows convenient integration of these processes. In Fig.~\ref{fig411}, we illustrate this through the example of the $F_0 = 3 \to F = 4$ closed transition in ${}^{85}$Rb. The crucial point for the effects is the presence and strong action of the control laser mode shifting the atoms from the background $F_0 = 3$ hyperfine sublevel to the signal $F_0 = 2$ sublevel via Raman interaction with the upper $F = 3, 2$ states. As a result, the quantum state of light can be mapped onto a disordered hyperfine coherence and collectivized by the atomic ensemble. Under ideal conditions, without relaxation and atomic losses, the stored state effectively performs a quantum hologram of the signal light, which can survive a relatively long time in the atomic spin subsystem. There is an evident analogy between this process and the process of Raman amplification under trapping conditions, which we discussed in section \ref{SecIVA2}. But now the state $F_0=2$ is initially empty such that we consider here the stimulated Raman scattering only as a tool for mapping the quantum state of a signal pulse onto the spin coherence and not for its amplification.\footnote{Reliability of any quantum memory protocol is often verified in example of a coherent signal pulse stored in the coherent state of a matter subsystem. It is important to recognize that in a linear transformation, performed by the protocol, any quantum states of light including its non-classical realizations such as single photon, entangled etc. can be expanded in a basis of coherent states via the quasi-probability distribution given by Glauber-Sudarshan $P$-type or Wigner symmetric-type representations, see \cite{WolfMandel}. Thus the transformation of the coherent-state is a crucial element for further estimating the efficiency of the quantum information processing.}

The transport of light under the described conditions is strongly affected by the presence of the control mode, which we will further assume as linearly polarized. The susceptibility tensor, given by Eq.(\ref{2.32}), can be structured into the following sum
\begin{equation}
\chi_{\mu\mu'}(\omega)\ =\ \chi_{0}(\omega)\,\delta_{\mu\mu'}+\chi^{(AT)}_{\mu\mu'}(\omega),\ \ \ \mu,\mu'=(x,y,z)%
\label{4.12}
\end{equation}
which, in this case, consists of a dominant linear isotropic term $\chi_{0}(\omega)$ and a nonlinear anisotropic correction $\chi^{(AT)}_{\mu\mu'}(\omega)$ associated with the "dressing" of the upper state by the control mode. The latter reflects the Autler-Townes modification of the quasi-energy structure of alkali-metal atom.

The most important ingredient contributing to the susceptibility (\ref{2.32}) and the scattering tensor (\ref{2.33}) is the retarded-type atomic Green's function of the excited state ${\cal G}_{nn'}^{(--)}(E)$. For the upper exited state $|n\rangle$ with $F=F_{\max}\equiv I+3/2=4$, which is not disturbed by the control mode, this function is simply defined as
\begin{equation}
{\cal G}_{nn'}^{(--)}(E)\ =\ \delta_{nn'}\frac{\hbar}{E-E_n+i\hbar\gamma/2}
\label{4.13}
\end{equation}
where $E_n$ is the energy of the state and $\gamma$ is its natural relaxation rate. The Green's functions for two other contributing states $|n\rangle,|n'\rangle$ form a matrix block of $2\times 2$ (with $F=I+1/2,I-1/2=3,2$ and with the same $M$). These functions obey the diagram equations (\ref{2.30}) and (\ref{2.31}), which after Fourier transform can be written as a system of algebraic equations. As a physical consequence the atomic resonances are strongly affected by the interaction with the control mode such that the full set of the coupled master equations generally includes a block of $3\times 3$ matrix components ${\cal G}_{nn'}^{(--)}(E)$ (with $F=I+1/2,I-1/2=3,2$ and additionally $F=I-3/2=1$ all with the same $M$ and this block of equations is given by
\begin{eqnarray}
\sum_{n''}\left[\left(E-E_n+i\hbar\frac{\gamma}{2}\right)\delta_{nn''}-
\frac{V_{nm'}V_{n''m'}^{*}}{E-\hbar\omega_c-E_{m'}}\right]\; {\cal G}_{n''n'}^{(--)}(E)&=&\hbar\delta_{nn'}%
\label{4.14}
\end{eqnarray}
where $V_{nm'}$ and $V_{n''m'}$ are the matrix elements of interaction with the control mode. This equation generalizes the approximation of the single upper state, performed by Eq.(\ref{4.10}), but with the requirement that any upper state $n,n',n''$ can be coupled by a control mode with only one ground Zeeman sublevel $m'$.  An example of the spectral behavior of the transverse component of the susceptibility tensor is shown by the red dashed curve in Fig.~\ref{fig42}.

Further Monte-Carlo simulation for the scattering process visualized in Fig.~\ref{fig411} shows that the control mode modifies it in two ways. First, this mode stimulates the Raman conversion of the scattered signal onto the spin subsystem and transforms the state of light into the disordered spin coherence. Second, the Autler-Townes resonance, for the same reasons as we discussed in section \ref{SecIVA2}, initiates losses via spontaneous anti-Stokes scattering. The results of \cite{GSKH} show that for a time delay comparable with the pulse duration the part of the pulse transmitted through the sample via elastic channels accumulates large losses of its initial energy. Nevertheless, in accordance with arguments given below, this is not a critical problem for the quantum memory.

As known from the studies of the Raman-type light storage scheme, performed for a standard one dimensional configuration, see \cite{Gorshkov,MKMP,SGSKMGL}, in the "write-in" stage of the memory protocol the spin distribution forms a quite inhomogeneous spatial profile. Those atoms, which are involved in the storage process, are mostly located near that edge where light entered the sample. That means that the spin polariton, created by the stimulated Raman process, has initially an extremely asymmetric shape along its propagation direction. The main losses, associated with the spontaneous scattering on the Autler-Townes resonance, are postponed and manifest themselves later while the pulse makes a complete passage and emerges from the sample. In one dimensional systems high efficiency can be attained if the pulse is retrieved in the direction backward to its original propagation or with an extremely weak read-out control mode minimizing the scattering losses.

\begin{figure}[tp]\center
{$\scalebox{0.6}{\includegraphics*{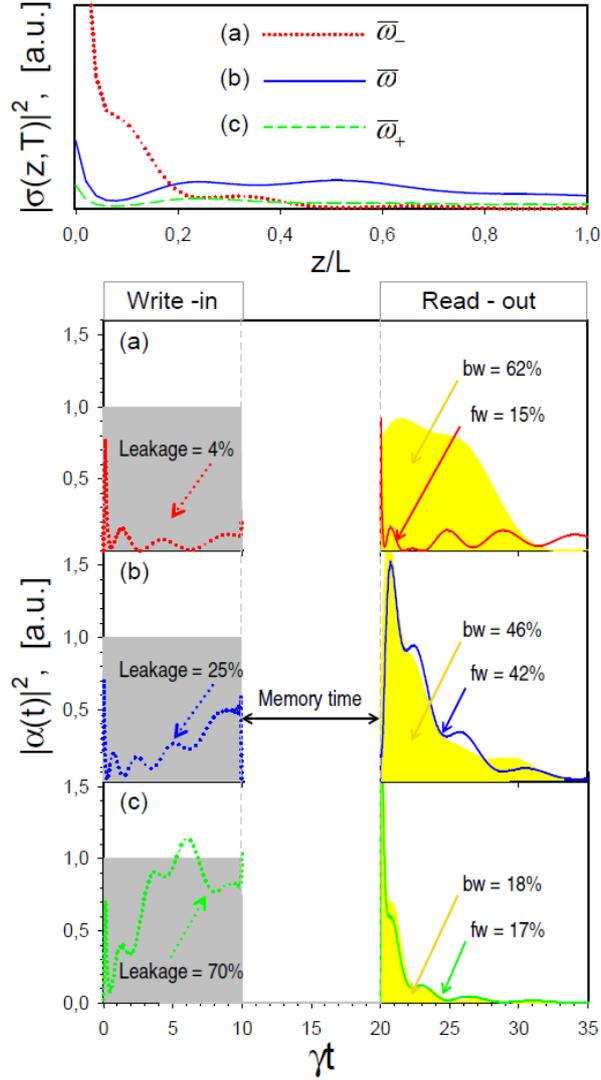}}$ }
\caption{(Color online) Quantum memory organized for rectangular signal pulse in the hyperfine manifold of $D_1$ line of ${}^{133}$Cs: one dimensional Raman-type protocol with data from \cite{SGSKMGL}. Upper panel shows the spin distribution $\sigma(z,T )$ in the sample of length $L$ at pulse duration time T ("stored light"). In the lower panel the signal pulse is retrieved on demand after a certain memory time via switching on the control field in either forward (solid lines) or backward (yellow filled area) direction. The leakage of each pulse, which is the light transmitted during the write-in stage, is indicated by the dotted line [(a) the input pulse carrier frequency is $\bar{\omega}_{-}$, (b) the input pulse carrier frequency is $\bar{\omega}$, and (c) the input pulse carrier frequency is $\bar{\omega}_{+}$, see text for definitions and explanation]. The read-out stage is characterized by the memory efficiency for the forward (fw) or backward (bw) retrieval.}
\label{fig412}
\end{figure}

This is illustrated in Fig.~\ref{fig412} where we reproduce the results of \cite{SGSKMGL} showing how a Raman-type memory protocol could be optimized in a one-dimensional configuration for the optically dense ensemble consisting of cesium atoms. The calculations were made for the hyperfine manifold of the $D_1$ line of ${}^{133}$Cs and for an optical depth on resonance around a few hundred. The control mode $\omega_c$ was assumed to be detuned between the upper state hyperfine sublevels and it created a quasi-energy Autler-Townes resonance located near $\omega_c$. Three lower panels in Fig.~\ref{fig412} correspond to the signal pulse storage and retrieval for three different carrier frequencies $\bar{\omega},\bar{\omega}_{\pm}$ and the upper panel indicates the spatial distribution of spin coherence $\sigma(z,T)$ created in the atomic subsystem at the time of the pulse duration $t=T$ ("stored light"). For a particular frequency $\bar{\omega}$ balancing losses, leakage and delay one obtains equal output for either forward or backward retrieval and a more or less homogenous distribution of the spin coherence. The other two frequencies $\bar{\omega}_{-}$ and $\bar{\omega}_{+}$ respectively shift the spectral location of the signal pulse closer to (more absorbing) and further from (more transparent) the Autler-Townes resonance. It may seem surprising but the optimal of the memory efficiency is observed in the protocol of backward retrieval for the carrier frequency $\bar{\omega}_{-}$ when the medium is more absorbing. The explanation of this is clearly seen from the highly asymmetric distribution of the spin coherence, which is shown in the upper panel.

It seems reasonable that in a three dimensional configuration we can similarly expect that the atoms located near that edge, where light enters the sample, are mostly involved in the diffuse storage, see Fig.~\ref{fig411}. Evidently the spatially-reversed retrieval scheme would be not so easy (or even impossible) to organize in the three-dimensional disordered configuration, see \cite{GSKOH}, and one can use an extremely weak control mode in the read-out stage of the protocol to solve the problem and minimize the spontaneous losses. But even in this case the profile of the retrieved pulse would be given by an unpredictable scattered mode and the described memory protocol would seem to be useless. Nevertheless there is a specific application example of entanglement where it is not necessary to retrieve the pulse in its original mode and it is only important to provide its effective storage, such that the protocol does not require the light transport throughout the whole sample. For this special situation, the above memory scheme is expected to be much more effective than its one-dimensional counterpart.

\subsubsection{Quantum memory for the macroscopic analog of the $\Psi^{(-)}$ Bell state}

The proposed memory scheme is applicable and adjusted to the situation where the quantum information is originally encoded in the total number of photons in the signal light beam(s), this number being considered as a quantum variable. Such variables are insensitive to either spatial or temporal mode structure of the signal light pulse. Physically this means that an unknown number of informative photons can be mapped onto the atomic subsystem via Raman-induced repopulation of the equivalent unknown number of the atoms to the signal level while light diffusely propagates through the sample. This gives us an example of a so-called topologically protected protocol of the quantum memory, which is insensitive to any particular atomic distribution.

Such a situation takes place with storage of the macroscopic analog of the $\Psi^{(-)}$ Bell state, this consisting of a pair of photons with either (orthogonal) horizontal (H) or vertical (V) polarizations having unknown but strongly correlated photon numbers, see \cite{ICRL}. This light, usually considered in a low energy single pair regime, seems a quite promising candidate for providing long-distance quantum correlations and communications. It probably gives us an example of a less fragile macroscopic quantum state among others since its entanglement observation does not need so sensitive detection as a phase dependent homodyne detection scheme.

We shall apply the above memory protocol to the following entangled quantum state of light
\begin{equation}
|\Psi^{(-)}\rangle\ =\ \sum_{m,n}\,\Lambda_{mn}^{(-)}\;|m\rangle_{H1}|n\rangle_{V1}%
|m\rangle_{V2}|n\rangle_{H2}%
\label{4.15}%
\end{equation}
where $m$ and $n$ are the occupation numbers of the field modes, specified by beam number either 1 or 2 and polarization type either horizontal $H$ or vertical $V$, and
\begin{equation}
\Lambda_{mn}^{(-)}\ =\ (-)^{n}\frac{\bar{n}^{\frac{m+n}{2}}}{[1+\bar{n}]^{\frac{m+n}{2}+1}}%
\label{4.16}%
\end{equation}
which possesses completely anti-correlated polarizations in the light beams 1 and 2, these beams propagating in different directions. The unique property of this state is that detection of a certain number of photons of any polarization (not only in $H$ or $V$ but also in any elliptical polarization mode) in beam 1 guarantees the detection of the same number of photons in beam 2 but always in an orthogonal polarization state. In \cite{ICRL} such a state was called a macroscopic analog of a singlet-type two-particle Bell state i.e. $\Psi^{(-)}$-state. Its Schmidt decomposition, given by Eqs. (\ref{4.15}) and (\ref{4.16}), can be found via basic expansion for the two mode squeezed state, see \cite{Braunstein}. The quantum state (\ref{4.15}) can be parameterized by the average number of photons in each light mode $\bar{n}$.

\begin{figure}[tp]\center
{$\scalebox{0.8}{\includegraphics*{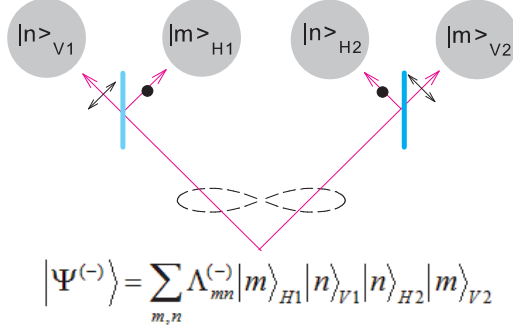}}$ }
\caption{(Color online) The quantum hologram of the macroscopically entangled state of light $|\Psi^{(-)}\rangle$, which can be prepared
by a SPDC process, see \cite{ICRL}. The unknown photon numbers in each polarization are subsequently stored in four memory units.}
\label{fig413}
\end{figure}

In Fig.~\ref{fig413} we illustrate how the quantum hologram of this light can be realized with the memory units, as described above. Each memory unit stores the photons of a particular polarization and frequency, which diffusely propagate through an atomic cloud and are substantially trapped on the closed transition and the stimulated Raman process is initiated by interaction with the control mode  $\omega_c$ on other hyperfine sublevels ($F=3, 2$).  This is illustrated in the transition diagram of Fig.~\ref{fig411}. The more subtle point of this process is that it creates a pure quantum state in the atomic subsystem, which is entangled among the four clouds and has an unknown but strongly correlated number of atoms repopulated onto the signal level $F_0=2$ in each cloud, such that their further measurement would demonstrate the presence of quantum non-locality in the matter subsystem.

\begin{figure}[tp]\center
{$\scalebox{1.1}{\includegraphics*{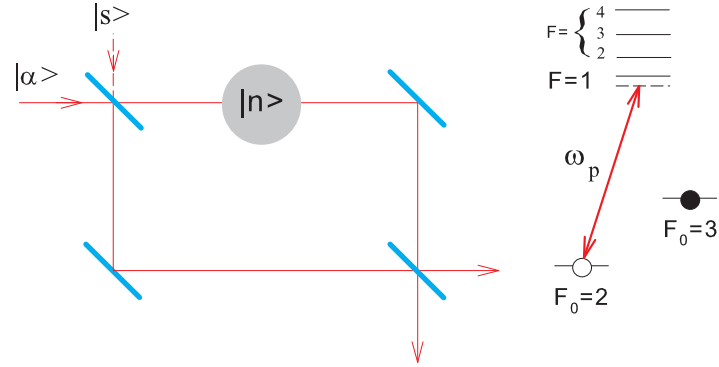}}$ }
\caption{(Color online) Schematic diagram
of the Mach-Zehnder interferometer for detecting a small number of atoms stored in a particular cloud.
The weak probe coherent mode $|\alpha\rangle$ of frequency $\omega_p$ is applied near
the resonance of the closed $F_0=2\to F=1$ transition
to avoid effects of Raman scattering. The sensitivity of the interferometer can be enhanced via sending
a portion of the squeezed light $|s\rangle$ to the second input port of the interferometer.}
\label{fig414}
\end{figure}

A standard strategy for a quantum memory normally aims towards a goal of recovery of the signal pulse in its original mode. In our situation there is no need to do that since all the quantum information is encoded into the numbers of repopulated atoms. The observation or detection of these numbers can be organized with a Mach-Zehnder interferometer, as is shown in Fig.~\ref{fig414}. The interferometer can be adjusted for balanced detection of the signal expressed by the difference of the photocurrents from the output ports. Then the measured signal associated with the small informative phase shift $\delta\phi$, induced in one arm of interferometer, is given by
\begin{equation}
i_{-}\ =\ i_{1}-i_{2}\propto \bar{i}\,\delta\phi,\ \ \ \delta\phi=\xi\, n%
\label{4.17}%
\end{equation}
where the phase shift is proportional to the number of detected atoms $n$ and to a small factor $\xi\ll 1$, which depends on geometry (sample size, aperture of the light beam, etc.) and reflects the weakness of the signal. The $D_2$-line energy structure of an alkali-metal atom allows one to tune the probe near the resonance associated with the closed transition and to avoid any negative presence of the Raman scattering channel. In the case of ${}^{85}$Rb, that is the $F_0=2\to F=1$ transition. Then the standard sensitivity of the measurement is limited by the shot-noise, Poissonian level, but it can be essentially improved via sending a portion of squeezed light to the second port of interferometer, see \cite{Kimble}. In principle the Mach-Zehnder interferometer allows approaching even the Heisenberg limit in its sensitivity while operating with two squeezed light beams in its input ports, see \cite{SokolovMZ}.

Let us make the following remark concerning the above detection scheme, which in an ideal situation would perform a certain type of quantum non-demolition measurement (QND), see \cite{WolfMandel}. At first sight the scheme seems to be specifically related to the polarization basis $|H\rangle$ and $|V\rangle$, which was used in expansion (\ref{4.15}) and in the storage protocol shown in Fig.~\ref{fig413}. However, an identical expansion could be rewritten in any other basis of arbitrary orthogonal elliptical polarizations and this would be described by the same expansion coefficients. In other words the state $|\Psi^{(-)}\rangle$ is insensitive to the type of polarization beamsplitters used for the hologram creation. Then the above QND operation with the hologram can be interpreted as delayed detection of the photons transmitted by the particular beamsplitters. With variation of the beamsplitter types the measurement statistics could demonstrate violation of the classical probability principles. In a particular case of a rare flux consisting of the photon pairs prepared in a "singlet state" the measurements would show violation of the Bell inequalities. We can also point out that the hologram yields various interferometric operations and could potentially be interesting as a logic element for the further quantum information processing based on a continuous variables scheme.

\subsubsection{Nanofiber-assisted quantum memory}

In this last subsection we shall briefly discuss the idea of a quantum interface as a combination of an atomic system and an optical fiber technique, proposed in \cite{BHKLM04,KBH04}, and point out its possible connection with the light trapping problem. In particular we follow the result, recently reported in \cite{GMNML15} and showing how the quantum memory can be organized for light transporting through an ensemble of cold atoms via a dielectric nanofiber. The field of the fundamental mode of a subwavelength dielectric nanofiber is mostly concentrated outside of it (evanescent mode) and has a transverse profile similar to a paraxial Gaussian mode in free space. In the experiment the signal pulse was originally converted into such an evanescent mode further interacting and scattering from the atoms of the ensemble. The transparency was recovered and the pulse was delayed and stored with applying a control field and realization of an EIT protocol. The potential advantage foreseen from this experiment is that the preliminary scheme could be essentially improved if the atoms involved in the light storage process were initially locked and arrayed along the fiber with a special fiber trap architecture. As has been earlier reported in \cite{GCADLPTSK12} the required trap can be realized with the aid of two standing waves associated with two far off-resonant red and blue detuned laser fields.  For an evanescent mode interacting with the atoms located in such a fiber trap the optical depth of $0.08\%$ per atom can be attainable. That means that for a signal pulse propagating in an evanescent mode the optical depth on the order of hundreds can be obtained for a standard MOT or cell experiments with typical optical depth on the order of ten. An example of significant enhancement of the scattering process and signature of collective effect of Bragg diffraction in one dimensional configuration has been recently demonstrated in experiments \cite{PolzikAppel16,KuprLaurat16}.

We can point out here that the soft cavity design, described in subsection \ref{SecIVA3} in the context of the random lasing phenomenon, can be also promising for improving the nanofiber-assisted quantum memory protocol. Indeed, if the frequency of the evanescent mode was resonant with a closed transition of a alkali-metal atom then the environment of elastic scatterers would prevent light losses associated with its scattering on atoms nested near the fiber. In other words the incoherent elastic reflection from the environment would preferably redirect the light into the fundamental fiber mode again. The physics of the entire process developing in a partly ordered atomic system is rather non-trivial and can have a signature of such subtle effects as Bragg diffraction and Anderson localization. The experimental verification as well as realistic theoretical description of such a complicated process of quasi one-dimensional light diffusion is quite challenging and needs further studies.

\section{Light scattering on dense and complex atomic systems}\label{Section V}
\setcounter{equation}{0}
\setcounter{figure}{0}

\subsection{Overview of the problem}\label{Section V.A}
The collective effects in radiation coupling of a chain of two level energy spins was first considered by Dicke in his seminal paper \cite{Dicke}. It was shown that a single mode of a radiation field interacts collectively with a sample of resonant two-level particles distributed on a spatial scale smaller than the wavelength of the field.  The negligible phase distribution of the field over the volume of the sample is responsible for the collectivity of the interactions. This may be achieved more readily in the radio or microwave region of the spectrum, but is challenging in the optical regime, because of the necessary very small size of the atomic samples required. Nevertheless preparation of atomic samples with a density $n_0\lambdabar^{3}\sim 0.1$ (that implies $n_0\lambda^{3}\gg 1$!) is now experimentally feasible, for example, for atoms loaded into the potential well of a quasistatic dipole trap. Under such conditions the density and disorder effects strongly affect the radiative coupling of atomic dipoles and as a consequence essentially modify the scattering process.

In studying dense and strongly disordered systems, the primary experimental efforts were applied to the spectroscopic observations of cooperative phenomena associated with the high density of atomic dipoles. The complex excitation spectra, which normally is a prerogative of condensed matter, can now be fabricated in cold atomic systems with mediating the optical trap parameters. At intermediate density, accessible for experimental verification, the intriguing issue for such a system is a collective light shift induced by interaction via the quantized transverse field between the resonant atoms in the medium \cite{Scully,RSSCR,KSKHSA,Manassah}. This phenomenon is partly masked by the existence of a local field Lorentz-Lorenz red shift. But, as we discuss below, both the shifts can have different signs and therefore can be resolved in experiment via its expected density dependence.

Another experimental tendency is to follow possible deviation from the standard behavior of light propagation under Beer's law and its diffuse scattering described by a sequence of independent and uncorrelated scattering events, as was discussed above in section \ref{2.F}. In the experiments the signature of cooperativity can be observed in the collectively scattered radiation in a particular directions and in the mechanical action of the driving field on the atomic system via the cooperative radiation pressure \cite{BBPK}. Rather recently in \cite{KRHSK} the enhanced transmission in measurements of the spectral dependence of forward light scattering by a high-density and cold ensemble of ${}^{87}$Rb atoms was reported.

An alternative option for preparation of a dense atomic system turns us to an experimental technique based on an optical lattice architecture \cite{Bloch05,BDZ08}. From the point of view of optical interaction such objects are quite closely related with disordered atomic systems which are primary focused on in the present review. The studies of optical properties of cold atomic ensembles arrayed in the optical lattices with particular addressing to the problem of a light-matter quantum interface and self organization have also generated significant research interest \cite{MekhovRitsch12,RDBE13}. One could say that any complex and either ordered or disordered atomic systems obtain as more similarities in their description as they are considered closer to the critical level of quantum degeneracy and self organization.

From a theoretical point of view the problem of high density and strong disorder is specifically interesting since it opens new physical capabilities of light control near the threshold between gaseous and condensed phases  of matter.  In the case of high density ensembles, containing a large number of atoms, the scattering problem, considered as a many-body and strongly interacting system, becomes quite difficult for ab-initio theoretical description. If we look more critically towards the basic definitions of the interaction Hamiltonian in the dipole gauge (\ref{2.1})-(\ref{2.3}) we have to pay attention to how the transformed variables of the atomic and field subsystems differ from their original, these being when atoms are separated by a distance on the order of a wavelength. The Hamiltonian of the dipole \textit{is not identical} to the Hamiltonian of the corresponding atom, which would result by unitary transformation of the original atomic Hamiltonian from the Coulomb to the dipole gauge. The same concept concerns the momentum operator of an atomic electron $-i\hbar\nabla$, which in the dipole gauge is a kinetic momentum not coinciding with its generalized momentum operator. As a consequence any dyadic operators between the dipole eigenstates (energy steady states) have a slightly different interpretation than in the original atomic Hamiltonian. Physically that means that the unitary transformation from Coulomb to dipole gauge \textit{entangles variables of atomic and field subsystems}, see \ref{Appendix.A} for details. In a single particle case the importance of such entanglement is clearly manifestable in the renormalization procedure of a long-wave dipole diverging contribution to the Lamb shift, see \cite{BrLfPt}.

As we see, in a dense, many-particle system clear interpretation of the calculated observables is evidently needed. The atomic and field variables transform to their original disentangled definitions only in the asymptotic regime of the scattering process when the field and atomic subsystems are uncoupled. Below we shall briefly discuss various approaches applied in the literature for description of the density and disorder effects in the light scattering problem.

\subsection{Scalar and other simplifying models of the atoms-light interaction}

The description of collective phenomena can be simplified if only the interaction of an atomic system with the transverse field (radiation modes) is taken into consideration and any presence of static interaction of the proximal dipoles via the near field is ignored. For the atoms considered as two-level quantum objects their coupling with the field subsystem can be treated as driven by the following Hamiltonian \cite{SvChSc}
\begin{equation}
\hat{H}_{\mathrm{int}}\ =\ \sum_{\mathbf{k}}\sum_{a=1}^{N}\hbar g_k \hat{\sigma}^{(a)\dagger} a_{\mathbf{k}}\mathrm{e}^{i\mathbf{k}\cdot\mathbf{r_{a}}} + H.c.%
\label{5.1}
\end{equation}
here $\hat{\sigma}^{(a)\dagger}=|e\rangle\langle g|$ is the raising operator acting between the ground $|g\rangle$ and excited $|e\rangle$ states of the $a$-th atom. Since, in such a description, an atomic transition has no angular dependence then by accepting the above Hamiltonian one should straightforwardly ignore any vector properties of light. The coupling constant $g_k$ is formally expressed by a transition dipole moment, similarly as in Eqs.(\ref{2.1}), (\ref{2.2}), but the difference between transverse electric and displacement fields as well as near field static interaction lies outside of the applied approximation. This Hamiltonian gives a resource for a certain generalization of the original Dicke system and in particular predicts anisotropic behavior of the radiation emission. The emission anisotropy and the light pressure associated with cooperative scattering has been recently discussed in \cite{BiPiKa,BBCRBCPK} in the context of their experimental verification.

In spite of the evident limitation in its entire applicability the scalar model gives a convenient way to perform complete microscopic calculation with keeping the main physical origins of the scattering process. In particular, in the scalar model, the strong field effects can be naturally incorporated in the calculation scheme via the  Heisenberg-Langevin approach \cite{OWLMK}. There the authors followed how the Mollow triplet is affected by cooperativity and predicted its asymmetrical structure. Such an asymmetry, if confirmed by experiment, could be a signature of washing out the subtle anti-bunching phenomenon and domination of the quasi-energy dressing mechanisms in formation of the fluorescence spectrum of a macroscopic atomic cloud. Another interesting prediction was made in \cite{GreWell}, where the authors claimed that the non-linear dipole dynamics enhanced by anti-collective disorder effects can lead to instability in radiation emission of the atomic ensemble at a certain density threshold. The experimental verification of these phenomena is still challenging but interesting as giving us new ways of coherent control of the radiation via nonlinear scattering processes.

Rather similar to the scalar model exists another approach based on the so called effective Hamiltonian approximation \cite{AkkGerKai,BelGerAkkKai}. In this case the near field effects associated with short range dipole-dipole interactions can be eliminated due to its specific angular dependence. Since for a random configuration of atoms the angular factor contributing to the interaction between any atomic pair vanishes after averaging $\langle P_2(\cos\theta)\rangle\to 0$, the interplay between quantized field modes and the atomic system can be mainly expressed via the radiation zone asymptote of the field Green's functions. That physically means that only coupling via interaction with the transverse field, as in Hamiltonian (\ref{5.1}), is taken into consideration. The effective Hamiltonian approach was applied for studying photon localization, stochastic dynamics and Dicke superradiance in dense atomic gases.

It is noteworthy to point out that description of an atom-field quantum interface in the dipole gauge is quite popular for atomic chain systems trapped in an optical lattice and interacting with an external cavity mode, see \cite{MekhovRitsch12,MCIM16}. In this situation a Hamiltonian in the form (\ref{5.1}) actually describes interaction of atomic dipoles with vector, not scalar, states of the field, where the field operators are associated with the cavity mode, and near field interaction can be safely incorporated into the entire parameters of the Bose-Hubbard model. Physically that means that in such a configuration optical coupling via cavity modes dominates the light-atom interaction process, but the cavity is considered as sufficiently large such that interaction via its open modes is assumed the same as in free space and, as a consequence, the interaction of proximal dipoles via the longitudinal field is also considered the same as in free space. Similarly the cavity cooling mechanisms as well as self organization of atoms, driven by a cavity field, can be basically described by interaction with the transverse field in the form (\ref{5.1}), which can be further used for derivation of an effective Hamiltonian with a description aimed towards optomechanical action on atomic system via Fokker-Planck  master equation, see \cite{RDBE13,Mishina14,SJM16,JSM16}.

\subsection{The self-consistent macroscopic approach}

The cooperative dynamics of a dense ensemble of atomic dipoles with an external driving field can be realistically described by the macroscopic Maxwell approach based on intrinsic self-consistency of the interaction process. In the self-consistent interpretation the proximal dipoles are physically indistinguishable on a scale of a wavelength and obey the coordinated time dynamics.  Let us consider, as an example, a dense ensemble of the simplest $V$-type two-level atoms with the minimal accessible number of quantum states, $\emph{i.e.}$ with angular momentum $F_0=0$ in the ground state and $F=1$ in the excited state. In the self-consistent approximation, see \cite{SKKH}, the steady state dynamics of an atomic dipole in a macroscopic ensemble consisting of many dipoles can be expressed by the following equation for its positive frequency component $d^{(+)}(\omega)$:
\begin{eqnarray}
-i\omega\, d^{(+)}\!(\omega)\!\!&=&\!\!-i\omega_{0}\,d^{(+)}\!(\omega)+\frac{i}{\hbar}d_0^2 \left[{\cal E}^{(+)}\!(\omega)\!+\!\frac{4\pi}{3}{\cal P}^{(+)}\!(\omega)\right]%
\ -\ i\Sigma(\omega)\,d^{(+)}(\omega)%
\label{5.2}
\end{eqnarray}
$\omega_0$ denotes the transition frequency, and $d_0=|(\mathbf{d}\cdot\mathbf{e})_{nm}|$ is the modulus of the transition matrix element and is equal for all transitions $m\equiv F_0,M_0\to n\equiv F,M$. This equation is surely valid for any polarization of excitation channel and we further omit the vector indices for the dipole and field. Then the probe driving amplitude ${\cal E}^{(+)}(\omega)$ is the macroscopic transverse electric field at frequency $\omega$, considered at the point of the dipole's location and ${\cal P}^{(+)}(\omega)$ is the local mesoscopically averaged polarization. The presence of this last term in Eq. (\ref{5.2}) corresponds to the so called Lorentz-Lorenz or local field correction associated with the longitudinal (quasi-static) interaction of the dipole with its local environment. Since in the self-consistent model the proximate dipoles are indistinguishable in the excitation process, the averaged polarization can be written as
\begin{equation}
{\cal P}^{(+)}(\omega)\ =\ n_0\, d^{(+)}(\omega)=\chi(\omega)\,{\cal E}^{(+)}(\omega)%
\label{5.3}
\end{equation}
where $n_0$ is the local density and $\chi(\omega)$ the macroscopic dielectric susceptibility of the sample.

The radiation losses, given by the last term in Eq. (\ref{5.2}), can be expressed via the retarded type Green's function for the transverse field  in the medium, see \cite{LfPtIX}.
\begin{eqnarray}
\Sigma(\omega)&=&-\frac{8\pi}{3\hbar}d_0^2\int\frac{d^3k}{(2\pi)^3}\,%
\frac{\omega^2}{c k^2-\epsilon(\omega)\omega^2}\ %
=\ \Delta_{\mathrm{Lamb}}-\frac{i}{2}\sqrt{\epsilon(\omega)}\gamma%
\label{5.4}
\end{eqnarray}
where $\epsilon(\omega)=1 + 4\pi\chi(\omega)$ is the dielectric susceptibility of the medium. The first term in the right hand side $\Delta_{\mathrm{Lamb}}$ selects a diverging contribution to the vacuum Lamb shift, which should be renormalized and incorporated into a physical energy of the atomic transition. The imaginary part of the second terms corresponds to the radiation damping created by the transverse field emitted in the scattering process and resulting in losses of the probe escaping the sample incoherently. Meanwhile, its real part (existing because $\epsilon$ is complex) is responsible for the correction to the radiative shift, which is induced by cooperative interaction of the neighboring dipoles.

As was shown in \cite{SKKH} the above formulas enable us to obtain a closed algebraic equation for the local dielectric permittivity $\epsilon(\omega)=\epsilon'(\omega)+i\epsilon''(\omega)$. This equation can be solved and reproduces the modification of the dispersion and absorption properties of the medium associated with collective effects treated self-consistently. The model can be straightforwardly generalized for a two-level atom with arbitrary ground and excited state angular momenta if the ground state is uniformly populated. To do this the density for the $V$-type two level atom should be rescaled as $n_0\lambdabar^3(2F+1)/[3(2F_0+1)]$.

\begin{figure}[tp]\center
{$\scalebox{0.9}{\includegraphics*{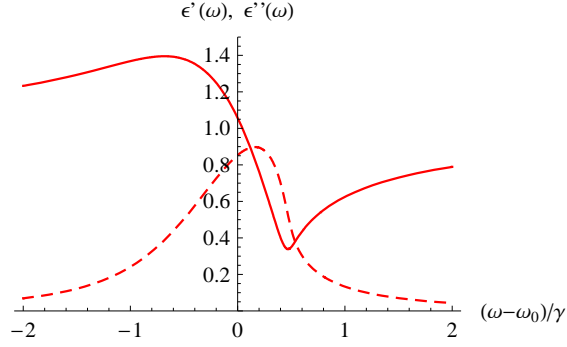}}$ }
\caption{(Color online) Real (solid) and imaginary (dashed) parts
of the dielectric permittivity for the sample with the scaled density
$n_0\lambdabar^3(2F+1)/[3(2F_0+1)]=0.05$. The origin of the plot corresponds
to the atomic resonance and the frequency $\omega$ is scaled by the natural
decay rate $\gamma$}
\label{fig51}
\end{figure}

\begin{figure}[tp]\center
{$\scalebox{0.9}{\includegraphics*{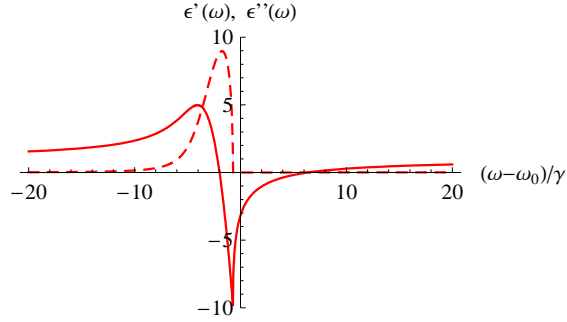}}$ }
\caption{(Color online) Same as in Fig.~\ref{fig51} but for the density $n_0\lambdabar^3(2F+1)/[3(2F_0+1)]=1$.}
\label{fig52}
\end{figure}

In figures \ref{fig51} and \ref{fig52} we show the spectral behavior for the dielectric susceptibility for two different densities. Even at relatively low density, as in Fig.~\ref{fig51}, the spectral profile demonstrates significant distortion of the standard Lorentzian profile normally observable in a dilute gas. The absorption resonance has a blue shift associated with the collective radiation correction of the light shift. At high density, as in Fig.~\ref{fig52}, the absorption profile changes dramatically and even qualitatively has no close similarity with a Lorentzian shape.  But its peak still has a clear signature of the Lorentz-Lorenz static red shift. The dispersive part becomes negative and emphasizes the strong role of the surface scattering effects.

\begin{figure}[tp]\center
{$\scalebox{0.7}{\includegraphics*{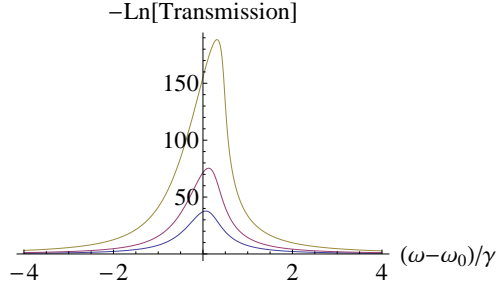}}$ }
\caption{(Color online) The transmission coefficient for an atomic slab of length $L=200 \lambdabar$ and for the densities $n_0\lambdabar^3=0.01, 0.02, 0.05$ (lower, middle and upper curves).}
\label{fig53}
\end{figure}

For a slab sample of length $L$ filled in by the medium with dielectric constant $\epsilon(\omega)$ the transmission amplitude at frequency $\omega$ is given by
\begin{equation}
{\cal T}_\omega\ =\ \frac{2\sqrt{\epsilon(\omega)}}{2\sqrt{\epsilon(\omega)}\cos\psi(\omega)-i(1+\epsilon(\omega))\sin\psi(\omega)}%
\label{5.5}%
\end{equation}
where $\psi(\omega)=L\sqrt{\epsilon(\omega)}\,\omega/c$. The transmission coefficient $|{\cal T}_\omega|^2$ can be calculated and shows the deviation from the standard Beer's law normally fulfilled in a dilute gas. In figure \ref{fig53} we show the calculations of the transmittance for three different densities $n_0\lambdabar^3=0.01, 0.02, 0.05$.  The existence of cooperativity in enhanced light transmission in cold atomic matter and a cooperative light shift (blue-shifted from the atomic resonance) has been recently observed in \cite{KRHSK,KaiserLukin16}.

\subsection{The microscopic approach}

Although the above macroscopic Maxwell approach realistically reproduces the main features in the cooperative dynamics of the disordered system of atomic dipoles, it should be critically reconsidered via applying the rigorous microscopic calculation scheme. The global motivation of it is that in a strongly disordered system its statistical description should be performed beyond the convenient conditions of the central limit theorem and reveal the extremely important role of the internal correlations for the wave propagation, see \cite{MullerDelande}. In the considered case of electromagnetic waves one can expect strong dependence of the scattering parameters on a particular configuration in distribution of atomic dipoles. This can be important if one has in mind to further propose such a dense and cold atomic system as a candidate for organization of light-matter quantum interface. In this section we briefly concern the quite difficult problem of microscopic description of light scattering from a collection of closely located atoms.

\subsubsection{The transition amplitude and the scattering cross section}
The quantum-posed description of the photon scattering problem is based on the formalism of the $T$ matrix, which is defined by
\begin{equation}
\hat{T}(E)\ =\ \hat{V}+\hat{V}\frac{1}{E-\hat{H}}\hat{V},%
\label{5.6}%
\end{equation}
where $\hat{H}$ is the total Hamiltonian consisting of the nonperturbed part $\hat{H}_0$ and an interaction term $\hat{V}$ such that $\hat{H}=\hat{H}_0+\hat{V}$. The energy argument $E$ is an arbitrary complex parameter in Eq.(\ref{5.6}). Then the scattering process, evolving from the initial state $|i\rangle$ to the final state $|f\rangle$, is expressed by the following relation between the differential cross section and the transition amplitude, given by the relevant $T$-matrix element considered as a function of the initial energy $E_i$:
\begin{equation}
d\sigma_{i\to f}\ =\ \frac{{\cal V}^2}{\hbar^2
c^4}\frac{\omega'^2}{(2\pi)^2}%
\left|T_{g'\mathbf{e}'\mathbf{k}',g\,\mathbf{e\,k}}(E_i+i0)\right|^2d\Omega%
\label{5.7}%
\end{equation}
Here the initial state $|i\rangle$ is specified by the incoming photon's wave vector $\mathbf{k}$, frequency $\omega\equiv\omega_k=c\,k$, and polarization vector $\mathbf{e}$, and the atomic system populates a particular ground state $|g\rangle$. The final state $|f\rangle$ is specified by a similar set of the quantum numbers, which are additionally superscripted by the prime sign, and the solid angle $\Omega$ is directed along the wave vector of the outgoing photon $\mathbf{k}'$. The presence of the quantization volume ${\cal V}$ in this expression is caused by the second quantized structure of the interaction operators. The scattering process conserves the energy of input and output channels, such that $E_i=E_f$. For numerical calculation it is convenient to make use of the optical theorem
\begin{equation}
Q_0\equiv\sum_f\sigma_{i\to f}=-\frac{2{\cal V}}{\hbar\,c}\;\mathrm{Im}\, T_{g\mathbf{e}\mathbf{k},g\,\mathbf{e\,k}}(E_i+i0)%
\label{5.8}
\end{equation}
which expresses the total cross section $Q_0=Q_0(\omega)$ via specific $T$-matrix element for the elastic channel of forward scattering.

The microscopic description for the scattering of a photon on an atomic system is here performed in the dipole approximation given by Eqs. (\ref{2.1}) and (\ref{2.2}). This means that the original Hamiltonian, introduced in the Coulomb gauge and valid for any globally neutral charge system, has been unitarily transformed to the dipole-type interaction with the assumption that the atomic size is much smaller than a typical wavelength of the field modes actually contributing to the interaction dynamics.  The bottom line in Eq.(\ref{2.2}) indicates the important difference between the actual transverse electric field denoted as $\hat{\mathbf{E}}_{\bot}(\mathbf{r})$ and the displacement field. The difference cannot be now ignored as far as we consider ensemble of atoms separated by a distance comparable or less than the radiation wavelength. As pointed out in the introductory part to this section, for such a dense configuration the definitions in the interaction Hamiltonian (\ref{2.1}) and (\ref{2.2}) should be fairly interpreted and the diverging terms that appear should be correctly renormalized.

The second term in Eq.(\ref{2.1}) reveals a diverging self-energy (self-action) of the dipoles. Let us recall our comment in the beginning, see section \ref{Section II.A}, that this term is often omitted in practical calculations since it does not principally affect the dipoles' dynamics, particularly when the difference between the transverse electric and displacement fields is small. It can be also formally incorporated into the internal Hamiltonian associated with the atomic dipoles, see \cite{ChTnDRGr}. However, as was pointed out in \cite{SKKH}, in a more precise vision based on tracking the complete Heisenberg dynamics of atomic variables the self-action term is actually compensated by the self-contact dipole interaction. The latter manifests itself in the dipoles' dynamics when $\mathbf{r}=\mathbf{r}_a=\mathbf{r}_b$ for interaction of a specific $a\!-\!\mathrm{th}$ dipole in the Hamiltonian (\ref{2.1}) with the longitudinal field created by the same dipole in the second term in Eq. (\ref{2.2}). Both these diverging self-action and self-contact interaction terms can be safely renormalized in evaluation of a single-particle contribution into the self-energy part of the resolvent operator, see below.

\subsubsection{The resolvent operator and $N$-particle Green's function}

The transition amplitude (\ref{5.6}) can be simplified if we substitute in it the interaction operator in the form (\ref{2.1}) by keeping only the terms with annihilation of the incoming photon in the input state and creation of the outgoing photon in the output state. Such a simplification is in accordance with the standard approach of the rotating wave approximation, which is surely fulfilled for a near-resonance scattering process. As a consequence of this approximation the transition amplitude is now determined by the complete resolvent operator projected onto the vacuum state for the field subsystem and onto the singly excited state for the atomic subsystem
\begin{equation}
\tilde{\hat{R}}(E)\ =\ \hat{P}\,\hat{R}(E)\,\hat{P}\ \equiv\ \hat{P}\frac{1}{E-\hat{H}}\hat{P}%
\label{5.9}%
\end{equation}
Here we defined the following projector
\begin{eqnarray}
\hat{P}&=&\sum_{a=1}^{N}\;\sum_{\{m_j\},j\neq a}\;\sum_{n}%
|m_1,\ldots,m_{a-1},n,m_{a+1},\ldots m_N\rangle%
\langle m_1,\ldots,m_{a-1},n,m_{a+1},\ldots,m_N|\,\times\,|0\rangle\langle 0|_{\mathrm{Field}}%
\label{5.10}%
\end{eqnarray}
which selects in the atomic Hilbert subspace the entire set of the states where any $j$-th of $N-1$ atoms populates a Zeeman sublevel $|m_j\rangle$ in its ground state and one specific $a$-th atom (with $a$ running from $1$ to $N$ and $j\neq a$) populates a Zeeman sublevel $|n\rangle$ of its excited state. The field subspace is projected onto its vacuum state and the operator $\tilde{\hat{R}}(E)$ can be further considered as a matrix operator acting only in the atomic subspace. The elements of the $T$ matrix can be directly expressed by the resolvent operator as follows:
\begin{eqnarray}
T_{g'\mathbf{e}'\mathbf{k}',g\,\mathbf{e\,k}}(E)&=&\frac{2\pi\hbar\sqrt{\omega'\omega}}{{\cal V}}%
\sum_{b,a=1}^{N}\;\sum_{n',n}%
(\mathbf{d}\mathbf{e}')_{n'm'_b}^{*}(\mathbf{d}\mathbf{e})_{nm_a}%
\mathrm{e}^{-i\mathbf{k}'\mathbf{r}_b+i\mathbf{k}\mathbf{r}_a}%
\nonumber\\%
&&\langle\ldots m'_{b-1},n',m'_{b+1}\ldots |\tilde{\hat{R}}(E)%
|\ldots m_{a-1},n,m_{a+1}\ldots \rangle%
\nonumber\\%
&&\label{5.11}%
\end{eqnarray}
This represents a generalization of the well-known Kramers-Heisenberg formula \cite{BrLfPt} for the scattering of a photon by a many-particle system consisting of atomic dipoles. The selected specific matrix element runs through all the possibilities when the incoming photon is absorbed by any $a$-th atom and the outgoing photon is emitted by any $b$-th atom of the ensemble, including the possible coincidence $a=b$. The initial atomic state is given by $|g\rangle\equiv|m_1,\ldots,m_N\rangle$ and the final atomic state is given by $|g'\rangle\equiv|m'_1,\ldots,m'_N\rangle$.

The projected resolvent operator contributing to Eq. (\ref{5.9}) is defined in the Hilbert subspace of a finite size with dimension $d_eN\,d_g^{N-1}$, where $d_e$ is the degeneracy of the atomic excited state and $d_g$ is the degeneracy of its ground state. The matrix elements of the operator $\tilde{\hat{R}}(E)$ can be linked with the $N$-particle causal Green's function of the atomic subsystem via the following Laplace-type integral transformation:
\begin{eqnarray}
\lefteqn{\hspace{-2cm}\langle\ldots m'_{b-1},n',m'_{b+1}\ldots |\tilde{\hat{R}}(E)%
|\ldots m_{a-1},n,m_{a+1}\ldots \rangle%
\,\times\,\delta\left(\mathbf{r}'_1-\mathbf{r}_1\right)\ldots\delta\left(\mathbf{r}'_b-\mathbf{r}_b\right)\ldots%
\delta\left(\mathbf{r}'_a-\mathbf{r}_a\right)\ldots\delta\left(\mathbf{r}'_N-\mathbf{r}_N\right)}%
\nonumber\\%
&&=-\frac{i}{\hbar}\int_0^{\infty}dt\,\,\exp\left[+\frac{i}{\hbar}E\,t\right]\;%
G^{(N)}\!\!\left(1',t;\ldots ;\!b',t;\ldots ;\!N',t|1,0;\ldots ;\!a,t;\ldots;\!N,0\right)%
\label{5.12}%
\end{eqnarray}
where in the right hand side we denoted $j=m_j,\mathbf{r}_j$ (for $j\neq a$) and $j'=m'_j,\mathbf{r}'_j$ (for $j'\neq b$), and for specific atoms $a=n,\mathbf{r}_a$ and $b'=n',\mathbf{r}'_b$. Here $\mathbf{r}_j=\mathbf{r}'_j$, for any $j=1\div N$, is the spatial location of $j$-th atom, which is assumed to be conserved in the scattering process. This circumstance is expressed by the sequence of $\delta$-functions in Eq. (\ref{5.12}). The causal Green's function is given by the vacuum expectation value of the following chronologically ($T$)-ordered product of atomic second quantized $\Psi$ operators introduced in the Heisenberg representation
\begin{eqnarray}
\lefteqn{G^{(N)}\left(1',t'_1;\ldots ;b',t'_b;\ldots ;N',t'_N|1,t_1;\ldots ;a,t_a;\ldots;N,t_N\right)}%
\nonumber\\%
&&=\langle T\, \Psi_{m'_1}(\mathbf{r}'_1,t'_1)\ldots\Psi_{n'}(\mathbf{r}'_b,t'_b)\ldots%
\Psi_{m'_N}(\mathbf{r}'_N,t'_N)\;%
\Psi_{m_N}^{\dagger}(\mathbf{r}_N,t_N)\ldots%
\Psi_{n}^{\dagger}(\mathbf{r}_a,t_a)\ldots\Psi_{m_1}^{\dagger}(\mathbf{r}_1,t_1)\rangle,%
\label{5.13}%
\end{eqnarray}
where $\Psi_{\ldots}(\ldots)$ and $\Psi_{\ldots}^{\dagger}(\ldots)$ are respectively the annihilation and creation operators for an atom in a particular state and at a particular coordinate. All the creation operators in this product contribute to the transform (\ref{5.12}) while being considered at an initial time "$0$" and all the annihilation operators are considered at a later time $t>0$. We can ignore effects of either bosonic or fermionic quantum statistics associated with the atomic subsystem as far as we neglect any possible overlap in atomic locations and consider the atomic dipoles positioned as classical objects randomly distributed in space. We ordered operators in Eq. (\ref{5.13}) in such a way that in the fermionic case (under the anticommutation rule) and without interaction it generates the product of independent individual single-particle Green's functions associated with each atom and with positive overall sign.

The perturbation theory expansion of the $N$-particle Green's function (\ref{5.13}) can be visualized by the series of  diagrams in accordance with the standard rules of the vacuum diagram technique; see \cite{BrLfPt}. In this case only $T$-ordered products of the second quantized operators can be generated by expansion of the evolution operator, which we originally marked by a "-" sign. For the sake of simplicity we omit this unnecessary mark in our notation below. After rearrangement, the diagram expansion can be transformed to the following generalized Dyson equation:
\begin{equation}
\scalebox{1.0}{\includegraphics*{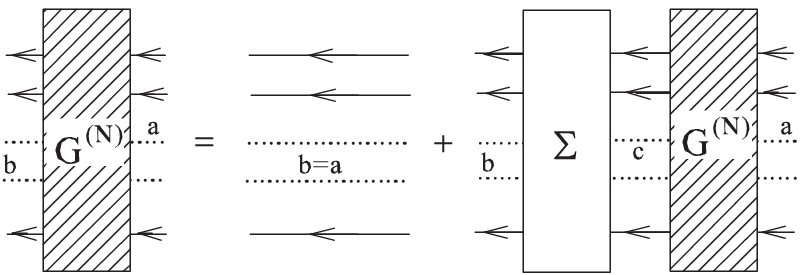}}%
\label{5.14}
\end{equation}
where the long straight lines with arrows correspond to individual causal single-particle Green's functions of each atom in the ensemble such that the first term in the right hand side performs the graph image of an undisturbed $N$-particle propagator (\ref{5.13}). The dashed block edged by short lines with arrows is the complete collective $N$-particle Green's function dressed by the interaction. In each diagram block of equation (\ref{5.14}) we indicated by $a,b,c$ (running from $1$ to $N$) the presence of one specific input as well as an output line associated with the single excited state equally shared by all the atoms of the ensemble. The sum of tight diagrams, which cannot be reduced to the product of lower order contributions linked by undisturbed atomic propagators, builds a block of the so-called self-energy part "$\Sigma$". The diagram equation (\ref{5.14}) in its analytical form performs an integral equation for $G^{(N)}(\ldots)$. With its transformation to the energy representation (\ref{5.12}) the integral equation can be recomposed to the set of algebraic equations for the matrix of the projected resolvent operator $\tilde{\hat{R}}(E)$, which can be further numerically solved. The crucial requirement for this is the knowledge of the self-energy part (a quasi-energy operator acting in the atomic subspace), which as we show below can be approximated by the lower orders in expansion of the perturbation theory.

\subsubsection{The self-energy}

In the lower order of perturbation theory the self-energy part consists of two contributions having single-particle and double-particle structures. Each specific line in the graph equation (\ref{5.14}) associated with excitation of an $a$-th atom generates the following irreducible self-energy diagram:
\begin{eqnarray}
\lefteqn{\scalebox{1.0}{\includegraphics*{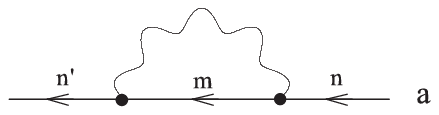}}}
\nonumber\\%
&&\Rightarrow\sum_{m}\int\frac{d\omega}{2\pi}\, d^{\mu}_{n'm}\,d^{\nu}_{mn}\,%
iD^{(E,c)}_{\mu\nu}(\mathbf{0},\omega)%
\;\frac{1}{E-\hbar\omega-E_m+i0}%
\ \equiv\ \Sigma^{(a)}_{n'n}(E),%
\label{5.15}
\end{eqnarray}
which is analytically decoded with applying transformation (\ref{5.12}) in the energy representation. Here the internal wavy line expresses the causal-type vacuum Green's function of the chronologically ordered polarization components of the field operators
\begin{equation}
iD^{(E,c)}_{\mu\nu}(\mathbf{R},\tau)\ =\ \left\langle T\hat{E}_{\mu}(\mathbf{r}',t')\,%
\hat{E}_{\nu}(\mathbf{r},t)\right\rangle,%
\label{5.16}%
\end{equation}
which depends only on the difference of its arguments $\mathbf{R}=\mathbf{r}'-\mathbf{r}$ and $\tau=t'-t$ and has the following Fourier image:
\begin{eqnarray}
D^{(E,c)}_{\mu\nu}(\mathbf{R},\omega)&=&\int_{-\infty}^{\infty} d\tau\,\mathrm{e}^{i\omega\tau}%
D^{(E,c)}_{\mu\nu}(\mathbf{R},\tau)%
\nonumber\\%
&=&-\hbar\frac{|\omega|^3}{c^3}\left\{i\frac{2}{3}h^{(1)}_0\left(\frac{|\omega|}{c}R\right)\delta_{\mu\nu}%
\ +\ \left[\frac{X_{\mu}X_{\nu}}{R^2}-\frac{1}{3}\delta_{\mu\nu}\right]%
ih^{(1)}_2\left(\frac{|\omega|}{c}R\right)\right\};%
\label{5.17}%
\end{eqnarray}
see \cite{BrLfPt}. Here $h^{(1)}_L(\ldots)$ with $L=0,2$ are the spherical Hankel functions of the first kind. As follows from Eq. (\ref{5.15}) the Green's function (\ref{5.17}) contributes in that expression in a self-interacting form with spatial argument $\mathbf{R}\to\mathbf{0}$. As a consequence, the expression (\ref{5.15}) tends to infinity in the limit $R\to 0$ and the integration over $\omega$ diverges. Part of the divergency should be associated with the longitudinal self-contact interaction. This is an artificial consequence of the dipole gauge and, as was explained in the related remark above, in the exact Heisenberg dynamics of the atomic dipole it is compensated by the dipolar self-action contra-term. The residual divergency has a radiative nature and demonstrates the general incorrectness of the Lamb-shift calculation in assumptions of the long-wavelength dipole approximation. Finally we follow the standard renormalization rule,
\begin{eqnarray}
\Sigma^{(a)}_{n'n}(E)&=&\Sigma^{(a)}(E)\delta_{n'n},%
\nonumber\\%
\Sigma^{(a)}(E)&\approx&\Sigma^{(a)}(\hbar\omega_0)=\hbar\Delta_{\mathrm{L}}-i\hbar\frac{\gamma}{2},%
\label{5.18}%
\end{eqnarray}
where $\Delta_{\mathrm{L}}\to\infty$ is incorporated into the physical energy of the atomic state. To introduce the single-atom natural decay rate $\gamma$ we applied the Wigner-Weiskopf pole approximation and replaced the energy $E=\hbar\omega_k+E_g$ by its near resonance mean estimate $E\approx E_n$ with the assumption that the atomic ground state is the zero-energy level such that $E_g=\sum_{j=1}^{N}E_{m_j}=E_m=0$. Then the energy of the excited state is given by $E_n=\hbar\omega_0$, where $\omega_0$ is the transition frequency.

In the lower order of perturbation theory, the double-particle contribution to the self-energy part consists of two complementary diagrams:
\begin{eqnarray}
\lefteqn{\scalebox{1.0}{\includegraphics*{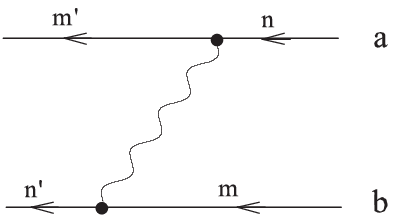}}}
\nonumber\\%
&&\hspace{-0.5cm}\Rightarrow\int\frac{d\omega}{2\pi}\, d^{\mu}_{n'm}\,d^{\nu}_{m'n}\,%
iD^{(E,c)}_{\mu\nu}(\mathbf{R}_{ab},\omega)%
\;\frac{1}{E-\hbar\omega-E_m-E_{m'}+i0}%
\ \equiv\ \Sigma^{(ab+)}_{m'n';nm}(E)
\label{5.19}
\end{eqnarray}
and
\begin{eqnarray}
\lefteqn{\scalebox{1.0}{\includegraphics*{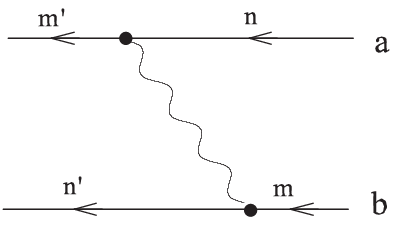}}}
\nonumber\\%
&&\hspace{-0.5cm}\Rightarrow\int\frac{d\omega}{2\pi}\, d^{\mu}_{n'm}\,d^{\nu}_{m'n}\,%
iD^{(E,c)}_{\mu\nu}(\mathbf{R}_{ab},\omega)%
\;\frac{1}{E+\hbar\omega-E_n-E_{n'}+i0}%
\ \equiv\ \Sigma^{(ab-)}_{m'n';nm}(E),
\label{5.20}
\end{eqnarray}
which are responsible for the excitation transfer from atom $a$ to atom $b$ separated by a distance $R_{ab}$. The vector components of the dipole matrix elements $d^{\nu}_{m'n}$ and $d^{\mu}_{n'm}$ are related with atoms $a$ and $b$ respectively. In the pole approximation $E\approx E_n=\hbar\omega_0$ the $\delta$-function features dominate in the spectral integrals (\ref{5.19}) and (\ref{5.20}) and the sum of both the terms leads to
\begin{eqnarray}
\Sigma^{(ab)}_{m'n';nm}(E)&\approx&\Sigma^{(ab+)}_{m'n';nm}(\hbar\omega_0)+%
\Sigma^{(ab-)}_{m'n';nm}(\hbar\omega_0)%
\ =\ \frac{1}{\hbar}\,d^{\mu}_{n'm}\,d^{\nu}_{m'n}\,D^{(E,c)}_{\mu\nu}(\mathbf{R}_{ab},\omega_0).%
\label{5.21}%
\end{eqnarray}
The derived expression has a clear physical meaning. For closely spaced atoms the real component of the double-particle contribution to the self-energy part reproduces the static interaction of the atomic dipoles. Its imaginary component is responsible for formation of cooperative dynamics of the excitation decay in the radiative process. For a dilute system, when atoms are separated by a substantial radiation zone, this term contributes to the amplitude of subsequent scattering on a pair of well-separated atoms, which weakly decreases with the interatomic separation. In a dense sample with small interatomic separations the strong coupling of the atomic dipoles, described by these terms, makes the entire scattering process strongly dependent on any particular atomic configuration.

It is a challenging problem to further improve the self-energy part by taking into consideration the higher orders of the perturbation theory expansion. Here we only substantiate the validity and sufficiency of the lower order approximation for the considered configuration. The main physical reason for this is the weakness of the interaction. This justifies ignoring any deviation from free dynamics of the atomic variables on a short time scale associated with the light retardation on distances of a few wavelengths. That yields the main cooperation in the radiative dynamics among neighboring dipoles which can effectively interact via the static longitudinal electric field. The diagram (\ref{5.20}), in contrast with (\ref{5.19}), is mostly important for evaluation of the static interaction such that in this graph the field propagator preferably links the points with coincident times on atomic lines. As a consequence, the presence of such diagram fragments as a crossing or overlapping part of any irreducible diagrams in higher orders would make the overall contribution small and negligible just because the static dipole-dipole interaction only weakly affects the dipoles' dynamics during the short retardation time, which can be roughly estimated by the wave period $2\pi/\omega_0$. For the same reason we can ignore any vertex-type corrections to the above self-energy diagrams. Another part of the self-energy diagrams in higher orders can be associated with correction of the static interaction itself. If the atomic system is so dense that the atoms are separated by a distance comparable with the atomic size (much shorter than the radiation wavelength) then the description of the static interaction in the simplest dipole model would be inconsistent and insufficient. This correction is evidently ignorable for an atomic ensemble with a density of a few atoms in a volume scaled by the cubic radiation wavelength. In this case the higher order static corrections are negligible as far as the dipole-dipole interaction is essentially less than the internal transition energy. As we can finally see, for the considered atomic systems, the self-energy part is correctly reproducible by the introduced lower order contributions.

\subsubsection{Illustrative examples}

The ab-initio microscopic calculations of light scattering on many particle systems have been done in \cite{SKKH,SMLK,SKGGLK} and an equivalent approach of microscopic stochastic simulations of recurrent scattering has been recently developed in \cite{RuostJavn16,Javanainen16}. In \cite{SKKH} the comparative group of self-consistent and microscopic numerical simulations for light scattering on an ensemble of two-level $V$-type atoms revealed good agreement in the area of applicability of both the calculation schemes. There was considered the macroscopic system of atoms with zero angular momentum in the ground state $F_0=0$ and with unit angular momentum in the excited states, $F=1$. For the densities not higher than $n_0\lambdabar^3\sim 0.1$ the total cross section demonstrated practically identical spectral behavior. However at higher densities the speckled microscopic spectrum significantly deviated from its self-consistent counterpart. This is explained by the non-trivial structure of the resolvent poles sensitive to any particular configurations of atomic dipoles. The proximal dipoles cannot be considered as indistinguishable objects forming only coordinated cooperative dynamics such that any slight changes in the location of atoms can change dramatically the global susceptibility of the system.

\begin{figure}[tp]\center
{$\scalebox{0.5}{\includegraphics*{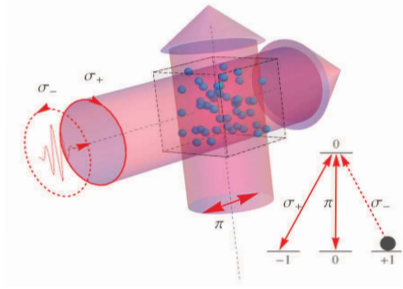}}$ }
\caption{(Color online) Excitation scheme of a dense and disordered atomic ensemble in the presence of control modes. The atoms have a total spin angular momentum $F_0=1$ in the ground state and $F=0$ in the excited state. A weak probe beam with $\sigma_-$ polarization propagates along the sample. The two control modes with equal Rabi frequencies $\Omega_c$ address the empty adjacent transitions with orthogonal polarizations $\sigma_+$ and $\pi$.}
\label{fig54}%
\end{figure}

The atomic ensembles consisting of two level atoms with a degenerate ground state is a natural and often discussed object in the literature since such systems physically follow from the main idea of the Dicke problem and associated super-radiance phenomenon, see \cite{BEMST}. Nevertheless in the context of a quantum interface and coherent control of the signal light the atomic systems with degenerate ground state are specifically interesting. One example of such system was considered in \cite{SKGGLK} aiming to test the capability of the EIT-based light storage protocol by using as few atoms as possible. The considered energy level diagram and excitation scheme are shown in Fig.~\ref{fig54}. There it was assumed a dense ensemble of the simplest $\Lambda$-type atoms with the minimal accessible number of quantum states, i.e. with angular momentum $F_0=1$ in the ground state and $F=0$ in the excited state. Initially, all the atoms populated only one Zeeman sublevel, $\{F_0=1,M_0=1\}$. Two coherent control modes were applied at the empty transitions and a weak and left-handed polarized signal probe propagated through the sample.

The main issue was how the effect of disorder, associated with the dense random distribution of atomic scatterers, can affect the EIT dynamics and make its spectral profile potentially sensitive to a specific atomic configuration. By evaluating the total cross-section via the optical theorem for densities, $n_0\lambdabar^3\sim 1$, in the self-energy diagrams (\ref{5.19}) and (\ref{5.20}) it was sufficient to keep the off-diagonal coupling only with the atoms separated by a distance of a few $\lambdabar$. That essentially reduced the number of inelastic scattering channels ($\emph{i.e.}$, Raman-type transitions repopulating atoms to other Zeeman states) coupled with the elastic one ($\emph{i.e.}$, saving atoms in their original Zeeman sublevel) and as a consequence the subspace dimension for the resolvent operator as well as the number of equations to be solved, which scaled as $d_e N d_g^{\mathrm{n}-1}$. Here "$\mathrm{n}-1$" corresponds to the effective number of the proximate neighbors, which mainly contributes in the recurrent diagrams responsible for "dressing" of an excited state propagator associated with any randomly selected atomic excitation in ensemble. This number was varying up and the entire calculation scheme with increasing "$\mathrm{n}$" showed rapidly self-converging such that in the limit of $\mathrm{n}\gg n_0\lambdabar^3$ it is expected to approach the exact result.

\begin{figure}[tp]\center
{$\scalebox{0.9}{\includegraphics*{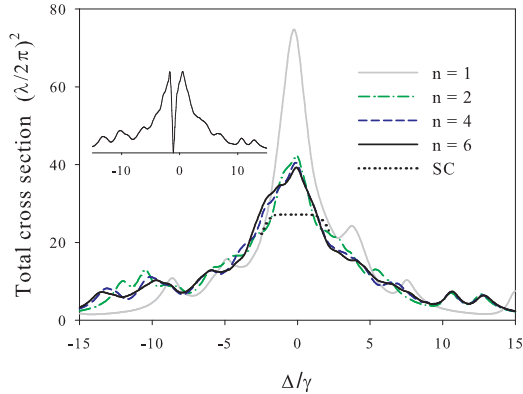}}$ }
\caption{(Color online) Total cross section for a single photon scattering on an ensemble consisting of fifty atoms randomly distributed with a density $n_0\lambdabar^3\sim 1$. The microscopic calculations have been performed for a particular configuration and for different numbers "$\mathrm{n}$" of proximate neighbors involved in the recurrent coupling, as described in the text. The inset shows the microscopic verification of EIT interaction with control pulses. The self-consistent (SC, dotted line) estimate of the cross section spectrum (black dotted) is scaled in accordance with the Babinet's principle and only the part of maximal absorption is shown.}
\label{fig55}%
\end{figure}

The spectral behavior of the cross section is demonstrated in Fig.~\ref{fig55} for an ensemble consisting of fifty atoms with density $n_0\lambdabar^3\sim 1$ and for a number of proximate neighbors contributing in the recurrent coupling up to five. The calculations were done for a particular configuration and reproduce a randomly created quasi-energy spectral structure of the resolvent operator for the chosen configuration. As a reference dependence we have shown here a fragment of the transmission spectrum calculated in the self-consistent approximation near the resonance point and applied Babinet's principle for the total cross section estimate at the resonance. Both the calculation schemes are in agreement in their general behavior and the microscopic result has a clear signature of the local field correction being slightly asymmetric towards the red wing. The fact that the microscopic cross section is a bit larger at the point of the maximum indicates that the macroscopic description is not so straightforwardly applicable to a system of mesoscopic size. Indeed, the sample size $L={}^{3}\!\!\!{\surd}50\, \lambdabar$ comes to the region $\lambdabar < L<\lambda$, so it is not large enough to follow the Fraunhofer diffraction and the Babinet's benchmark. The EIT phenomenon can be involved in the microscopic calculation scheme by adding the relevant self-energy part in the atomic propagators of unoccupied states. The dip of the transparency window, shown in the inset, is only slightly sensitive to a particular atomic configuration that makes such a system potentially interesting for developing of memory units with mesoscopic size.

\section{Summary and outlooks}
In this review we have analyzed a number of recent accomplishments of experimental and theoretical efforts in the frontier research areas of atomic physics and quantum optics. The systems of trapped cold atoms makes a unique experimental platform for elaboration of quantum interfaces between light and matter subsystems with many practical implementations. Just a few decades ago, via operating with optical pumping technique, adjusted for experiments with atomic cells, the ground state spin dynamics performed the main observation channel for the system control and manipulation \cite{Happer}. At that time it had been an unaccessible option for elimination of any external perturbations such as collisions with a buffer gas, cell walls, etc. At the present level of experimental technical development, the atomic ensembles frozen below the Doppler limit give us natural access for mediating with practically undisturbed atomic multilevel structure and open various new mechanisms of coherent control of the scattering processes.

In the broadest sense, we are concerned with the realistic description of evolution of a particular physical system, taken in its entirety, as various physical parameters are varied. In this review we have mainly focused on the optical response of a cold atomic ensemble, and considered mesoscopic physical parameters such as the sample size, volume, atom density, optical depth, and spectral detuning from resonance excitation with an external light source. These, in turn, are effected by underlying microscopic interactions, including radiative and near field dipole-dipole interaction inside the atomic cluster. At the highest dilutions, the sample is accurately considered as a collection of individual noninteracting atoms driven by external coherent fields and re-scattering the near resonant trapped probe radiation. As discussed in this review, even dilute regime reveals a number of interesting physical phenomena and non-trivial manifestations of various coherent effects in mesoscopic systems. However, as parameters are varied and the atomic density evolves to higher levels, significant spatial disorder, quantum degeneracy, and other new physical properties emerge.

Among the large number of emergent qualities of interest is the development of superradiant and subradiant excitations associated with certain types of the collective atomic superposed and entangled states. Although superradiant Dicke states \cite{Dicke} for inverted systems have been observed and studied experimentally and theoretically by a large number of investigators, experimental research on single photon superradiance and the relatively fragile subradiant states in the optical domain has been more limited by experimental capabilities. Original consideration of such states in the optical domain was of high-density microscopic samples for which there are a number of atoms in a volume of radius $\lambdabar$. However, in cold atomic systems generation of superradiant and subradiant states is not limited to this case. For instance, recent research has shown that, through off-resonance optical excitation of an atomic system, a so-called timed-Dicke state may be created \cite{Scully,RSSCR,ScullySvidz09,ScullySvidz10}. Such a state is a short-lived superradiant excitation distributed through the sample as a whole, and demonstrates the possibility of many-body, but one-photon superradiance and a collective Lamb shift \cite{KaiserLukin16}. Further, the timed-Dicke state may be mixed through so-called Fano couplings, or through external control fields, into subradiant modes, which may then be experimentally observable \cite{BBPK,BiPiKa}. Additionally, it was recently shown that the diatomic molecular equivalent of subradiant and superradiant states, traditionally termed gerade and ungerade states in diatomic molecular physics, may be directly studied through spectroscopy of weakly bound pairs of atoms trapped in optical lattices \cite{TSTBCJ12,MMISMZ15,MMITSMZ15}.

Another example, where effects of coherency and cooperativity in the complex mesoscopic system attract significant research attention, has naturally appeared in the area of quantum information. The current tendencies in quantum optics follow the light-matter interface, measurements and manipulations with physical variables beyond the standard quantum limit, which is as challenging as technologically important for quantum metrology, information processing, and computing.  Entanglement and swapping of the microscopic and mesoscopic states, considered as elementary logic units of a quantum network, reveal many new and important physical issues for quantum electrodynamics of cold atoms and complex quantum systems. This challenging research area is only in the beginning and is a rapidly evolving branch of mesoscopic physics.

As a possible extension of the cold atom technique it seems very promising to integrate nanophotonic technologies with atomic systems existing in collective quantum states such as optical lattices or dense systems of trapped atoms. The recent experiments with a nanofiber trap, where atoms interact with the evanescent field transporting through a MOT, reported in \cite{GCADLPTSK12,GMNML15}, and experiments on Bragg scattering in one-dimensional systems, reported in \cite{PolzikAppel16,KuprLaurat16}, open a really promising new architecture for the light-matter quantum interface. However, many physical problems immediately arise with relevant description of the scattered modes and atomic system near the nanofiber performing an open cavity. This concerns the atomic energy structure distorted by the nanofiber, manifestation of the Purcell effect, modification of the radiation shift, and so on.  All these areas require further comprehensive development, including extensive experimental investigation and theoretical description of coherent light scattering via combining the microscopic and macroscopic approaches.

Degenerate states of matter give another natural example of a physical object where the scattering process should have a complete quantum description. After first observations of the Bose-Einstein condensate (BEC) in \cite{AEMWC95} and Fermi-gas of atoms in \cite{DeMarcoJin99} the optical properties of these intriguing macroscopic states of matter had attracted attention of many experimental and theoretical research groups. The coherent joint propagation of light and matter waves through a degenerate quantum gas existing in the BEC phase has been predicted and discussed in \cite{Politzer91} even before the condensate was created in a laboratory. Evident signatures of coherent and cooperative dynamics in light scattering on a condensate have been later observed in experiments \cite{STBSPK03,HKTOPM08}. The optical properties of the degenerate macroscopic systems such as a BEC or a Fermi-gas cannot be simply framed by a macroscopic Maxwell approach and requires a subsequent second-quantized quantum description for light scattering, coherent interaction and quantum interface \cite{Ezhova16}. This has a significant potential for improving the traditional interface protocols, for example based on EIT design \cite{HHDB99}, as well as for developing the optomechanical schemes using superfluidity as a main macroscopic quantum property of such matter state. As far as the atomic condensate scatters light strongly in a broad spectral domain it can effectively scatter the correlated light beams, normally created in a spontaneous down-conversion process and also characterized by a broad spectrum. This potentially valuable property could be used for developing optomechanical interface schemes between light and the condensate.

\section*{Acknowledgements}
We thank Maria Chekhova and Sergei Kulik for fruitful and supporting discussions during many conferences, which partly motivated us for preparation of this review. We specially thank Alexandra Sheremet and Leonid Gerasimov for productive discussions and for providing the graphical materials in sections \ref{Section IV} and \ref{Section V}. We also thank the participants of the Colloquium on Quantum Optics in St-Petersburg, and particularly Yuri Golubev, Alexander Troshin, and Ivan Sokolov, the seminar organizers, where the materials presented here have been reported and discussed in an excellent scientific atmosphere.

This work was supported by RFBR (Grant No. 15-02-01060) and by the National Science Foundation (Grant Nos. NSF-PHY- 0654226 and NSF-PHY-1068159). D.V.K. would like to acknowledge support from the External Fellowship Program of the Russian Quantum Center and the Foundation "Dynasty".

\appendix

\section[\hspace{1.5cm}Entanglement of the atom-field variables in the dipole gauge]{Entanglement of the atom-field variables in the dipole gauge}\label{Appendix.A}
Transformation of an atom-field system Hamiltonian to its dipole long-wavelength approximation (dipole gauge) is fairly done in \cite{ChTnDRGr}. It can be introduced as a unitary transformation of the original Hamiltonian
\begin{equation}
\hat{H}'=U\hat{H}U^{\dagger}
\label{a.1}
\end{equation}
Suppose we have only one single electron atom, with its center of mass (in our approximations coinciding with the nuclear position) located at the origin of the coordinate frame. Then the transformation operator is given by
\begin{eqnarray}
U&=&\exp\sum_{s}\left\{-i\left(\frac{2\pi}{{\cal V}\hbar\omega_s}\right)^{1/2}\left[(\hat{\textbf{d}}\!\cdot\!\textbf{e}_s)\,a_s+(\hat{\textbf{d}}\!\cdot\!\textbf{e}^*_s)\,a_s^{\dagger}\right]\right\}%
\ =\ \exp\left[-\frac{i}{\hbar c}\hat{\textbf{d}}\!\cdot\!\hat{\mathbf{A}}(\mathbf{0})\right]%
\label{a.2}
\end{eqnarray}
where "$s$" specifies any field mode, which can have complex polarization in the general case. The transformed Hamiltonian is given by the following expansion
\begin{eqnarray}
\hat{H}'&=&\sum_s\hbar\omega_s\,a^{\dagger}_sa_s + \hat{H}_{\mathrm{dip}}-\hat{\mathbf{d}}\hat{\mathbf{E}}(\mathbf{0})+\hat{H}_{\mathrm{self}}%
\nonumber\\%
\hat{H}_{\mathrm{dip}}&=&\frac{\hat{\mathbf{p}}^2}{2m_e}+\hat{V}_{\mathrm{Cl}}(\mathbf{r})%
\label{a.3}%
\end{eqnarray}
where $\hat{\mathbf{p}}=-i\hbar\nabla$ is the operator of kinetic momentum of the atomic electron and $\hat{V}_{\mathrm{Cl}}(\mathbf{r})$ is the operator of its Coulomb interaction with the nuclear and $m_e$ is the electronic mass. Other terms are explained and commented on in the main text, see section \ref{Section II.A}.

The subtle point is that the Hamiltonian $\hat{H}_{\mathrm{dip}}$ does not coincide with the transformed atomic Hamiltonian i.e.
\begin{eqnarray}
\hat{H}_{\mathrm{dip}}&\neq &\hat{H}_{\mathrm{atom}}'=U\hat{H}_{\mathrm{atom}}U^{\dagger}%
\nonumber\\%
\hat{H}_{\mathrm{atom}}&=&\frac{\hat{\mathbf{p}}^2}{2m_e}+\hat{V}_{\mathrm{Cl}}(\mathbf{r})%
\label{a.4}%
\end{eqnarray}
where again $\hat{\mathbf{p}}=-i\hbar\nabla$ but has another physical meaning of generalized momentum. The canonical variables of the electron obey the following transformations
\begin{eqnarray}
\mathbf{r}'&=&U\mathbf{r}U^{\dagger}\ =\ \mathbf{r}%
\nonumber\\%
\hat{\mathbf{p}}'&=&U\hat{\mathbf{p}}U^{\dagger}\ =\ \hat{\mathbf{p}}+\frac{e}{c}\hat{\mathbf{A}}(\mathbf{0})%
\label{a.5}%
\end{eqnarray}
such that the transformed generalized momentum is displaced on a potential of the field at the point of the dipole's location.

As a consequence, the atomic dipole operator of itself $\hat{\mathbf{d}}=e\mathbf{r}$ is not modified by the transformation but its frequency components are sensitive to which Hamiltonian either $\hat{H}_{\mathrm{dip}}$ or $\hat{H}_{\mathrm{atom}}'$ is associated with the atomic energy spectrum. In the basis set of any of these Hamiltonians, reduced to two level system and parameterized by transition frequency $\omega_0$, the frequency components of the atomic dipole are defined as
\begin{equation}
\hat{\mathbf{d}}\ =\ \sum_{n>m}\left[\mathbf{d}_{nm}\,|n\rangle\langle m|\,+\,\mathbf{d}_{mn}\,|m\rangle\langle n|\right]\ \equiv\ \hat{\mathbf{d}}^{(-)}+\hat{\mathbf{d}}^{(+)}%
\label{a.6}
\end{equation}
One can verify that for $\hat{H}_{\mathrm{dip}}$ and for $\hat{H}_{\mathrm{atom}}$ (before the transformation) the frequency components are given by identical mathematical operators
\begin{equation}
\hat{\mathbf{d}}^{(\pm)}\ =\ \frac{e}{2}\left[\mathbf{r}\pm\frac{\hbar}{m_e\omega_0}\nabla\right]
\label{a.7}%
\end{equation}
but, as pointed above, with different physical meaning for the right-hand side. The following relation is fulfilled
\begin{equation}
\left[\hat{\mathbf{d}}^{(\pm)}\right]_{\mathrm{dip}}\ =\ \left[\hat{\mathbf{d}}^{(\pm)}\right]'_{\mathrm{atom}}\mp i\frac{e^2}{2m_ec\,\omega_0}\hat{\mathbf{A}}(\mathbf{0})%
\label{a.8}%
\end{equation}
which emphasizes the difference between these quantities.

In the dipole approximation of the light-atom interaction just the definition in the left-hand side of (\ref{a.8}), associated with Hamiltonian $\hat{H}_{\mathrm{dip}}$, is assumed. As was commented in section \ref{Section V.A} and follows from (\ref{a.4}) and (\ref{a.5}) the dipole Hamiltonian actually entangles atomic and field variables such that it adds small but not negligible correction to the radiation part of atomic spectrum. This correction can make difficult the renormalization of that contribution to the radiation (Lamb) shift, which is associated with the non-relativistic coupling of the atomic electron with long-wavelength field modes. The difference between the Hamiltonians can be also important for precise resolution of the atomic and field variables in the dense multiatomic configuration with an aim towards describing them in the standard definitions of the macroscopic Maxwell approach.

Let us also point out that the dipole gauge has one quite important advantage in comparison with the original Coulomb gauge. As can be verified for the configuration consisting of many atoms, considered as a globally neutral charge particles, the description of them as a system of dipoles eliminates a fictitious non-retarding part of the Coulomb interaction between those atoms, which are separated by a distance of the radiation zone.

\section[\hspace{1.5cm}The basic Green's functions of the Keldysh's diagram approach]{The basic Green's functions of the Keldysh's diagram approach}\label{Appendix.B}
Here we give a set of the vacuum-type Green's functions for the field subsystem and unperturbed Green's functions for an atomic subsystem, which are defined by Eqs. (\ref{2.13}) and (\ref{2.14}) respectively.
\subsection[\hspace{1.5cm}Field subsystem]{Field subsystem}
\noindent Each of the functions $D_{\mu_1\mu_2}^{(\sigma_1\sigma_2)}(\mathbf{r}_1,t_1;\mathbf{r}_2,t_2)$ depends only on the difference of its arguments $\mathbf{R}=\mathbf{r}_1-\mathbf{r}_2$ and $\tau=t_1-t_2$. Then they can be expanded in their frequency component parts
\begin{equation}
D_{\mu_1\mu_2}^{(\sigma_1\sigma_2)}(\mathbf{R},\tau)\ =\ D_{\mu_1\mu_2;+}^{(\sigma_1\sigma_2)}(\mathbf{R},\tau)\;+\;D_{\mu_1\mu_2;-}^{(\sigma_1\sigma_2)}(\mathbf{R},\tau)%
\label{b.1}
\end{equation}
With substituting (\ref{2.3}) (in the interaction representation) in (\ref{2.13}) one straightforwardly gets for positive frequency component of the causal function
\begin{eqnarray}
iD_{\mu_1\mu_2;+}^{(--)}(\mathbf{R},\tau)&=&\left[A_{\mu_1}^{(0,+)}(\mathbf{r}_1,t_1),A_{\mu_2}^{(0,-)}(\mathbf{r}_2,t_2)\right]\,\theta(\tau)%
\nonumber\\%
&=& 2\pi\hbar c\int\frac{d^3k}{(2\pi)^3}\frac{1}{k}\,\mathrm{e}^{-i\omega_k\tau+i\mathbf{k}\cdot\mathbf{R}}\,%
\left[\delta_{\mu_1\mu_2}-\frac{k_{\mu_1}k_{\mu_2}}{k^2}\right]\theta(\tau)%
\label{b.2}
\end{eqnarray}
where $\omega_k=c\, k$. Its negative frequency component is given by transposition of the positive one
\begin{equation}
D_{\mu_1\mu_2;-}^{(--)}(\mathbf{R},\tau)\ =\ D_{\mu_2\mu_1;+}^{(--)}(-\mathbf{R},-\tau)%
\label{b.3}
\end{equation}
Similar relations for the anti-causal function read
\begin{eqnarray}
iD_{\mu_1\mu_2;+}^{(++)}(\mathbf{R},\tau)&=&\left[A_{\mu_1}^{(0,+)}(\mathbf{r}_1,t_1),A_{\mu_2}^{(0,-)}(\mathbf{r}_2,t_2)\right]\,\theta(-\tau)%
\nonumber\\%
&=& 2\pi\hbar c\int\frac{d^3k}{(2\pi)^3}\frac{1}{k}\,\mathrm{e}^{-i\omega_k\tau+i\mathbf{k}\cdot\mathbf{R}}\,%
\left[\delta_{\mu_1\mu_2}-\frac{k_{\mu_1}k_{\mu_2}}{k^2}\right]\theta(-\tau)%
\label{b.4}
\end{eqnarray}
and
\begin{equation}
D_{\mu_1\mu_2;-}^{(++)}(\mathbf{R},\tau)\ =\ D_{\mu_2\mu_1;+}^{(++)}(-\mathbf{R},-\tau)%
\label{b.5}
\end{equation}
Two other functions have either positive or negative components only, such that
\begin{eqnarray}
iD_{\mu_1\mu_2;+}^{(+-)}(\mathbf{R},\tau)&=&iD_{\mu_2\mu_1;-}^{(-+)}(-\mathbf{R},-\tau)%
\nonumber\\%
&=& 2\pi\hbar c\int\frac{d^3k}{(2\pi)^3}\frac{1}{k}\,\mathrm{e}^{-i\omega_k\tau+i\mathbf{k}\cdot\mathbf{R}}\,%
\left[\delta_{\mu_1\mu_2}-\frac{k_{\mu_1}k_{\mu_2}}{k^2}\right]%
\label{b.6}%
\end{eqnarray}
and
\begin{equation}
D_{\mu_1\mu_2;-}^{(+-)}(\mathbf{R},\tau)\ =\ D_{\mu_2\mu_1;+}^{(-+)}(\mathbf{R},\tau)=0%
\label{b.7}%
\end{equation}
which is a direct consequence of their definition (\ref{2.13}).

The retarded type photon propagator has its positive frequency component coinciding with $D_{\mu_1\mu_2;+}^{(--)}(\mathbf{R},\tau)$ and negative component coinciding with $D_{\mu_1\mu_2;-}^{(++)}(\mathbf{R},\tau)$. Respectively the advanced type photon propagator has its positive component coinciding with $D_{\mu_1\mu_2;+}^{(++)}(\mathbf{R},\tau)$ and negative one coinciding with $D_{\mu_1\mu_2;-}^{(--)}(\mathbf{R},\tau)$.

\subsection[\hspace{1.5cm}Atomic subsystem]{Atomic subsystem}

As pointed out in the main text, the definition of the atomic Green's functions (\ref{2.14}) is based on assumption of Gaussian-type factorization for the expectation value of any operators' product. In the macroscopic description this is justified by the central limit theorem and by consideration of the atomic gas as a non-degenerate quantum system. To define the non-disturbed Green's functions we assume the atomic system in equilibrium such that its semiclassical single particle density matrix $\rho_{mm}(\mathbf{p},\mathbf{r})$ (introduced in the Wigner representation) has only diagonal components, is time independent but can have spatial dependence in the case when external fields confine atoms with a trap.

Then the averaged time-ordered product of $\Psi$-operators in the ground state is given by
\begin{eqnarray}
iG_{m_1m_2}^{(--)}(\mathbf{r}_1,t_1;\mathbf{r}_2,t_2)&=&%
\left\langle T\,\hat{\Psi}_{m_1}^{(0)}\!(\mathbf{r}_1,t_1)\,%
\hat{\Psi}_{m_2}^{(0)\dagger}(\mathbf{r}_2,t_2)\right\rangle%
\nonumber\\%
&=&\delta_{m_1m_2}\int\!\frac{d^3p}{(2\pi)^3}\exp\left[\frac{i}{\hbar}\mathbf{p}\cdot(\mathbf{r}_1-\mathbf{r}_2)%
-\frac{i}{\hbar}\left(\frac{p^2}{2m}+E_{m_1}\right)(t_1-t_2)\right]%
\nonumber\\%
&&\times\;\left[\theta(t_1-t_2)\;\pm\;\rho_{m_1m_1}\left(\mathbf{p},\frac{\mathbf{r}_1+\mathbf{r}_2}{2}\right)\right]%
\label{b.8}%
\end{eqnarray}
Here the expansion $\mathbf{r}_1-\mathbf{r}_2$ in the right-hand side is limited by a distance less than the scale of spatial inhomogeneity associated with the spatial dependence of the density matrix. The internal sign either $+$ or $-$ depends on statistics, these being either Bosonic or Fermionic respectively.

The time-antiordered product can be expressed via Hermitian conjugation of the time-ordered function
\begin{eqnarray}
iG_{m_1m_2}^{(++)}(\mathbf{r}_1,t_1;\mathbf{r}_2,t_2)&=&%
\left\langle \tilde{T}\,\hat{\Psi}_{m_1}^{(0)}\!(\mathbf{r}_1,t_1)\,%
\hat{\Psi}_{m_2}^{(0)\dagger}(\mathbf{r}_2,t_2)\right\rangle%
\ =\ -iG_{m_2m_1}^{(++)\ast}(\mathbf{r}_2,t_2;\mathbf{r}_1,t_1)%
\label{b.9}%
\end{eqnarray}
The remaining Green's functions are given by
\begin{eqnarray}
iG_{m_1m_2}^{(+-)}(\mathbf{r}_1,t_1;\mathbf{r}_2,t_2)&=&%
\left\langle \hat{\Psi}_{m_1}^{(0)}\!(\mathbf{r}_1,t_1)\,%
\hat{\Psi}_{m_2}^{(0)\dagger}(\mathbf{r}_2,t_2)\right\rangle%
\nonumber\\%
&=&\delta_{m_1m_2}\int\!\frac{d^3p}{(2\pi)^3}\exp\left[\frac{i}{\hbar}\mathbf{p}\cdot(\mathbf{r}_1-\mathbf{r}_2)%
-\frac{i}{\hbar}\left(\frac{p^2}{2m}+E_{m_1}\right)(t_1-t_2)\right]%
\nonumber\\%
&&\times\;\left[1\;\pm\;\rho_{m_1m_1}\left(\mathbf{p},\frac{\mathbf{r}_1+\mathbf{r}_2}{2}\right)\right]%
\label{b.10}%
\end{eqnarray}
and
\begin{eqnarray}
iG_{m_1m_2}^{(-+)}(\mathbf{r}_1,t_1;\mathbf{r}_2,t_2)&=&%
\left\langle \hat{\Psi}_{m_2}^{(0)\dagger}(\mathbf{r}_2,t_2)\,%
\hat{\Psi}_{m_1}^{(0)}\!(\mathbf{r}_1,t_1)\right\rangle%
\nonumber\\%
&=&\pm\,\delta_{m_1m_2}\int\!\frac{d^3p}{(2\pi)^3}\exp\left[\frac{i}{\hbar}\mathbf{p}\cdot(\mathbf{r}_1-\mathbf{r}_2)%
-\frac{i}{\hbar}\left(\frac{p^2}{2m}+E_{m_1}\right)(t_1-t_2)\right]%
\nonumber\\%
&&\times\;\rho_{m_1m_1}\left(\mathbf{p},\frac{\mathbf{r}_1+\mathbf{r}_2}{2}\right)%
\label{b.11}%
\end{eqnarray}
and they are specifically important for the macroscopic diagram approach

For excited atomic states we can assume that initially there is neither population of these states nor their coherent coupling with the ground states. That means that their non-disturbed Green's functions are reproduced by similar expressions (\ref{b.8})-(\ref{b.11}) excepting the density matrix contributions and in particular $G_{12}^{(-+)}=0$ for the excited states.

In the interaction processes initiated by external driving fields the complicated dynamics of the atomic system strongly affect the above functions and additionally generate the coherent coupling between all the involved states. As example, as shown by Eq.(\ref{2.23}), the actual time evolved density matrix components contribute in the product of Heisenberg operators dressed by the interaction process. In some specific configurations the assumptions of the central limit theorem can be violated with emphasizing the cooperative phenomena in the system dynamics.

\section[\hspace{1.5cm}Matrix elements of atomic dipole and magnetic moment operators]{Matrix elements of atomic dipole and magnetic moment operators}\label{Appendix.C}
The atomic dipole moment is a real vector and physical observable, but it is convenient to express this quantity via its complex spherical components. The complex basis set of spherical unit vectors is given by
\begin{eqnarray}
\mathbf{e}_{0}&=&\mathbf{e}_{z}%
\nonumber\\
\mathbf{e}_{\pm 1}&=&\mp (\mathbf{e}_{x}\pm i\mathbf{e}_{y})/\sqrt{2}%
\label{c.1}
\end{eqnarray}
Then the spherical components of the dipole operator are given by its projections on these vectors
\begin{eqnarray}
d_q&=&\textbf{d}\cdot\textbf{e}_q%
\nonumber\\
d_{0}&=&d_{z}%
\nonumber\\
d_{\pm 1}&=&\mp (d_{x}\pm id_{y})/\sqrt{2}%
\label{c.2}
\end{eqnarray}
and their angular dependence is equivalent to $Y_{1q}(\theta,\phi)$ spherical functions.

The basic matrix element of the dipole operators is off-diagonal in the basis of the ground and excited atomic states specified by the quantum numbers of total angular momentum and its projection
\begin{equation}
\left(\textbf{d}\cdot\textbf{e}_q\right)_{nm}\ \equiv\ \langle F,M|d_q|F_0,M_0\rangle%
\label{c.3}
\end{equation}
The key step in calculation of this matrix element is in its interpretation in terms of the transformation rules. From the point of view of rotational transformation the action of the dipole operator on the ground state can be treated as
\begin{equation}
d_q|F_0,M_0\rangle\ \propto\ Y_{1q}(\theta,\phi)|F_0,M_0\rangle\sim |F_0,M_0;1,q\rangle%
\label{c.4}
\end{equation}
where by the latter relation we emphasize its similarity with the product of basis functions for any angular momentum composition. Then in accordance with the angular momentum algebra the projector
\begin{equation}
\langle F_0,1;F,M|F_0,M_0;1,q\rangle\ =\ C^{FM}_{F_0M_0\,1q},
\label{c.5}
\end{equation}
 where the bra-vector of $F,M$-state is assumed as a state of the integrated system, directly defines the Clebsch-Gordan coefficient, see \cite{LaLfIII,VMK}.  One can expect that the transition matrix element of an atomic dipole operator has to be proportional to the same Clebsch-Gordan coefficient and can be factorized in the following product
\begin{equation}
\langle F,M|d_q|F_0,M_0\rangle\ =\ (-)^{2\cdot 1}\frac{\langle F\!\parallel\! d_1\!\parallel\! F_0\rangle}{\sqrt{2F+1}}\;C^{FM}_{F_0M_0\,1q}%
\label{c.6}
\end{equation}
Such intuitive arguments can be supported by the Wigner-Eckart theorem, rigorously proven in the group theory, and the factor $\langle F\!\parallel\! d_1\!\parallel\! F_0\rangle$ is known as the reduced matrix element of the dipole (vector) operator of rank $k=1$, see \cite{LaLfIII,VMK}. The reader can straightforwardly generalize the obtained result for any irreducible tensor operator of rank $k$, which can be even half-integer for complex objects (spinors) represented in the real coordinate space. So we left the phase factor in Eq.(\ref{c.6}) in its general form.

The quantum numbers $F,M$ and $F_0,M_0$ are the values of the total angular momenta for the composition of electronic (orbital and spin) and nuclear (spin) states into a coupled hyperfine state.  In the decoupled basis the dipole operator does not affect the nuclear subsystem. In this case it is convenient to eliminate the nuclear subsystem according to the weakness of the hyperfine interaction with respect to the spin-orbital interaction. Omitting the derivation details we reproduce here the final result. The reduced matrix element of the dipole operator can be factorized as follows
\begin{eqnarray}
\langle F\!\parallel\! d_1\!\parallel\! F_0\rangle &=& (-)^{F_0+J+I-1}\left[(2F+1)(2F_0+1)\right]^{1/2}%
\;\left\{\begin{array}{ccc} S & I & F_0\\ F & 1 & J \end{array}\right\}%
\langle J\!\parallel\! d_1\!\parallel\! S\rangle %
\label{c.7}
\end{eqnarray}
where the factor $\langle J\parallel d_1\parallel S\rangle$ performs the reduced matrix element when the nuclear subsystem is completely ignored. Here $J$ is the total (spin and orbital) angular momentum of the excited state and $S\equiv J_0=1/2$ is the electronic spin coinciding with the total angular momentum of the ground state. The table-factor in curly braces is so called  $6j$-symbol appearing due to decomposition of the coupled state in the decoupled basis of the electronic and nuclear spin subsystems, see \cite{BrLfPt,VMK}.

The performed factorization of the transition matrix element for an atomic dipole operator allows us to express it by an experimentally measurable parameter, namely, by the spontaneous radiation decay rate, which is given by
\begin{equation}
\gamma_J\ =\ \frac{4\omega_{J0}^3}{3\hbar c^3}\;\frac{|\langle J\!\parallel\! d_1\!\parallel\! S\rangle|^2}{2J+1}\sim\gamma%
\label{c.8}
\end{equation}
where $\omega_{J0}$ is the transition frequency for either $J=1/2$ ($D_1$-line) or $J=3/2$ ($D_2$-line). In reality, in the case of not too highly excited alkali-metal atoms, the decay rate $\gamma_J$ is weakly sensitive to the fine structure splitting in the upper state such that it is practically the same for both the lines. Thus the expressions (\ref{c.6})-(\ref{c.8}) allow us to scale all the set of the transition matrix elements for an atomic dipole via one and well known experimental parameter $\sqrt{\gamma}$. Such scaling was applied in derivation of the repumping terms of the master equation (\ref{2.24})-(\ref{2.28})

A similar scheme is applicable for evaluation of the matrix elements for the magnetic moment, which we associate only with electronic spin $\mathbf{m}=2\mu_\mathrm{B}\mathbf{S}$, where $\mu_\mathrm{B}$ is the Bohr magneton (sign included) and "$2$" is approximately the gyromagnetic ratio for a single electron atom. Then the matrix element of any spherical component of the magnetic moment is given by
\begin{equation}
\langle F_0',M_0'|m_q|F_0,M_0\rangle\ =\ \frac{\langle F_0'\!\parallel\! m_1\!\parallel\! F_0\rangle}{\sqrt{2F_0'+1}}\;C^{F_0'M_0'}_{F_0M_0\,1q}
\label{c.9}
\end{equation}
where
\begin{eqnarray}
\langle F_0'\!\parallel\! m_1\!\parallel\! F_0\rangle &=& (-)^{F_0+S+I-1}\left[(2F_0'+1)(2F_0+1)\right]^{1/2}%
\;\left\{\begin{array}{ccc} S & I & F_0\\ F_0' & 1 & S \end{array}\right\}%
\langle S\!\parallel\! m_1\!\parallel\! S\rangle %
\label{c.10}
\end{eqnarray}
Here for a single electron atom $\langle S\!\parallel\! m_1\!\parallel\! S\rangle=\sqrt{6}\,\mu_\mathrm{B}$. But in the case when the system is driven by a microwave mode the matrix elements is more natural to scale by a specific transition such as the clock transition, which couples the Zeeman sublevels $F_0'=I+1/2,M_0'=0$ and $F_0=I-1/2,M_0=0$ in the ground state.





\end{document}